\documentclass[12pt]{article}
\pdfoutput=1
\usepackage{jheppub}

\usepackage{slashed}
\usepackage{color, verbatim}
\usepackage{latexsym}
\usepackage{epsfig}
\usepackage{amsmath,accents}

\newcommand*{\dt}[1]{%
  \accentset{\mbox{\bfseries .}}{#1}}
\newcommand*{\ddt}[1]{%
  \accentset{\mbox{\bfseries .\hspace{-0.25ex}.}}{#1}}
\usepackage{amssymb}
\usepackage{graphicx}
\usepackage{bm}
\usepackage{stackrel}

\usepackage[font={small}]{caption}

\usepackage{hyperref}
\usepackage{epstopdf}
\definecolor{myred}{rgb}{0.7, 0, 0}
\definecolor{myblue}{rgb}{0, 0, 0.7}
\definecolor{mygreen}{rgb}{0.04, 0.7, 0.5}
\hypersetup{colorlinks,citecolor=myred,linkcolor=myblue,urlcolor=myblue,linktocpage=true}

\makeatletter

\@addtoreset{equation}{section}
\makeatother

\newcommand{\be}{\begin{equation}}
\newcommand{\ee}{\end{equation}}
\newcommand{\bea}{\begin{eqnarray}}
\newcommand{\eea}{\end{eqnarray}}

\newcommand{\tr}{\operatorname{tr}}
\newcommand{\diag}{\operatorname{diag}}

\newcommand{\A}{\mathcal A}
\newcommand{\BB}{\mathcal B}

\newcommand{\K}{\mathcal K}

\newcommand{\Q}{\mathcal Q}

\newcommand{\Z}{\mathcal Z}

\newcommand{\GeV}{{\textrm{GeV}}}
\newcommand{\TeV}{{\textrm{TeV}}}

\newcommand{\Imaginary}{{\textrm{Im}}}
\newcommand{\PP}{{\mathcal{P}}}

\newcommand{\un}{{\textrm{un}}}
\newcommand{\res}{{\textrm{res}}}

\begin{document}

\thispagestyle{empty}

\begin{center}


\begin{center}

\vspace{.5cm}

{\Large\sc
Analytical Green's Functions for \\ \vspace{0.3cm}Continuum Spectra
}\\

\end{center}

\vspace{1.cm}

\textbf{
Eugenio Meg\'ias$^{\,a}$, Mariano Quir\'os$^{\,b}$
}\\

\vspace{.1cm}
${}^a\!\!$ {\em {Departamento de F\'{\i}sica At\'omica, Molecular y Nuclear and \\ Instituto Carlos I de F\'{\i}sica Te\'orica y Computacional, Universidad de Granada,\\ Avenida de Fuente Nueva s/n,  18071 Granada, Spain}}

${}^b\!\!$ {\em {Institut de F\'{\i}sica d'Altes Energies (IFAE) and\\ The Barcelona Institute of  Science and Technology (BIST),\\ Campus UAB, 08193 Bellaterra, Barcelona, Spain
}}

\end{center}

\vspace{0.8cm}

\centerline{\bf Abstract}
\vspace{2 mm}

\begin{quote}\small
 Green's functions with continuum spectra are a way of avoiding the strong bounds on new physics from the absence of new narrow resonances in experimental data. We model such a situation with a five-dimensional model with two branes along the extra dimension $z$, the ultraviolet (UV) and the infrared (IR) one, such that the metric between the UV and the IR brane is AdS$_5$, thus solving the hierarchy problem, and beyond the IR brane the metric is that of a linear dilaton model, which extends to $z\to\infty$. This simplified metric, which can be considered as an approximation of a more complicated (and smooth) one, leads to analytical Green's functions (with a mass  gap $m_g \sim \TeV$ and a continuum for $s > m_g^2$) which could then be easily incorporated in the experimental codes. The theory contains Standard Model gauge bosons in the bulk with Neumann boundary conditions in the UV brane. To cope with electroweak observables the theory is also endowed with an extra custodial gauge symmetry in the bulk, with gauge bosons with Dirichlet boundary conditions in the UV brane, and without zero (massless) modes. All Green's functions have analytical expressions and exhibit poles in the second Riemann sheet of the complex plane at $s=M_n^2-i M_n\Gamma_n$, denoting a discrete (infinite) set of broad resonances with masses ($M_n$) and widths ($\Gamma_n$). For gauge bosons with Neumann or Dirichlet boundary conditions, the masses and widths of resonances satisfy the (approximate) equation $s= - 4 m_g^2 \mathcal W_n^2[\pm (1+i)/4]$, where $\mathcal W_n$ is the $n$-th branch of the Lambert function.
\end{quote}

\vfill

\newpage

\tableofcontents

\newpage

\section{Introduction}
\label{sec:introduction}

The Standard Model (SM) of electroweak (EW) and strong interactions has been put on solid grounds by past and current experimental data, collected at e.g.~the Large Electron Positron (LEP) or the Large Hadron Collider (LHC)~\cite{ALEPH:2005ab,Olive:2016xmw}. In spite of the lack of clear deviations in particle physics experiments, there is a number of observational facts (dark matter, baryon asymmetry of the universe, $\dots$), and theoretical drawbacks (hierarchy problem, $\dots$) which cannot be coped by the SM and demand some ultraviolet (UV) completion of the theory. This has motivated a plethora of beyond the SM (BSM) models.

One of the most successful BSM models is the Randall-Sundrum (RS) model, proposed in 1999~\cite{Randall:1999ee}, where the hierarchy between the four-dimensional (4D) Planck scale $M_{\rm Pl}$ and the $\TeV$ scale is solved by means of a warped fifth dimension and two branes, the UV brane and the infrared (IR) brane. Associated with each SM field, the theory predicts a discrete spectrum made out of towers of composite discrete states known as Kaluza-Klein (KK) states, with masses in the TeV range. The elusiveness of isolated and narrow resonances in direct searches at the LHC~\cite{Sirunyan:2018ryr,Aaboud:2019roo}, led people to explore different solutions to the hierarchy problem that could escape present detection, as e.g.~the presence of broad resonances~\cite{Escribano:2021jne}. Some other exploring scenarios include the clockwork models, as well as their 5D continuum limit~\cite{Giudice:2017suc,Giudice:2017fmj}, the linear dilaton models (LDM)~\cite{Antoniadis:2011qw,Cox:2012ee} and the Little String theories~\cite{Antoniadis:2001sw}, which predict discrete spectra with a TeV mass gap and a mass separation between modes $\sim 30\, \GeV$.

A new scenario has been recently proposed, in which there appears a
TeV mass gap followed by a continuum of resonances heavier than the
mass
gap~\cite{Csaki:2018kxb,Megias:2019vdb,Megias:2020cpw,Megias:2021mgj,Csaki:2021gfm}. These
models are characterized by the absence of the IR boundary (replaced
by an admissible singularity of the metric)~\footnote{Identifying the
  IR brane with the manifold boundary gives rise to a different class
  of models with a different phenomenology, which has been explored in
  a number of papers,
  Refs.~\cite{Cabrer:2010si,Cabrer:2011fb,Cabrer:2011vu,Cabrer:2011mw,Cabrer:2011qb}.},
and the gapped continuum spectrum is present when the behavior of the
bulk potential of the stabilizing (canonically normalized) 5D scalar
field $\phi$ is given, in the limit $\phi \to \infty$, by the critical
behavior, $V(\phi) \propto \exp\left( \sqrt{\frac{2}{3 M_5^3}} \phi
\right)$, where $M_5$ is the 5D Planck scale~\cite{Cabrer:2009we}. The
behavior of the metric near the UV boundary is AdS$_5$, thus giving a
connection with the RS model in this regime, and allowing for a
holographic interpretation of the model and relating it with
unparticles~\cite{Georgi:2007ek,Georgi:2007si}. The model includes an
IR brane, where the Higgs (a mesonic doublet) is localized and which
triggers EW symmetry breaking, while the fifth dimension extends
beyond the IR brane till the singularity.

In a recent publication we have presented the results of the Green's
functions by focusing on the holographic method, which is convenient
for the computation of UV-to-UV brane
propagators~\cite{Megias:2019vdb}. In this paper we will use a
different approach based on the direct computation of the Green's
functions from the inhomogeneous equations of motion with appropriate
boundary conditions. We will use for that a simplified metric which
behaves like AdS$_5$ between the UV and the IR branes, and like the
metric of the LDM, between the IR brane and the singularity.
This will allow us to compute the Green's functions of fields
propagating at any point in the bulk. In addition, the model presented
in this work is simple enough to lead to analytical formulas while, at
the same time, sharing all the desirable features of a model which
leads to a gapped continuum spectrum as discussed in
Ref.~\cite{Megias:2019vdb}.

This theory can be considered as a modelization of 4D theories with continuum spectra and a mass gap, as can be the case of unparticle theories and Unhiggs theories, which share similar features and whose phenomenology has been extensively studied in a number of papers~\cite{Delgado:2007dx,Delgado:2008rq,Delgado:2008gj,Delgado:2008px,Stancato:2008mp,Falkowski:2008yr,Falkowski:2009uy,Bellazzini:2015cgj}. In order to protect EW precision observables we will need to introduce an extra custodial gauge symmetry. Although more realistic models can be introduced, we will just consider the simplest model~\cite{Agashe:2003zs} where the gauge symmetry for the EW sector in the bulk is $SU(2)_L\times SU(2)_R\times U(1)_{B-L}$ which breaks to $U(1)_Y$ by the UV boundary conditions, while it remains unbroken in the IR brane.

The outline of this paper is as follows. We introduce in Sec.~\ref{sec:general} the general formalism for the 5D action, including the gravitational background and the gauge sector which will be used throughout the rest of the paper. The Green's functions and the spectral functions for the massless gauge bosons are studied in Sec.~\ref{sec:gauge_bosons}. In particular the Green's functions in the complex $s$ plane are studied, which lead to complex poles in the second Riemann sheet, interpreted as broad resonances. The similar analysis for the SM massive gauge bosons $W,Z$, including the Green's functions, spectral functions and resonances, is postponed to Appendix~\ref{sec:massive_gauge_bosons}.
The computation of the Green's functions for gauge bosons with Dirichlet boundary condition in the UV brane is addressed in Sec.~\ref{sec:Dirichlet}. Finally we present in Sec.~\ref{sec:EWPO} the prediction of the model for the electroweak precision observables. We conclude with a discussion of our results, and an outlook toward future directions in Sec.~\ref{sec:conclusions}.

\section{The five-dimensional model}
\label{sec:general}

We consider a slice of 5D space-time between a brane at the value $y = y_0=0$ in proper coordinates, the UV brane, and an admissible singularity placed at $y = y_s$, a value which is determined dynamically. In addition, we will introduce an IR brane, at $y = y_1<y_s$, responsible for electroweak breaking, where we will assume the Higgs sector is localized. 

The 5D action of the model, including the stabilizing bulk scalar $\phi(x,y)$, with mass dimension $3/2$, reads as
\begin{eqnarray}
S &=& \int d^5x \sqrt{|\det g_{MN}|} \left[ -\frac{1}{2\kappa^2} R + \frac{1}{2} g^{MN}(\partial_M \phi)(\partial_N \phi) - V(\phi) \right]\nonumber \\
&-& \sum_{\alpha} \int_{B_\alpha} d^4x \sqrt{|\det \bar g_{\mu\nu}|} \lambda_\alpha(\phi)  
 -\frac{1}{\kappa^2} \int_{B_0} d^4x \sqrt{|\det \bar g_{\mu\nu}|} K_0  \,, \label{eq:action}
\end{eqnarray}
where $\kappa^2=1/(2M_5^3)$, with $M_5$ being the 5D Planck scale, $V(\phi)$ and $\lambda_\alpha(\phi)$ are the bulk and brane potentials of the scalar field $\phi$, and the index $\alpha=0 \; (\alpha=1)$ refers to the UV (IR)  brane. We will assume a $\mathbb Z_2$ symmetry ($y\to -y$) across the UV brane, which translates into boundary conditions on the fields, while we will impose matching conditions for bulk fields across the IR brane. Note that the fifth dimension continues beyond the IR brane until the singularity. The IR brane is responsible for the generation of the $\sim \TeV$ scale, and contains the brane Higgs potential which spontaneously breaks the electroweak symmetry, thus solving the hierarchy problem, as we will see. 

The parameter $\kappa^2$, can be traded by the parameter $N$ in the dual theory by the relation~\cite{Gubser:1999vj}
$
N^2\simeq \frac{8\pi^2\ell^3}{\kappa^2}\,,
$
where $\ell\equiv 1/k$ is a parameter of the order of the Planck length, which determines the value of the 5D curvature. 
The metric $g_{MN}$ is defined in proper coordinates by
\begin{eqnarray}
ds^2 &=&g_{MN}dx^M dx^N\equiv e^{-2A(y)} \eta_{\mu\nu} dx^\mu dx^\nu - dy^2 \,,  \label{eq:metric}  
\end{eqnarray}
so that in Eq.~(\ref{eq:action}) the 4D induced metric is $ \bar g_{\mu\nu}=e^{-2A(y)}\eta_{\mu\nu}$, where the Minkowski metric is given by $\eta_{\mu\nu} =\diag(1,-1,-1,-1)$. 
The last term in Eq.~(\ref{eq:action}) is the usual Gibbons-Hawking-York boundary term~\cite{York:1972sj,Gibbons:1976ue}, where $K_{0}$ is the extrinsic UV curvature. In terms of the metric of Eq.~(\ref{eq:metric}) the extrinsic curvature term reads as~\cite{Megias:2018sxv} $K_{0} = - 4 A^\prime(y_{0})$.

The equations of motion (EoM) read then as~\footnote{From here on the prime symbol $(\,{}^\prime\,)$ will stand for the derivative of a function with respect to its argument, and the dot symbol $(\dt{\phantom{a}})$ derivative only with respect to the conformal coordinate $z$ related to $y$ by $dy=e^{-A}dz$.}
\begin{eqnarray}
&&A^{\prime\prime}
= \frac{\kappa^2}{3} \phi^{\prime \, 2} + \frac{\kappa^2}{3} \sum_\alpha \lambda_\alpha(\phi) \delta(y - y_\alpha)  \,, \label{eq:eom1}\\
&&A^{\prime\, 2} 
= -\frac{\kappa^2}{6} V(\phi) + \frac{\kappa^2}{12} \phi^{\prime\, 2} \,,  \label{eq:eom2}\\
&&\phi^{\prime\prime} - 4 A^\prime \phi^\prime = V^\prime(\phi) + \sum_\alpha \lambda_\alpha^\prime(\phi) \delta(y - y_\alpha)  \,. \label{eq:eom3}
\end{eqnarray}
The EoM in the bulk can also be written in terms of the superpotential $W(\phi)$ as~\cite{DeWolfe:1999cp}
\begin{equation}
\phi^\prime(y) = \frac{1}{2} \frac{\partial W}{\partial \phi} \,, \qquad A^\prime(y) = \frac{\kappa^2}{6} W \,, \label{eq:phiA}
\end{equation}
and
\begin{equation}
V(\phi) = \frac{1}{8} \left( \frac{\partial W}{\partial \phi} \right)^2 - \frac{\kappa^2}{6} W^2(\phi) \,. \label{eq:V}
\end{equation}
Due to the $\mathbb Z_2$ symmetry across the UV brane, the localized terms impose the following boundary conditions in the UV
\begin{equation}
W(\phi(y_0)) = \lambda_0(\phi(y_0)) \,, \qquad W^\prime(\phi(y_0)) = \lambda_0^\prime(\phi(y_0)) \,. \label{eq:W_bcUV}
\end{equation}
In addition, the IR brane leads to the following jumping conditions
\begin{equation}
\Delta W(\phi(y_1)) = 2 \lambda_1(\phi(y_1)) \,, \qquad \Delta W^\prime(\phi(y_1)) = 2 \lambda_1^\prime(\phi(y_1))  \,, \label{eq:W_jcIR}
\end{equation}
where $\Delta X$ is the jump when crossing the brane. 

In the following we will impose continuity conditions for $W(\phi)$ and $W^\prime(\phi)$. Simple brane potentials satisfying the boundary conditions of Eq.~(\ref{eq:W_bcUV}), the jumping conditions of Eq.~(\ref{eq:W_jcIR}) with $\Delta W(\phi(y_1))  = \Delta W^\prime(\phi(y_1))=0$, and fixing dynamically the values $v_\alpha$ of $\phi$ at the branes, i.e.~$v_\alpha = \phi(y_\alpha)$, are given by
\be
\lambda_0(\phi)=W(\phi)+\frac{1}{2}\gamma_0(\phi-v_0)^2\,, \qquad  \lambda_1(\phi)=\frac{1}{2}\gamma_1(\phi-v_1)^2  \,.
\label{eq:lambda01}
\ee
This formalism has been extensively discussed in e.g.~Refs.~\cite{Csaki:2000zn,Csaki:2004ay}.

\subsection{The gravitational background}

We will provide in this section a particular realization of the gravitational background that will be used in the present work. As we will see, the model is simple enough to obtain analytical results for the Green's function in the forthcoming sections, but it contains all the ingredients needed to study the physics of the gapped continuum spectra, while solving the hierarchy problem \textit{\`a la} RS. 

A simple model solving the hierarchy problem and with a continuum spectrum was already characterized in Ref.~\cite{Megias:2019vdb} by the superpotential
\be
W(\phi)=\frac{6 k}{\kappa^2}\left(1+e^{\kappa \phi/\sqrt{3}}\right) \,,
\ee
or the corresponding bulk potential $V(\phi)$
\be
V(\phi)=-\frac{6k^2}{\kappa^2}\left[1+2e^{\kappa\phi/\sqrt{3}}+\frac{3}{4}e^{2\kappa\phi/\sqrt{3}}\right] \,.
\ee
After solving the EoM, the background value of the scalar field $\phi$ and warp factor $A(y)$ are given by
\be
\phi(y)=-\frac{\sqrt{3}}{\kappa}\log\left[ k(y_s-y) \right] \,, \qquad  A(y) = ky - \log\left( \frac{y_s - y}{y_s} \right) \,, \label{eq:Abar}
\ee
where $y_s$ is the location of the singularity in proper coordinates,
such that after fixing the value of the field $\phi$ in the branes, at $y=y_\alpha$, by brane potentials $\lambda_\alpha(\phi)$, dynamically fixing  $\phi(y_\alpha)=v_\alpha$, the brane and singularity distances are fixed by
\be
ky_s=e^{-\kappa v_0/\sqrt{3}},\quad ky_1=e^{-\kappa v_0/\sqrt{3}}-e^{-\kappa v_1/\sqrt{3}} \,.
\ee

As we have seen in Ref.~\cite{Megias:2019vdb}, the value of the gap and the warped $k$ scale at the IR brane
\be
\rho \equiv e^{-A(y_1)}k \,,  \label{eq:rhok}
\ee
of the TeV size, as required to solve the hierarchy problem, should be of the same order of magnitude. The further requirement of identification of both scales leads to the extra condition
\be
k(y_s-y_1)=1 \,,  \label{eq:ysy1}
\ee
which amounts to the choice $v_1=0$, which can be taken without loss of generality. In the rest of this paper the relation (\ref{eq:ysy1}) will be adopted. 

\begin{figure}[htb]
\centering
\includegraphics[width=8.5cm]{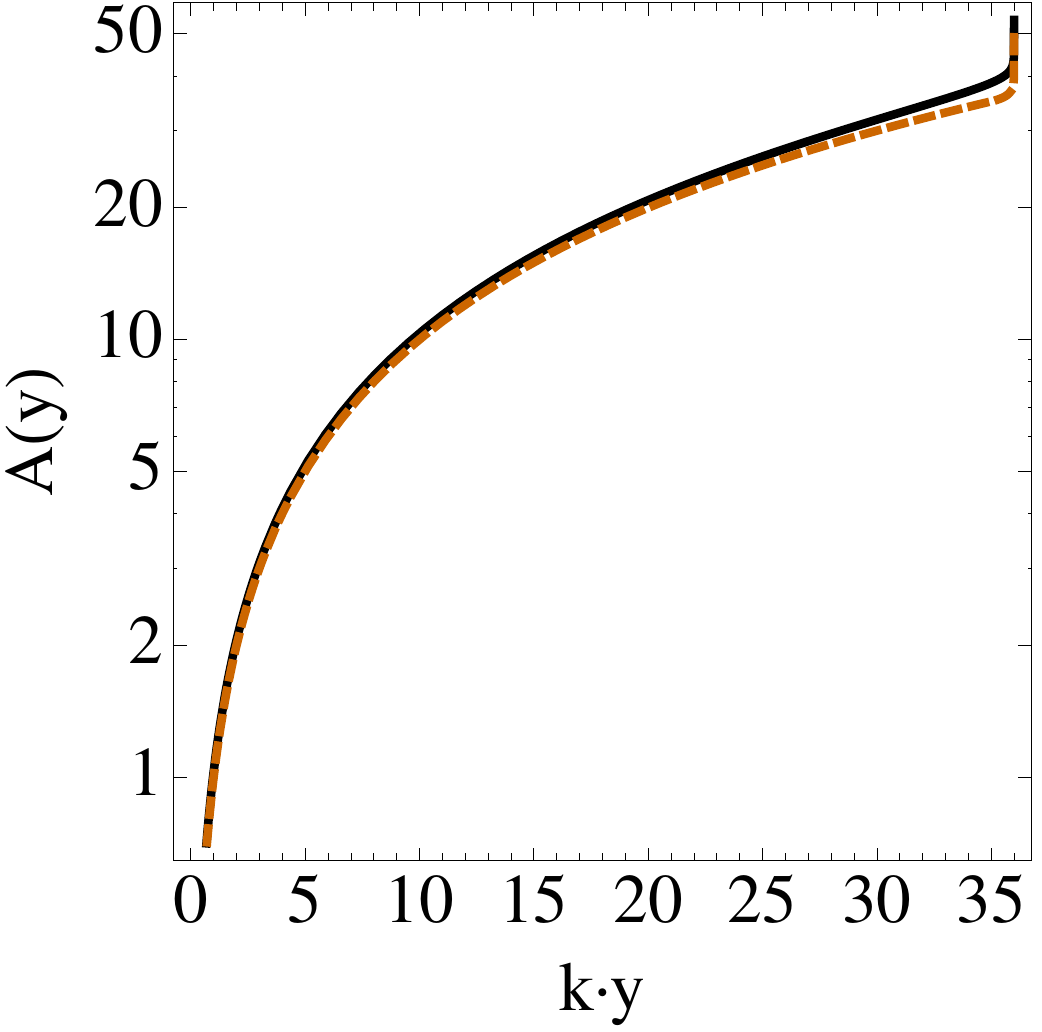}
\caption{\it The warp factor $A(y)$ for the exact solution (solid line) of Eq.~(\ref{eq:Abar}) and the approximate solution (dashed line) of Eq.~(\ref{eq:Aapp}). We have considered $ky_s=36$.}
\label{fig:Aapp}
\end{figure}

In particular the warp factor $A(y)$ behaves like the RS-metric between the UV and IR branes ($0<ky\lesssim ky_1$), $A(y)\simeq ky$, while it behaves like 
$A(y)\simeq -\log\left( \frac{y_s - y}{y_s} \right)$ between the IR brane and the singularity ($y\gtrsim y_1$). We can then approximate the exact metric by the approximate one
\be
A(y)\simeq ky\ \Theta(y_1-y)+\left[k y_1 - \log \left( k y_s - k y \right)\right]\Theta(y-y_1) \,,
\label{eq:Aapp}
\ee
where the step function is $\Theta(x)=1$ (0) for $x>0$ ($x\leq 0$). Comparison between both, the exact (\ref{eq:Abar}) and approximate (\ref{eq:Aapp}), metrics is done in the plot of Fig.~\ref{fig:Aapp}.
As we can see  the approximate solution of Eq.~(\ref{eq:Aapp}) behaves like the exact one in the relevant regions of the parameter space and moreover, as we will see in the next sections, it will allow for analytical solutions to the Green's functions which, otherwise, could only be computed numerically~\cite{Megias:2019vdb}.

The approximate metric of Eq.~(\ref{eq:Aapp}) can arise from the (approximate) superpotential 
\be
W(\phi)\simeq \frac{6 k}{\kappa^2}\ \Theta(v_1-\phi)+\frac{6 k}{\kappa^2}e^{\kappa(\phi-v_0)/\sqrt{3}}\ \Theta(\phi-v_1) \,,
\label{eq:Wapp}
\ee
which leads to the following profile for the field $\phi$
\be
\phi(y)\simeq v_0\ \Theta(y_1-y)+\left[ v_0-\frac{\sqrt{3}}{\kappa}\log(k y_s-ky)\right]\ \Theta(y-y_1) \,.
\label{eq:phiapp}
\ee

It can be useful to define the metric in conformally flat coordinates defined by
$
ds^2 = e^{-2A(z)} \left( \eta_{\mu\nu} dx^\mu dx^\nu - dz^2 \right) .
$
Comparison with Eq.~(\ref{eq:metric}) leads to the relation between conformal and proper coordinates, $z^\prime(y) = e^{A(y)}$, that can be solved to obtain
\begin{equation}
z(y) \simeq  \frac{e^{ky}}{k} \ \Theta(y_1 - y)  + \frac{e^{k y_1}}{k} \left[  1 - \log \left( k y_s - k y \right)  \right] \  \Theta(y - y_1) \,, \label{eq:zy}
\end{equation}
so that $z_s=\lim_{y\to y_s} z(y)=\infty$. Similarly one finds the following expression for the scalar field as a function of the conformal coordinate
\be
\phi(z) \simeq v_0\ \Theta(z_1-z)+\left[ v_0 + \frac{\sqrt{3}}{\kappa}\rho (z - z_1)  \right]\ \Theta(z-z_1) \,,
\label{eq:phiz}
\ee
where we have defined the quantities
\begin{equation}
z_0 \equiv 1/k  \qquad \textrm{and} \qquad z_1 \equiv 1/\rho \,. \label{eq:rho_z1}
\end{equation}
Note that we have fixed the integration constant by considering in the UV brane $z(y=0) = 1/k$. Finally,  one can write the warp factor in terms of the conformal coordinate, leading to
\begin{equation}
A(z) \simeq  \log(kz) \ \Theta(z_1 - z)  +  \left[ \log(k z_1) + \rho (z - z_1) \right] \ \Theta(z - z_1) \,.  \label{eq:Az}
\end{equation}
In this paper we will indistinctly use proper ($y$) or conformal ($z$) coordinates.

The value of $M_5$ is determined by the relation of $M_5$ and $k$ with the 4D rationalized Planck mass, $M_{\textrm{Pl}} = 2.4 \times 10^{18} \, \textrm{GeV}$, via the expression
\begin{equation}
\kappa^2 M_{\textrm{Pl}}^2 = \int_{0}^{y_s} dy \, e^{-2A(y)} \qquad \Longrightarrow \qquad M_5^3 \simeq k M_{\textrm{Pl}}^2 \,.
\end{equation}
Taking into account that $k \lesssim M_5$, this yields $M_5 \simeq {\mathcal O}(M_{\textrm{Pl}})$ so that the Planck scale turns out to be the fundamental scale of the theory, and the TeV is a derived scale after warping. This situation contrasts with the discrete LDM, in which $M_5 \simeq \rho \simeq \TeV$ are fundamental scales~\cite{Antoniadis:2011qw,Cox:2012ee} (see the discussion in Ref.~\cite{Megias:2021mgj}).

\subsection{The gauge sector}
\label{sec:custodial_model}

As it will become clear in this paper, as the theory is RS between the UV and IR branes, in order to protect electroweak observables from going out of control, the SM gauge group has to be extended with an extra custodial gauge symmetry $SU(2)_R$~\cite{Agashe:2003zs}. 
The custodial model is then based on the bulk gauge group~\cite{Mohapatra:1974hk,Mohapatra:1974gc,Senjanovic:1975rk,Agashe:2003zs,Carena:2018cow}
\be
SU(3)_c\otimes SU(2)_L\otimes SU(2)_R\otimes U(1)_X ,
\ee
where $X\equiv B-L$, with 5D gauge bosons $(G,W_L,W_R,X)$, with mass dimension $3/2$, and 5D couplings $(g_c,g_L,g_R,g_X)$, with mass dimension $-1/2$.

The breaking $SU(2)_R\otimes U(1)_X \to U(1)_Y$, where $Y$ is the SM hypercharge, with gauge boson $B$ and coupling $g_Y$, is done in the UV brane by boundary conditions. Therefore the gauge fields $(W_L^a,W_R^a,X)$ define $(W_L^{a},W_R^{1,2},B,Z_R)$, with (UV, IR) boundary conditions, as
\begin{align}
&W_L^a\ (a=1,2,3) \,, & (+,+)\\
B&=\frac{g_X W_R^3+g_RX}{\sqrt{g_R^2+g_X^2}} \,, &(+,+)\\
&W_R^{1,2}, & (-,+)\\
Z_R&=\frac{g_R W_R^3-g_X X}{\sqrt{g_R^2+g_X^2}} \,. &(-,+)
\end{align}
The $SU(2)_L\otimes SU(2)_R$ symmetry is unbroken in the IR brane, where all composite states are localized, such that the \textit{custodial} symmetry is exact.

The covariant derivative for fermions is
\be
\slashed{D}=\slashed\partial-i\left[g_L\sum_{a=1}^3 \slashed{W}_L^{a} T_L^a+g_R\sum_{b=1}^2 \slashed{W}_R^{b} T_R^b+
g_Y \slashed{B}\,Y+g_{Z_R}\slashed{Z}_R\,Q_{Z_R} 
\right] \,,
\ee
where $g_Y$ and $g_{Z_R}$ are defined in terms of $g_R$ and $g_X$ as
\be
g_Y \equiv \frac{g_Rg_X}{\sqrt{g_R^2+g_X^2}} \,, \quad g_{Z_R} \equiv \sqrt{g_R^2+g_X^2}\,, \label{eq:gY}
\ee
and the hypercharge $Y$ and the charge $Q_{Z_R}$ are defined by
\be
Y \equiv T_R^3+Q_X \,, \quad Q_{Z_R} \equiv \frac{g_R^2T_R^3-g_X^2 Q_X}{g_R^2+g_X^2} \,,
\ee
with $Q_X \equiv (B-L)/2$. 

Electroweak symmetry breaking is triggered in the IR brane by the bulk Higgs bi-doublet
\be
\mathcal H=\left( \begin{array}{cc}
H^{0}_2 & H_1^+ \\
H_2^- & H_1^0
\end{array}\right),\quad Q_X=0 \,,
\ee
where the rows transform under $SU(2)_L$ and the columns under $SU(2)_R$. We will denote their VEVs as $\langle H_2^0\rangle\equiv v_2/\sqrt{2}$ and $\langle H_1^0\rangle \equiv v_1/\sqrt{2}$, so that we will introduce the angle $\beta$ as, $\cos\beta=v_1/v$ and $\sin\beta=v_2/v$, with $v \equiv \sqrt{v_1^2+v_2^2}$.

One can rotate to the gauge boson mass eigenstates by considering the angle $\theta_L \equiv \theta_W$, which is the usual weak mixing angle, and $\theta_R$, defined as
\be
\cos\theta_R \equiv \frac{g_R}{\sqrt{g_R^2+g_X^2}},\quad \sin\theta_R \equiv \frac{g_X}{\sqrt{g_R^2+g_X^2}} \,. \label{eq:cR}
\ee
Using Eq.~(\ref{eq:gY}) and (\ref{eq:cR}) one finds $\sin \theta_R = g_Y / g_R < 1$.

As for fermions, left-handed ones are in $SU(2)_L$ bulk doublets as in the SM
\be
Q_{L}^i=\left(\begin{array}{c} u_{L}\\ d_L\end{array}\right)^i,\quad L_L^i=\left(\begin{array}{c} \nu_L\\ e_L\end{array}\right)^i \,,
\ee
where the index $i$ runs over the three generations. On the other hand, as $SU(2)_R$ is a symmetry of the bulk, right-handed fermions $F_R^i$ ($F=e,u,d$) should appear in doublets of $SU(2)_R$, $F_R^i=(f_R,f'_R)^i$. However, as $SU(2)_R$ is broken by the orbifold conditions on the UV brane it means, for bulk right-handed fermions, that one component of the doublet must be even, under the orbifold $\mathbb Z_2$ parity, and has a zero mode, while the other component of the doublet must be odd, and thus without any zero mode. We thus need to double the SM right-handed fermions in the bulk.

\section{Standard model massless gauge bosons}
\label{sec:gauge_bosons}

In this section we will compute Green's functions for massless SM gauge bosons $A_\mu$ (i.e.~the SM photon and gluon). The Lagrangian for massless gauge bosons is~\footnote{We are using in this section the gauge $A_5=0$.}
\be
\mathcal L= \int_0^{y_s} dy\left[ -\frac{1}{4}\tr F_{\mu\nu}F^{\mu\nu}-\frac{1}{2}e^{-2A}\tr A'_\mu A'_\mu  \right]\,,
\label{eq:Lagrangian_GB}
\ee
where the trace is over gauge indices. After Fourier transforming the coordinates $x^\mu$ into momenta $p^\mu$ we can make the field decomposition
$A_\mu(p,y)=f_A(y) A_\mu(p)/\sqrt{y_s}$, and
the EoM of the fluctuations is given by~\cite{Cabrer:2010si}
\begin{equation}
p^2 f_A(y) + \partial_y(e^{-2A} \partial_y f_A(y)) = 0 \,.  \label{eq:fAy}
\end{equation}

In conformal coordinates, and after rescaling the field by ${ f}_A(z) = e^{A(z)/2} \hat f_A(z)$, we obtain the Schr\"odinger like form for the EoM
\begin{equation}
-\ddt{{\hat f}}_A(z) + V_A(z) {\hat f_A}(z) = p^2 {\hat f_A}(z) \,, \label{eq:ftA}
\end{equation}
where the effective Schr\"odinger potential is
\begin{equation}
V_A(z) =  \frac{1}{4} {\dt A}^2(z) - \frac{1}{2} {\ddt A}(z)  \,. \label{eq:VA}
\end{equation}
Plugging Eq.~(\ref{eq:Az}) into this equation, we find the following result for the effective potential
\begin{equation}
V_A(z) =  \left\{ 
\begin{array}{cc}
3/(4z^2)  & \qquad z  \le   z_1  \\
\rho^2/4 &  \qquad   z_1 < z  
\end{array} \,, \right. \label{eq:VAz}
\end{equation}
where $z_1 \equiv 1/\rho$ and $\rho$ is defined in Eq.~(\ref{eq:rhok})~\footnote{Note that $V_A(z)$ is discontinuous at $z=z_1$, and given Eq.~(\ref{eq:ftA}) this induces a discontinuity in $\ddt{{\hat f}}_A(z_1)$. Nevertheless $\hat f_A(z)$ and $\dt{{\hat f}}_A(z)$ are continuous functions.}. We can see that in the IR regime the potential is constant
\begin{equation}
V_A(z) \stackrel[z > z_1]{=}{} m_g^2  \,, \qquad \textrm{with} \qquad m_g = \frac{\rho}{2}  \,.
\end{equation}
We thus find the existence of a mass gap of the potential, which will translate into a gap followed by a continuum KK spectrum.

\subsection{General Green's functions}
\label{sec:gauge_bosons_z0z1}

We will now compute the Green's functions for gauge bosons propagating in the bulk of the 5D space-time from $y$ to $y^\prime$, where both $y$ and $y^\prime$ are considered arbitrary. To compute the Green's function we have to solve an inhomogeneous version of the EoM Eq.~(\ref{eq:fAy}). This is given by
\begin{equation}
p^2 G_A(y,y^\prime;p) + \partial_y \left( e^{-2A} \partial_y G_A(y,y^\prime;p) \right) = \delta(y-y^\prime) \,,  \label{eq:GAy}
\end{equation}
where the derivatives are with respect to the variable~$y$. After fixing the value of $y^\prime$, we can divide the $y$ space into the following domains: \textit{i)} Region I: $0 \le y \le y^\prime$, \textit{ii)} Region II: $y^\prime < y \le y_1$, and, \textit{iii)} Region III: $y_1 < y < y_s$; where we are assuming $y^\prime < y_1$~\footnote{For the case $y_1 < y^\prime$, one should consider as domains:  $0 \le y \le y_1$, $y_1 < y \le y^\prime$ and $y^\prime < y < y_s$; leading to a general solution $G_A(y,y^\prime;p)$ whose expression differs slightly from Eq.~(\ref{eq:solGA_homogeneous}) due to the different definitions of the domains.}. When doing so, we find the general solution
\begin{equation}
G_A(y,y^\prime;p) =  \left\{ 
\begin{array}{cl}
C^I_1 \cdot e^{ky} J_1\left( \frac{e^{ky}}{k} p \right) + C^I_2 \cdot e^{ky} Y_1\left( \frac{e^{ky}}{k} p \right) & \quad \textrm{Region I} 
 \\
C^{II}_1 \cdot e^{ky} J_1\left( \frac{e^{ky}}{k} p \right) + C^{II}_2 \cdot e^{ky} Y_1\left( \frac{e^{ky}}{k} p \right) & \quad \textrm{Region II}
\\
C^{III}_1 \cdot (y_s - y)^{\frac{1}{2} \Delta_A^-} + C^{III}_2 \cdot (y_s - y)^{\frac{1}{2}\Delta_A^+ } &  \quad \textrm{Region III}
\end{array} \,, \right. \label{eq:solGA_homogeneous}
\end{equation}
where $J_1(x)$ and $Y_1(x)$ are Bessel functions of the first and second kind, respectively, and
\begin{equation}
\Delta_A^{\pm}\equiv \pm \delta_A-1,\quad    \delta_A= \sqrt{1 - 4p^2/\rho^2} \,. \label{eq:deltaA_prescription}
\end{equation}
Unless otherwise stated, the square root will be considered in the first Riemann sheet~\footnote{Given the square root function, $f(z) = \sqrt{z}$, we will define the first Riemann sheet in the complex plane $z = |z| e^{i \varphi} \in \mathbb C$ as the one corresponding to $\varphi \in (-\pi,\pi]$, so that this function has a branch cut along the negative real axis. The second Riemann sheet is reached by shifting $\varphi \to \varphi + 2\pi$, i.e.~it corresponds to $\varphi \in (\pi,3\pi]$.  Then, the relation between the square root in the first, $f_{\rm I}(z)$, and second, $f_{\rm II}(z)$, Riemann sheets is 
$
f_{\rm II}(z) = -f_{\rm I}(z) \,,
$
cf.~e.g.~Ref.~\cite{Wolkanowski:2013qca} and references therein.}. For time-like momenta, $p^2>0$, we will adopt the usual prescription $p^2 \to p^2 + i\epsilon$, so that for real values of $p$ above the mass gap, $p > m_g$, $\delta_A = -i \sqrt{4p^2/\rho^2 - 1}$. For space-like momenta $p^2 < 0$, $p \equiv i|p|$, then $\delta_A = \sqrt{ 1 + 4 |p|^2/\rho^2}$ which is always positive. 

The solution of Eq.~(\ref{eq:solGA_homogeneous}) involves six arbitrary constants $C_i^{I,II,III} \; (i = 1,2)$, i.e. two constants per region. The Green's functions are subject to the following boundary and matching conditions
\begin{align}
\begin{split}
& (\partial_y G_A)(y_0) = 0 \,, \qquad\hspace{0.35cm} \Delta G_A(y^\prime) = 0 \,, \qquad \Delta (\partial_y G_A)(y^\prime) = e^{2A(y^\prime)} \,, \\
& \Delta G_A(y_1) = 0 \,, \qquad \Delta (\partial_y G_A)(y_1)  = 0 \,,  \label{eq:GA_bc}
\end{split}
\end{align}
where only the behavior on the first variable $y$ is shown in the Green's functions, and
$\Delta f(y) \equiv  \lim_{\epsilon \to 0} \left( f(y+\epsilon) - f(y-\epsilon) \right)$. In addition, we should impose regularity at the singularity $y=y_s$, i.e. $C_1^{III} = 0$. This corresponds to outgoing wave boundary condition in Lorentzian AdS, which follows from the analytical continuation of the IR regular solution for Euclidean AdS~\cite{Son:2002sd}, i.e. in conformal coordinates
\begin{equation}
G_A(z,z^\prime;p) \stackrel[z_1 \ll z]{\simeq}{} e^{- \frac{1}{2}\Delta_A^+ \rho z} \propto e^{- \sqrt{m_g^2- p^2} \, z} = e^{i \sqrt{-m_g^2 + p^2} \, z}  \,, \label{eq:outgoing}
\end{equation}
where in the last equality we have assumed that $p > m_g$, and adopted the prescription mentioned above~\footnote{Notice that we are not considering in Eq.~(\ref{eq:outgoing}) the incoming wave $\propto e^{-i \sqrt{-m_g^2 + p^2} \, z}$ as this is singular.}.
Then, all the integration constants are fixed. After conveniently defining the variables
\begin{equation}
y_\downarrow \equiv \min(y,y^\prime) \,, \qquad y_\uparrow \equiv \max(y,y^\prime) \,,
\end{equation}
and after implementing the boundary and matching conditions in the general solution of Eq.~(\ref{eq:solGA_homogeneous}) for $y^\prime < y_1$ (and in the equivalent general solution for $y_1 < y^\prime$), one finds
\begin{equation}
G_A(y,y^\prime;p) =  \left\{ 
\begin{array}{cl}
\frac{\pi}{2k} e^{k (y + y^\prime)} \frac{\PP(y_\downarrow) \Z(y_\uparrow)}{\Phi(p)}  & \quad   y_\downarrow, y_\uparrow \le y_1   \\
-\frac{2}{\rho} e^{k y_\downarrow} \left( k(y_s-y_\uparrow) \right)^{\Delta_A^+/2} \frac{\PP(y_\downarrow)}{\Phi(p)} & \quad   
y_ \downarrow \le y_1 < y_\uparrow   \\
  \left(\frac{y_s - y_\uparrow}{y_s - y_\downarrow} \right)^{\Delta_A^+/2}  \delta_A^{-1}  \cdot \frac{\Q(y_\downarrow) }{\Phi(p)} &  \quad 
   y_1 < y_\downarrow  , y_\uparrow
\end{array} \,, \right. \label{eq:GAypy1}
\end{equation}
a solution valid for $0 \le y,y^\prime < y_s$. The functions $\Phi(p)$, $\PP(y)$, $\Z(y)$ and $\Q(y)$ are defined as
\begin{align}
\Phi(p)&= Y_0(p/k) \cdot J_+(p/\rho) - J_0(p/k) \cdot Y_+(p/\rho) \,, \nonumber \\
\Psi(p) &= Y_0(p/k)  \cdot J_-(p/\rho) - J_0(p/k) \cdot Y_-(p/\rho) \,, \nonumber\\
\PP(y)&=  Y_0(p/k) \cdot J_1\left( e^{ky} p/k  \right)   -  J_0(p/k)  \cdot Y_1 \left( e^{ky} p/k \right)  \,, \nonumber\\
\mathcal Z(y)&= J_+(p/\rho) \cdot Y_1\left( e^{k y}p/k \right) - Y_+(p/\rho) \cdot J_1\left( e^{k y}p/k \right) \,, \nonumber\\
\Q(y) &= - \frac{k}{\rho^2}\frac{1}{k(y_s-y)} \left[\Phi(p) -   (k(y_s-y))^{\delta_A} \Psi(p)\right]  \label{eq:Q} \,,
\end{align}
where we define
\begin{equation}
J_\pm(p/\rho) = 2\frac{p}{\rho}J_0(p/\rho)+\Delta_A^\pm J_1(p/\rho),\quad Y_\pm(p/\rho) =  2\frac{p}{\rho}Y_0(p/\rho)+\Delta_A^\pm Y_1(p/\rho) \,. \label{eq:Jpm}
\end{equation}
Up to now we have not made any approximation. However, some of these functions can be slightly simplified by assuming $p \ll k$. In this case $J_0(p/k) \simeq 1 + {\mathcal O}\left((p/k)^2\right)$ and $Y_0(p/k) \simeq \mathcal K + {\mathcal O}\left((p/k)^2\right)$ with  
\begin{equation}
\K \equiv \frac{2}{\pi} \left(\gamma_E -\log(2) + \log(p/\rho)-ky_1 \right) \,.
\label{eq:K}
\end{equation}
Then, the approximate expressions of $\Phi(p)$, $\Psi(p)$ and $\PP(y)$ for $p \ll k$ turn out to be
\begin{align}
\Phi(p)&\simeq \mathcal K \cdot J_+(p/\rho)-Y_+(p/\rho) \,, \qquad \Psi(p) \simeq \mathcal K \cdot J_-(p/\rho)-Y_-(p/\rho) \,, \nonumber\\
\PP(y)&\simeq  \K \cdot J_1\left( e^{ky} p/k  \right)   -   Y_1 \left( e^{ky} p/k \right)  \,.  \label{eq:Phi_approx}
\end{align}
When $y = y_0$ we can consider further simplifications in $\PP(y)$ as well as in $\mathcal Z(y)$. While all the computations in this paper will be performed by using the exact Green's function given by Eqs.~(\ref{eq:GAypy1})-(\ref{eq:Jpm}), we  will provide sometimes in the text approximate formulas to make the explicit expressions simpler~\footnote{Notice that when considering momenta $p \sim \mathcal O(\rho)$, which will be the case throughout this paper, we are neglecting in Eqs.~(\ref{eq:K})-(\ref{eq:Phi_approx}) corrections of order $\mathcal O((\rho/k)^2) = \mathcal O(10^{-30})$, so that the approximation made in these expressions turns out to be extremely good.}.

Let us study some of the properties of the Green's functions. In the limit $y \to y_0$ the Green's function can be written in the simplified form
\begin{equation}
G_A(y_0,y^\prime;p) =  \left[ e^{ky^\prime} \Z(y^\prime) \Theta(y_1 - y^\prime) - \frac{4}{\pi} \frac{k}{\rho} (k(y_s - y^\prime))^{\Delta_A^+/2}  \Theta(y^\prime - y_1) \right]  \frac{1}{p} \frac{1}{\Phi(p)}  \,. \label{eq:GAz0zp}
\end{equation}

Notice that the Green's function~(\ref{eq:GAypy1}) can be expressed as the product of two functions in the form $G_A(y,y^\prime;p) = \A(y_\downarrow) \BB(y_\uparrow)$, and this can also be written as
\begin{equation}
\A(y_\downarrow) \BB(y_\uparrow) = \A(y) \BB(y^\prime) \Theta(y^\prime - y) +  \A(y^\prime) \BB(y) \Theta(y - y^\prime)  \,. \label{eq:AB}
\end{equation}
Then, it is clear that the Green's function is symmetric under the exchange of $y$ and $y^\prime$, i.e.~it fulfills the property
\begin{equation}
G_A(y,y^\prime;p) = G_A(y^\prime,y;p) \,. \label{eq:GA_sim}
\end{equation}
This property is not obvious from the EoM, Eq.~(\ref{eq:GAy}). 

Another property is
\begin{equation}
\Imaginary \left( \A(y) \BB(y^\prime) \right) = \Imaginary \left( \A(y^\prime) \BB(y) \right) \,, \qquad y, y^\prime \le y_1 \quad \textrm{or} \quad y, y^\prime \ge y_1 \,,  \label{eq:ImAB}
\end{equation}
for $p^2 > 0$, which follows from the explicit expressions of Eq.~(\ref{eq:GAypy1}), and taking into account the following relations
\begin{eqnarray}
&&\left( \Delta_A^{\pm}(p) \right)^\ast = \Delta_A^{\mp}(p) \,, \quad \Phi^\ast(p) = \Psi(p) \,, \nonumber \\
&&J_{\pm}^\ast(p/\rho)= J_{\mp}(p/\rho)  \,, \quad Y_{\pm}^\ast(p/\rho) = Y_{\mp}(p/\rho) \,, \qquad (p^2 \ge m_g^2) \,, \label{eq:prop_conj}
\end{eqnarray}
which are valid for time-like momenta. The properties given by Eqs.~(\ref{eq:ImAB}) and (\ref{eq:prop_conj}) will be relevant for the study of the spectral functions in Sec~\ref{subsec:spectral_function}.

\subsection{Brane-to-brane Green's functions}
\label{subsec:brane_to_brane}

Using the general result for $G_A(y,y^\prime;p)$, one can obtain the particularly interesting cases of brane-to-brane Green's functions for gauge bosons. There are three relevant cases: \textit{i)} UV-to-UV, \textit{ii)} UV-to-IR, and \textit{iii)} IR-to-IR Green's functions; and they are obtained by considering the limits
\begin{equation}
G_A(y_\alpha,y_\beta;p) = \lim_{\substack{y\to y_\alpha \\ y^\prime \to y_\beta}} G_A(y,y^\prime; p) \,.
\end{equation}
In particular, the UV-to-UV Green's function can be computed as well by using the holographic formalism, see Ref.~\cite{Megias:2019vdb}. It is interesting and useful to provide the explicit analytical expressions for the brane-to-brane Green's functions. These are
\begin{eqnarray}
G_A^{-1}(y_0,y_0;p) &=& \frac{p \Phi(p)}{\mathcal Z(y_0)} \simeq -\frac{\pi p^2}{2k} \cdot  \frac{  \Phi(p)  }{ J_+(p) }    \label{eq:GA_z0z0_asymp}  \,, \\
G_A^{-1}(y_0,y_1;p) &=&  -\frac{\pi}{4} \frac{\rho}{k} p \Phi(p)  \,, \label{eq:GA_z0z1_asymp}  \\
G_A^{-1}(y_1,y_1;p) &=& -\frac{\rho^2}{2k} \cdot \frac{\Phi(p)}{\PP(y_1)} \simeq -\frac{\rho^2}{2k}\cdot \frac{\Phi(p)}{ \K \cdot J_1\left( \frac{p}{\rho} \right) - Y_1\left( \frac{p}{\rho} \right) }   \,, \label{eq:GA_z1z1_asymp} 
\end{eqnarray}
where in the second equality of Eqs.~(\ref{eq:GA_z0z0_asymp}) and (\ref{eq:GA_z1z1_asymp}) we have assumed $p \ll k$. All Green's functions include the zero-mode contribution which behaves as
\be
G_A^0=\frac{1}{y_s p^2}=\lim_{p\to 0}G_A(y,y^\prime;p)
\ee
which, after coupling to two fermions lines, with strength $g_5^2$, yields the usual 4D behavior $g_4^2/p^2$. Then, we can define Green's functions contributed only by the continuum KK modes, with the zero-mode contribution subtracted out, as
\begin{equation}
\mathcal G_A(y,y^\prime;p) = G_A(y,y^\prime;p) - G_A^0 \,.
\end{equation}

\begin{figure}[t]
\centering
\includegraphics[width=5.2cm]{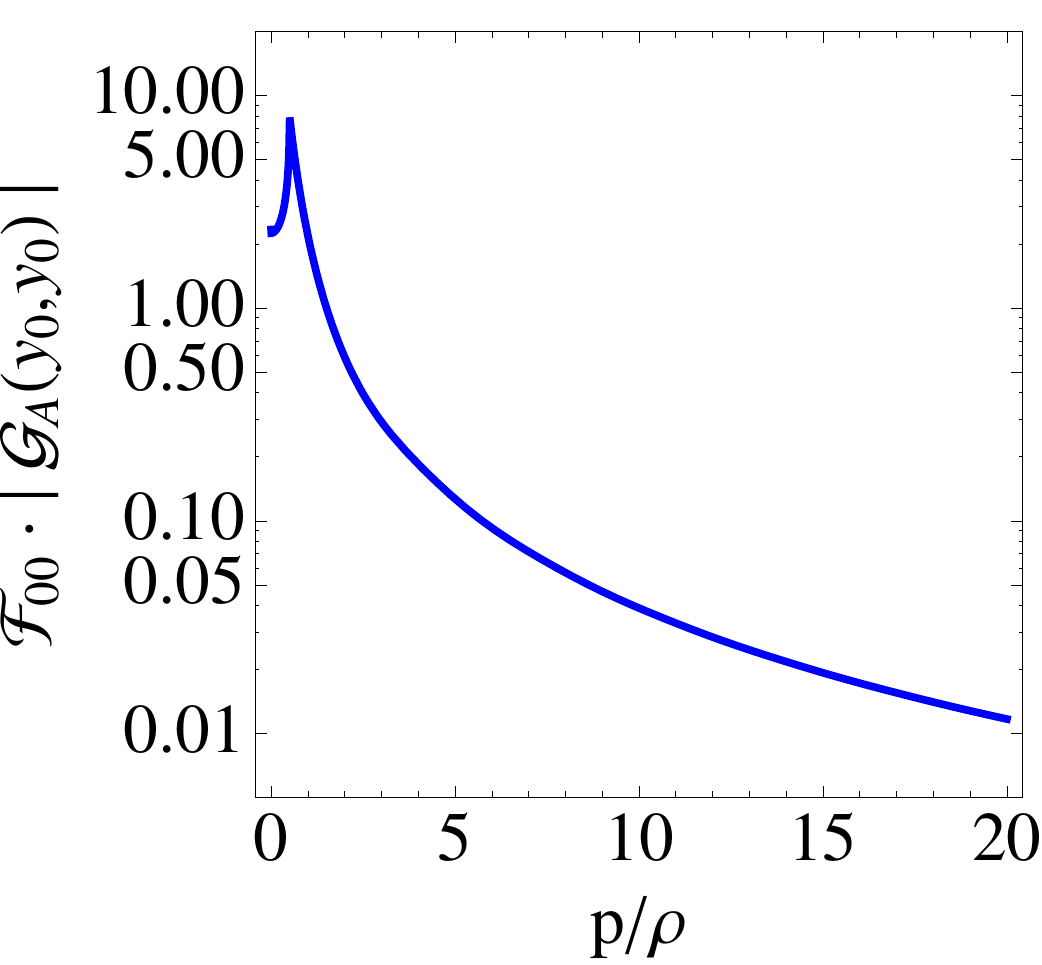}
\includegraphics[width=5.0cm]{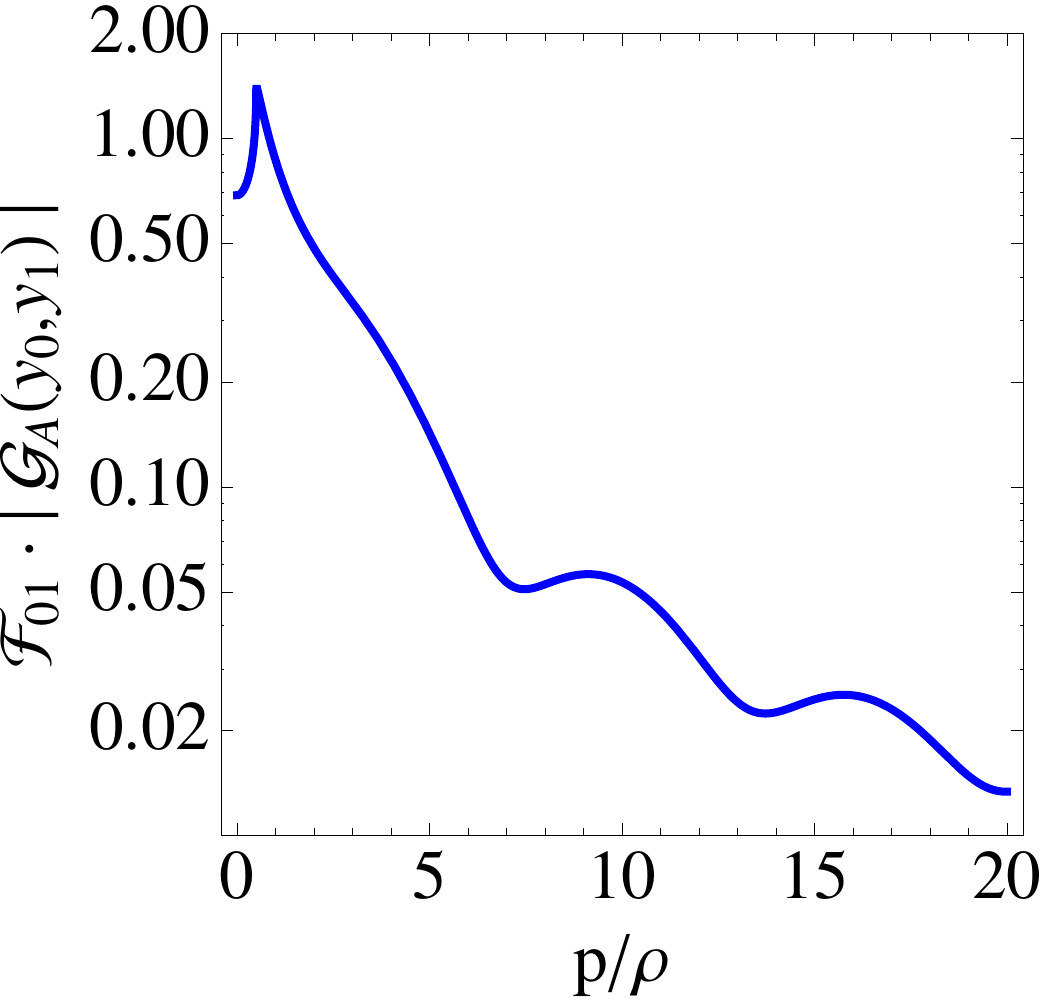}
\includegraphics[width=5.1cm]{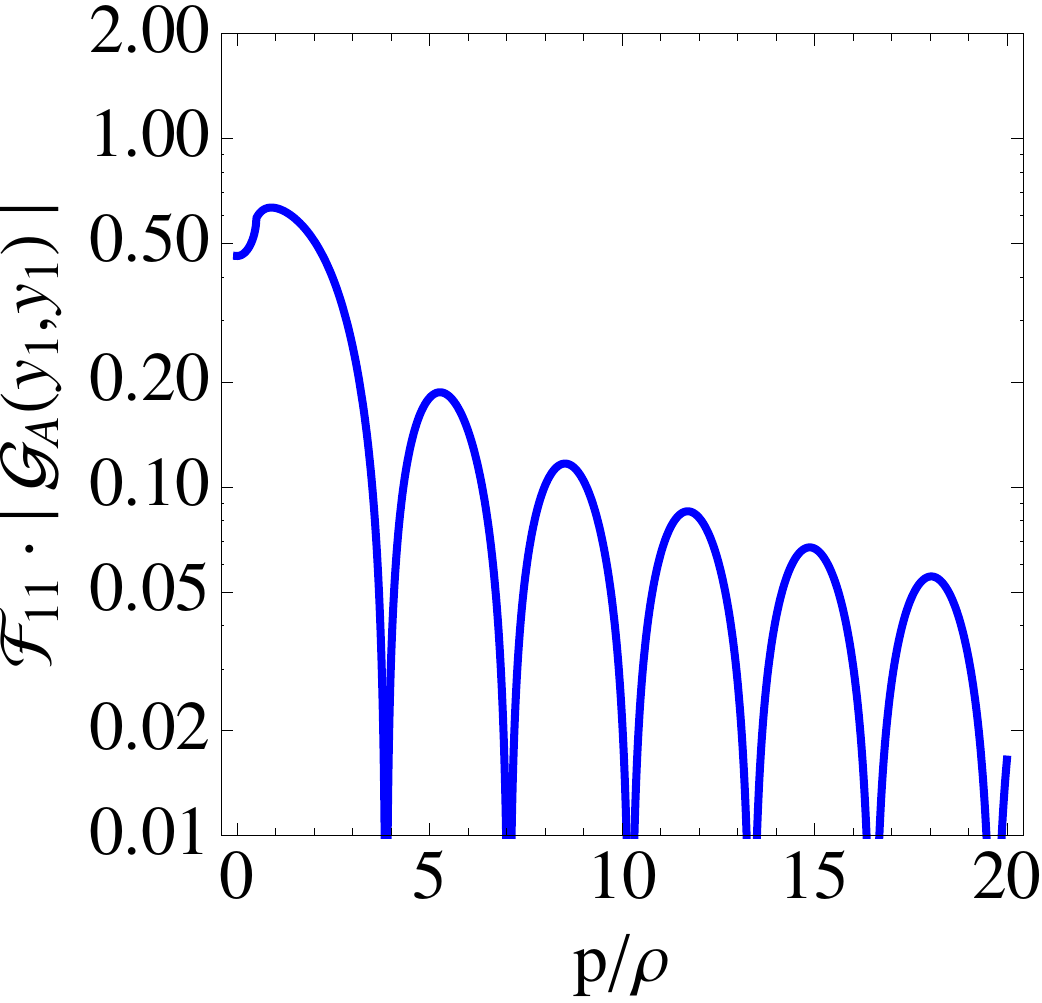}
\caption{\it Plots of the rescaled Green's functions $\mathcal F_{00} \cdot |\mathcal G_A(y_0,y_0;p)|$ (left panel), $\mathcal F_{01} \cdot |\mathcal G_A(y_0,y_1;p)|$ (middle panel), and $\mathcal F_{11} \cdot |\mathcal G_A(y_1,y_1;p)|$ (right panel) as functions of $p/\rho$. We have used $A_1 = 35$ in all panels and assume time-like momenta $p^2>0$.
}
\label{fig:spectral_gaugebosons}
\end{figure}

\begin{figure}[t]
\centering
\includegraphics[width=5.2cm]{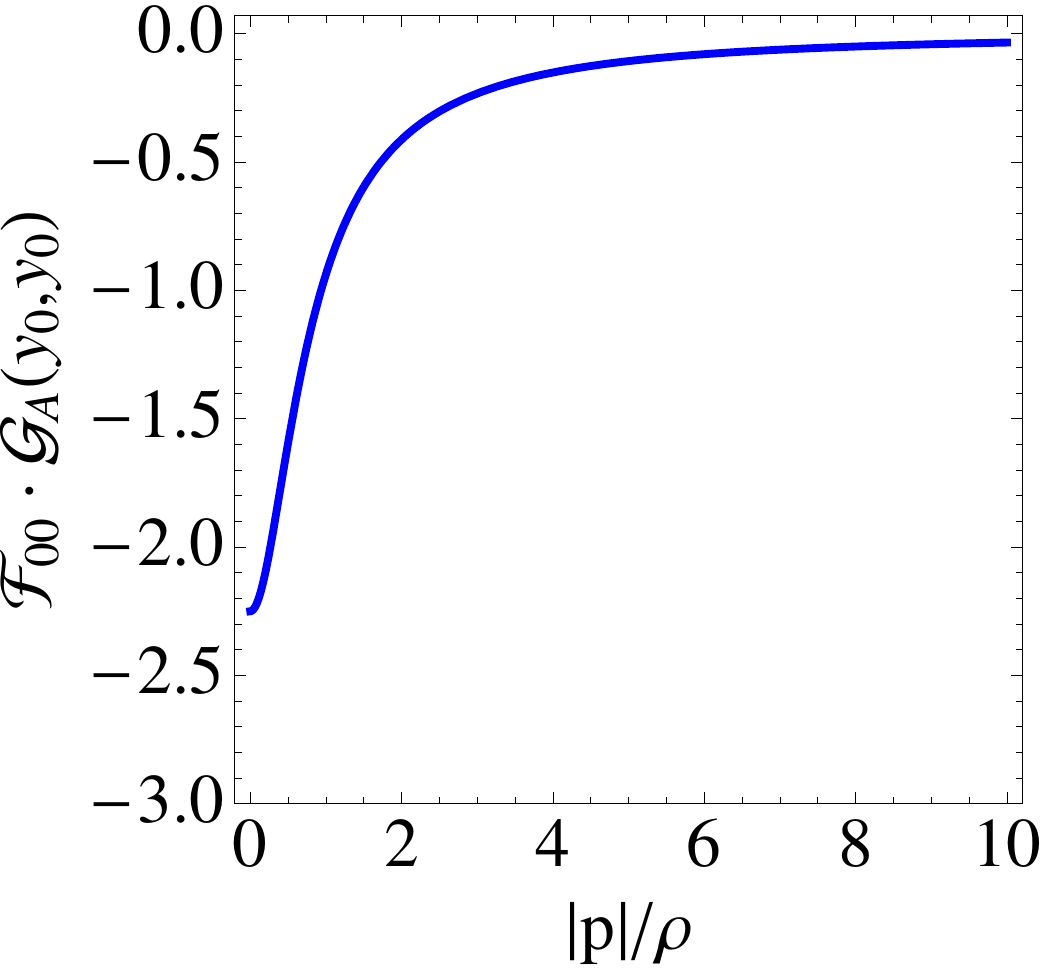}
\includegraphics[width=4.9cm]{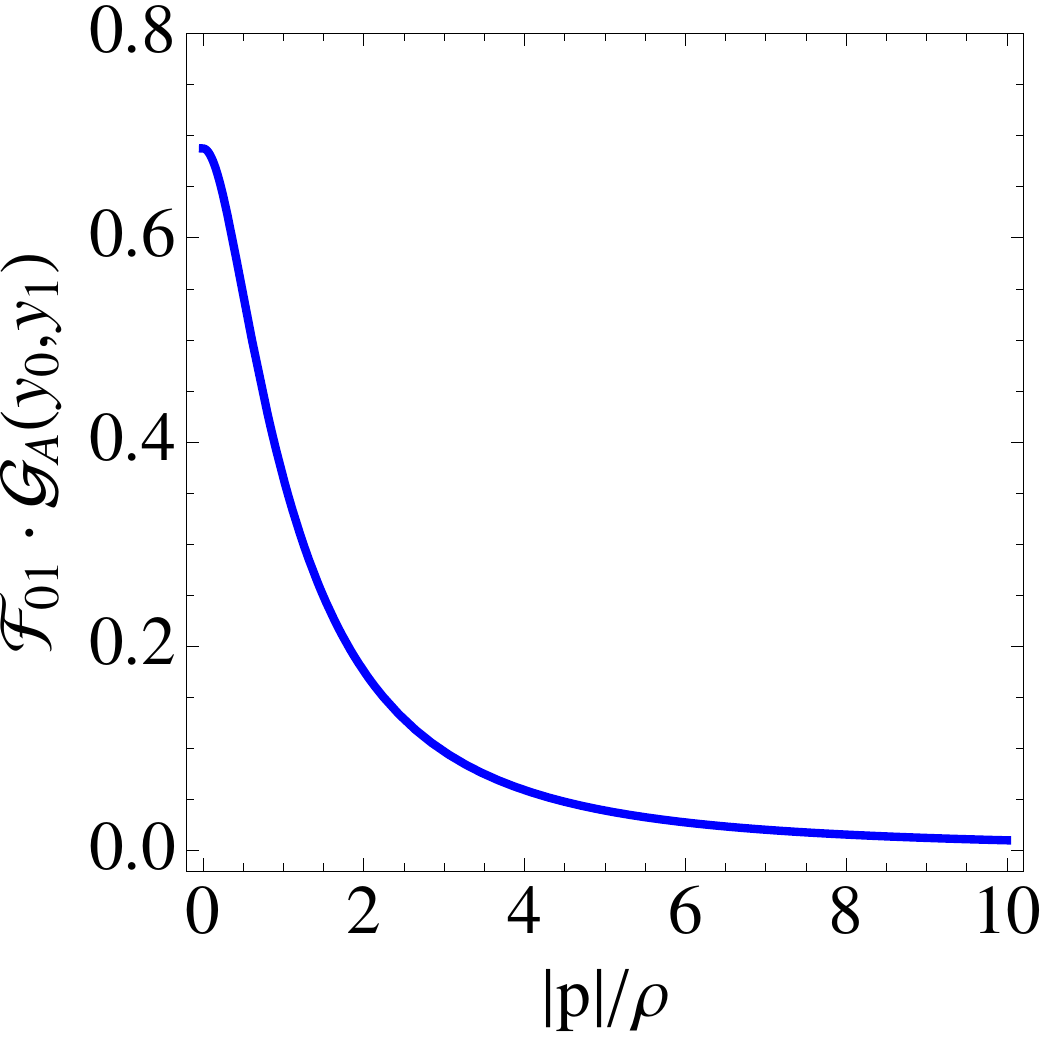}
\includegraphics[width=5.1cm]{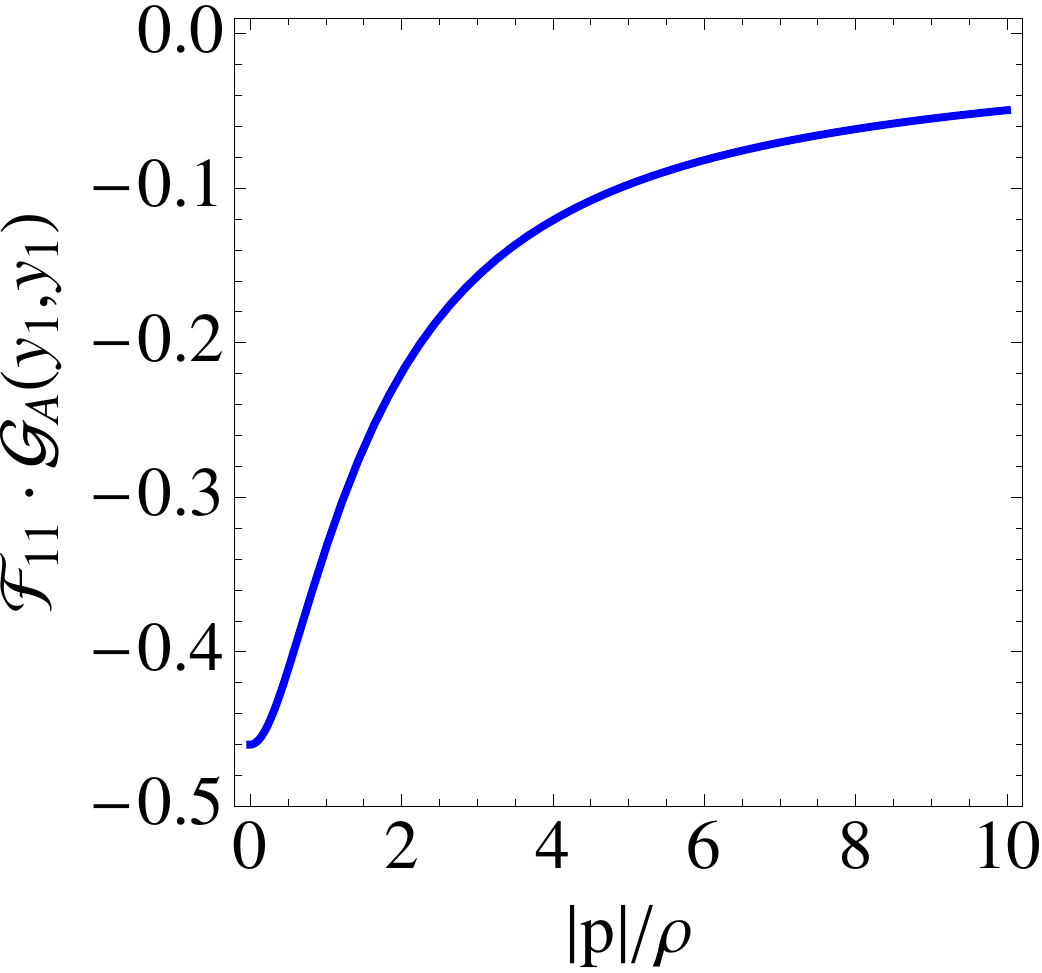}
\caption{\it Plots of $\mathcal F_{00} \cdot \mathcal G_A(y_0,y_0;|p|)$ (left panel), $\mathcal F_{01} \cdot \mathcal G_A(y_0,y_1;|p|)$ (middle panel) and $\mathcal F_{11} \cdot \mathcal G_A(y_1,y_1;|p|)$ (right panel) as a function of $|p|/\rho$.  We have used $A_1 = 35$ in all panels and assume space-like momenta $p^2<0$.
}
\label{fig:spacelike_momenta}
\end{figure}

Note that while scale invariance is explicitly broken by the scales $\rho$ and $k$, it is possible to define rescaled Green's functions $\mathbb G_A(y,y^\prime;p) = \mathcal F_G \cdot G_A(y,y^\prime;p)$ that turn out to be dimensionless, and their dependence on momenta and scales is through dimensionless products and ratios $\mathbb G_A(ky,ky^\prime;p/\rho,\rho/k)$ (cf. Ref.~\cite{Megias:2019vdb}). The required scaling factor
\begin{equation}
\mathcal F_G = \rho \left(\frac{\rho}{k}\right)^a (ky_s)^b \,, \label{eq:FG}
\end{equation}
where $a$ and $b$ are real numbers, is independent of the momentum $p$. We will be interested in the effect on the brane-to-brane Green's functions of a change of the scale~$\rho$. To this end, we will consider a rescaling of the form
\begin{equation}
p \to p^\prime = c p \,, \qquad \rho \to \rho^\prime = c \rho \qquad \textrm{and} \qquad k \to k^\prime = \bar{c} k \,,  \label{eq:prhok_rescaling}
\end{equation}
with $c \ne \bar c$. While the ratio $p/\rho$ is not affected by the rescaling, one has $\rho/k \to (c/\bar c) \cdot \rho/k$. Using that $k y_s = 1 - \log(\rho/k)$, cf. Eqs.~(\ref{eq:rhok})-(\ref{eq:ysy1}), one can see that Eq.~(\ref{eq:prhok_rescaling}) implies a shift of $ky_s$ (or $A(y_1)$), i.e. $k y_s \to k y_s - \log(c/\bar c)$. When considering $a = 1$ in Eq.~(\ref{eq:FG}), the dimensionless brane-to-brane Green's functions $\mathbb G_A  = \mathbb G_A(p/\rho,\rho/k)$ turn out to have a smooth logarithmic dependence on $\rho/k$, so that the effect of the scaling parameters $c$ and $\bar c$ is also logarithmic. In addition, it is possible to choose the parameter $b$ in Eq.~(\ref{eq:FG}) so that the corresponding scaling factor $\mathcal F_G$ removes the dominant dependence in $\log(\rho/k)$ of the respective Green's function, making it almost invariant under shifts of $k y_s$. While the factor that makes $G_A^0$ invariant is $\mathcal F_G = \frac{\rho^2}{k} (ky_s) $, the brane-to-brane Green's functions with the zero-mode subtracted out $\mathcal G_A(y_\alpha,y_\beta;p)$ will be approximately invariant under shifts of $ky_s$ when multiplying them by the factors $\mathcal F_{\alpha\beta}$, where
\be
\mathcal F_{00} = \frac{\rho^2}{k}(k y_s)^2  \,, \qquad  \mathcal F_{01} = \frac{\rho^2}{k}(k y_s)  \,, \qquad \mathcal F_{11} = \frac{\rho^2}{k} \,.  \label{eq:Fab}
\ee
These factors will be used in the rest of the manuscript, in particular in Sec.~\ref{subsec:spectral_function} for the spectral functions, as well as in Sec.~\ref{sec:Dirichlet} and Appendix~\ref{sec:massive_gauge_bosons}.
We plot, in Fig.~\ref{fig:spectral_gaugebosons}, $|\mathcal G_A(y_0,y_0;p)|$ (left panel), $|\mathcal G_A(y_0,y_1;p)|$ (middle panel), and $|\mathcal G_A(y_1,y_1;p)|$ (right panel), normalized by the factors $\mathcal F_{\alpha\beta}$ of Eq.~(\ref{eq:Fab}), as functions of $p/\rho$, for time-like momenta $p^2>0$.
For space-like momenta $p^2<0$ the Green's functions are purely real. In Fig.~\ref{fig:spacelike_momenta} we plot the Green's functions $\mathcal G_A(y_0,y_0;|p|)$, $\mathcal G_A(y_0,y_1;|p|)$ and $\mathcal G_A(y_1,y_1;|p|)$, normalized by $\mathcal F_{\alpha\beta}$, as functions of $|p|/\rho$, for space-like momenta $p^2 < 0$. 

It is also interesting to provide the limits $p \ll \rho$ of the Green's functions. This leads to the following Taylor series expansions
\begin{eqnarray}
G_A^{-1}(y_0,y_0;p) &\stackrel[p \ll \rho]{\simeq}{}& y_s p^2 + \frac{9}{4k} \frac{p^4}{\rho^2} + \mathcal{O}(p^6) \label{eq:GA_z0z0_p0}  \,, \\
G_A^{-1}(y_0,y_1;p)  &\stackrel[p \ll \rho]{\simeq}{}&   y_s p^2 + \frac{(9-3 ky_s)}{4k} \frac{p^4}{\rho^2} + \mathcal{O}(p^6)  \,,   \label{eq:GA_z0z1_p0}  \\
G_A^{-1}(y_1,y_1;p)  &\stackrel[p \ll \rho]{\simeq}{}&  y_s p^2 + \frac{(9 + 2ky_s (-3 + ky_s))}{4k} \frac{p^4}{\rho^2} \nonumber \\
&&+ \frac{(113 + 4 k y_s [63 + 2 k y_s (-19 + 4 k y_s)])}{128k}  \frac{p^6}{\rho^4} + \mathcal{O}(p^8)   \,.  \label{eq:GA_z1z1_p0} 
\end{eqnarray}
The $p^2$ behavior is valid for any Green's function, i.e. $G_A^{-1}(y,y^\prime;p) \stackrel[p \ll \rho]{\simeq}{} y_s p^2 + \cdots$~\footnote{Notice that the rescaled Green's functions behave as $\mathbb G_A^{-1}(y,y^\prime;p) \stackrel[p \ll \rho]{\simeq}{} (p/\rho)^2 + \cdots$, and they turn out to be functions of $p/\rho$ with ``power-like'' corrections in $ky_s = 1 -\log(\rho/k)$.}. We keep terms up to $\mathcal O(p^6)$ in the IR-to-IR Green's function as these will be needed in the computation of electroweak precision observables of Sec.~\ref{sec:EWPO}. 

Finally, the behavior in the regime $\rho \ll p$ (and $p \ll k$) is, for time-like momenta, $p^2>0$
\begin{eqnarray}
G_{A}^{-1}(y_0,y_0;p) &\stackrel[\rho \ll p]{\simeq}{}&    \left(  k y_1 - \log\left( \frac{p}{\rho} \right) + i\frac{\pi}{2}  \right)  \left(\frac{p}{\rho}\right)^2\, \frac{\rho^2}{k}  \,, \label{eq:GA_z0z0_largep}   \\
G_{A}^{-1}(y_0,y_1;p) &\stackrel[\rho \ll p]{\simeq}{}&  \sqrt{\frac{2}{\pi}} e^{-i(p/\rho - \pi/4)} \left( k y_1 - \log\left( \frac{p}{\rho} \right) + i\frac{\pi}{2} \right)\left(\frac{p}{\rho}\right)^{3/2}\, \frac{\rho^2}{k} \,,   \label{eq:GA_z0z1_largep}  \\
G_{A}^{-1}(y_1,y_1;p) &\stackrel[\rho \ll p]{\simeq}{}&  \frac{e^{-i(p/\rho - \pi/4)}}{\cos\left( \frac{p}{\rho} + \frac{\pi}{4} \right)} \left(\frac{p}{\rho}\right)\, \frac{\rho^2}{k} \label{eq:GA_z1z1_largep}  \,,
\end{eqnarray}
and for space-like momenta $p^2<0,\,p\equiv i|p|$,
\begin{eqnarray}
G_{A}^{-1}(y_0,y_0;|p|)&\stackrel[\rho \ll |p|]{\simeq}{}&   \left( \log\left(\frac{|p|}{\rho}\right) - k y_1 \right) \left|\frac{p}{\rho}\right|^2\, \frac{\rho^2}{k} \,, \label{eq:GA_z0z0sl_largep}   \\
G_{A}^{-1}(y_0,y_1;|p|)&\stackrel[\rho \ll |p|]{\simeq}{}&  \sqrt{\frac{2}{\pi}} e^{|p|/\rho}\, \left( \log\left(\frac{|p|}{\rho}\right) -k y_1 \right) \left|\frac{p}{\rho}\right|^{3/2}\, \frac{\rho^2}{k}
 \,,   \label{eq:GA_z0z1sl_largep}  \\
G_{A}^{-1}(y_1,y_1;|p|)&\stackrel[\rho \ll |p|]{\simeq}{}&  -2\left|\frac{p}{\rho}\right|\, \frac{\rho^2}{k} \,.  \label{eq:GA_z1z1sl_largep} 
\end{eqnarray}
Notice that for space-like momenta the Green's function $G_A(y_0,y_1;|p|)$ goes exponentially to zero as $e^{-|p|/\rho}$ for $|p|\gg\rho$, a property which was recently noticed in Ref.~\cite{Costantino:2020vdu}. The general asymptotic behavior for the Green's function $G_A(z_0,z^\prime;|p|)$, with $z^\prime < z_1$, is $\sim\exp(-z^\prime |p|/z_1\rho)$.

\subsection{Green's functions in the complex plane and resonances}
\label{sec:Resonances}

Although the spectrum of excitations is a continuum, starting from the mass gap $m_g=\rho/2$, which is characteristic of a conformal theory,
as the conformal invariance is explicitly (spontaneously) broken by the UV (IR) brane it is worth exploring the structure of the Green's functions in the complex $s\equiv p^2$ plane, with
\be
s \equiv M^2-i M \Gamma = M^2(1-i r), \quad r\equiv\Gamma/M\,,
\label{eq:poles}
\ee
as it is well known in Quantum Field Theory that resonances with mass $M$ and decay width $\Gamma$ are associated to the presence of poles in the unphysical Riemann sheet. Needless to say, in ordinary Quantum Field Theory, the presence of poles in the complex plane are associated to production processes corresponding to decays of the resonance into other particles of mass $m$, for energies above the threshold $s>4 m^2$. Nevertheless our Green's functions, even considered at the classical level, have an imaginary part, unrelated to any decay process, similarly to the case of unparticles. Still exploring the complex $s$ plane is worth given that, as we stated above, conformal invariance is broken, which makes a fundamental difference with respect to the case of unparticles.

Let us study the possible existence of poles of the Green's functions $G_A(y,y^\prime;s)$ in the complex $s$-plane. As the origin of the non-vanishing imaginary part of Green's functions is the threshold function $\delta_A(s)=\sqrt{1-4s/\rho^2}$ which has two Riemann sheets, similar to the threshold function of the decay into two particles,  
in order to perform an analysis of the resonances, one should compute the Green's functions in the second Riemann sheet of the square root function. It can be easily seen that a change from the first (I) to the second (II) Riemann sheet is equivalent to the replacement $\delta_A ^{\rm II}(s)\to -\delta_A^{\rm I}(s)=-\delta_A(s)$, i.e.
\begin{equation}
\delta_A^{\textrm{II}} (s)= - \sqrt{1 - 4s/\rho^2} \,,
\end{equation}
where the square root in this formula is the one in the first Riemann sheet. Let us point out that the function $\delta_A(s) $, and then the Green's functions~$G_A(y,y^\prime;s)$, have a branch cut from the mass gap $s = \rho^2/4$ to infinity along the real axis with the first Riemann sheet corresponding to $\varphi \in [0,2\pi)$, where $\varphi$ is defined as $s - \rho^2/4 = |s - \rho^2/4| \, e^{i\varphi}$. The second Riemann sheet corresponds to $\varphi \in [2\pi, 4\pi)$. The Green's functions are continuous when changing from the first to the second Riemann sheets, but there appears a discontinuity at the branch cut if one approaches it using the same Riemann sheet. This discontinuity is accounted by the spectral function.

\begin{figure}[t]
\centering
\includegraphics[width=7cm]{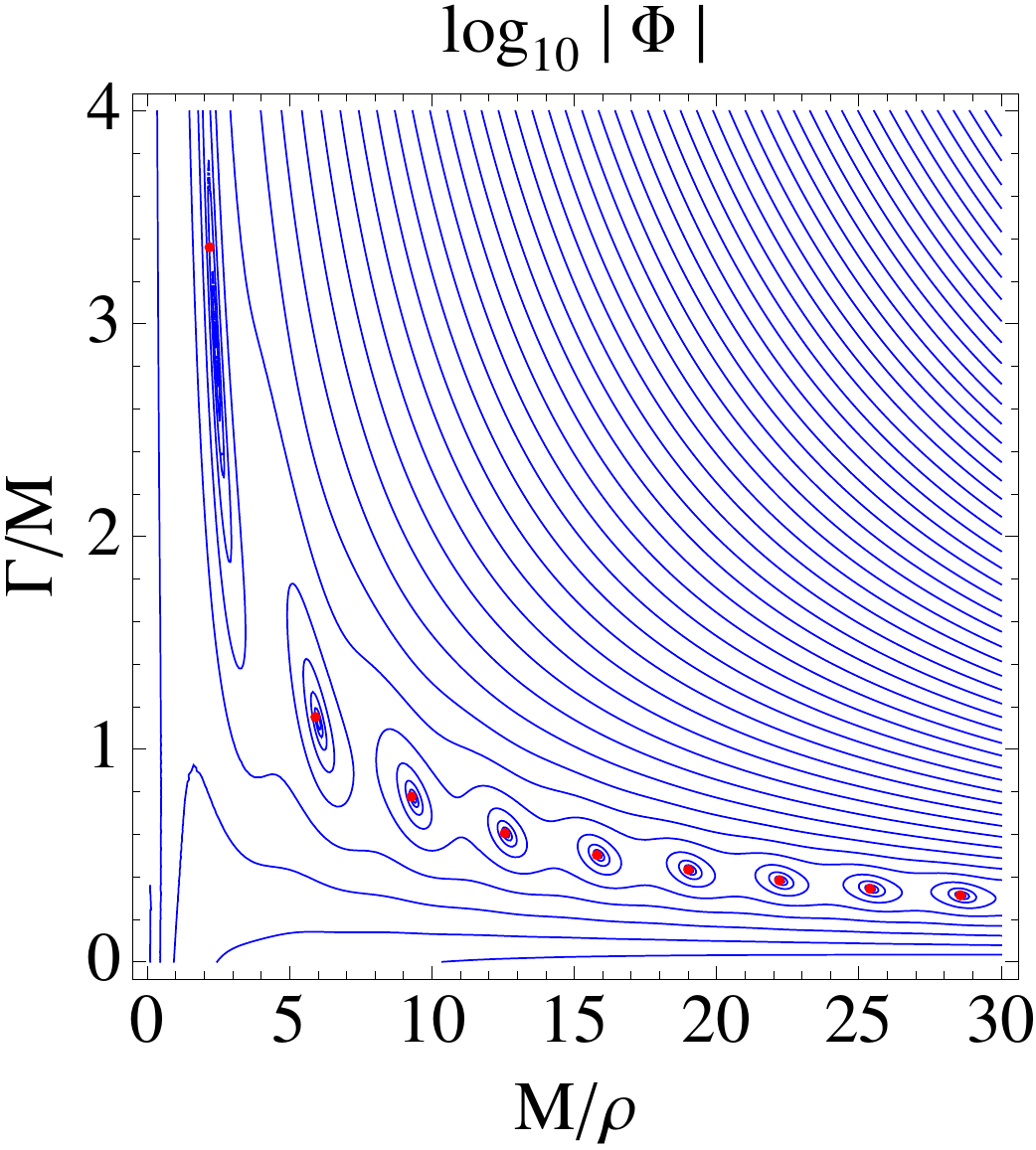} \hspace{0.5cm} \includegraphics[width=7cm]{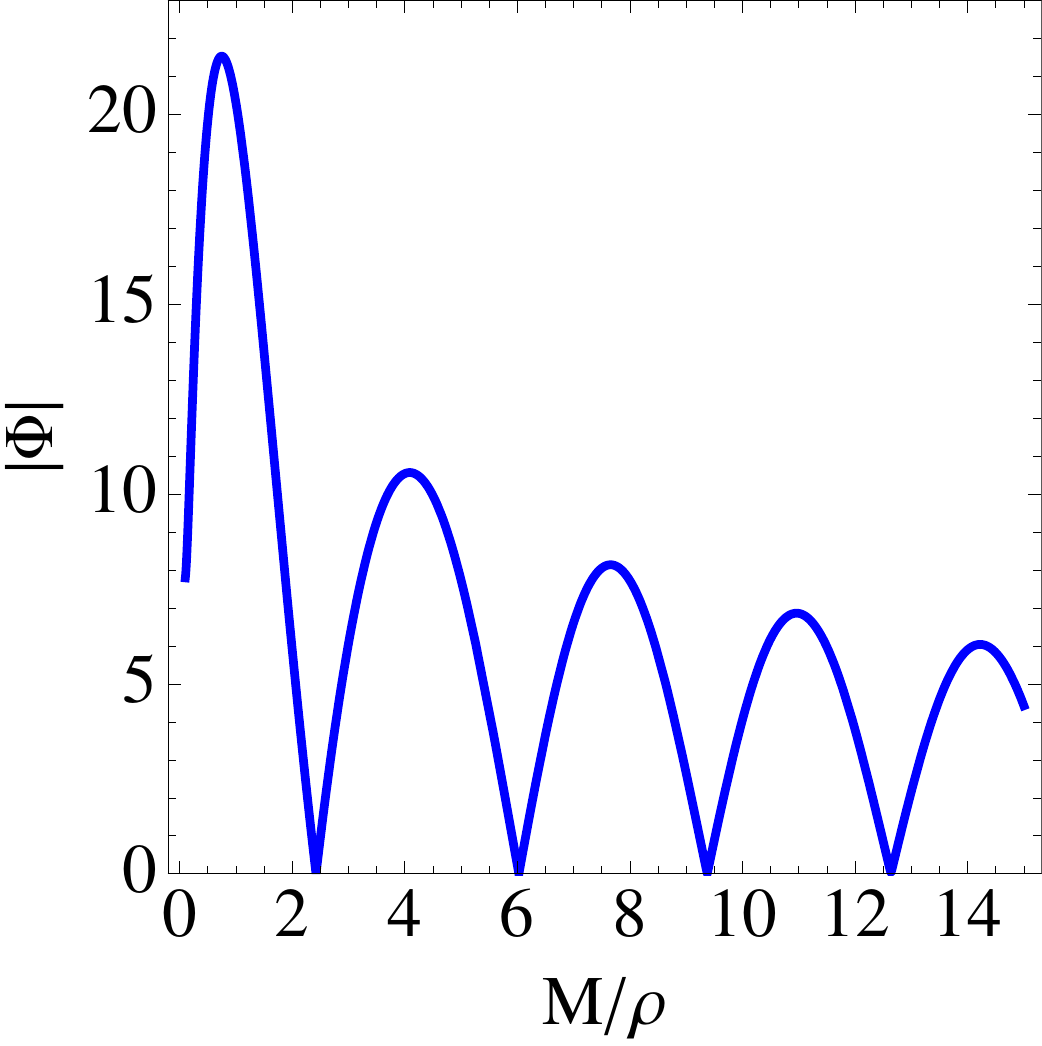} 
\caption{\it Left panel: Contour plot of $\log_{10} |\Phi(p)|$ in the plane $(M/\rho,\Gamma/M)$. The contour lines form small circles around the zeros of $\Phi(p)$. The (red) dots stand for the positions of the zeros of $\Phi(p)$ as given by the analytical formula of Eq.~(\ref{eq:p2Lambert}). Right panel: Plot of $|\Phi(p)|$ along an interpolating curve connecting the zeros of $\Phi(p)$ appearing in the left panel. We have used the variables $M/\rho$ and $r \equiv \Gamma/M$ defined in Eq.~(\ref{eq:poles}). We have considered $A_1 = 35$.}
\label{fig:Phi_2D}
\end{figure}

From Eqs.~(\ref{eq:GA_z0z0_asymp})-(\ref{eq:GA_z1z1_asymp}), one realizes that the possible poles (excluding the zero-mode) should appear as zeros of the function $\Phi(p)$. Following this idea, we display in Fig.~\ref{fig:Phi_2D} a contour plot of $\log_{10} |\Phi(p)|$ computed in the second Riemann sheet. For convenience, we have expressed the squared complex momenta $s$ in the plane $M/\rho$ and $r$ given in Eq.~(\ref{eq:poles}).
One can see from the left panel of Fig.~\ref{fig:Phi_2D} that there appears an intriguing structure of zeros of the function $\Phi(p)$ in the $s$ plane. In the right panel of Fig.~\ref{fig:Phi_2D} we plot the function $|\Phi|$ along an interpolating curve connecting the zeros of $\Phi(p)$ (the red points in the left panel of Fig.~\ref{fig:Phi_2D}), as a function of $M/\rho$. We can check that the $\Phi$ function indeed vanishes at those points. 

All resonances appear for both positive and negative values of~$\Gamma$. The latter are unphysical shadow poles, required by Hermitian analyticity~\cite{Delgado:2008gj,Landshoff:1963}. The lightest resonances 
appear at the values
\begin{equation}
(M/\rho,r) = (2.42,2.87), (6.03,1.12), (9.37,0.768), (12.64,0.601), (15.87,0.500) , \cdots  \,.   \label{eq:MG}
\end{equation}
The values of $M/\rho$ follow a pattern similar to the KK modes in the RS model, for which the eigenvalues $M_n/\rho$ are close to the zeros of the $J_0(M_n/\rho)$ function. However, contrary to the RS model, the resonances in the gapped continuum model have a finite width. This width increases slowly with energy, but the relative width $\Gamma/M$ decreases, so that the resonances tend to a distribution closer to Dirac delta functions at high energies.

We can study analytically the location of these zeros in the following way. If one performs an expansion of the function $\Phi(p)$ at large momentum $\rho \ll |p|$ $(|p| \ll k)$ one finds
\begin{equation}
\Phi(p) \stackrel[\rho \ll |p|]{\simeq}{}  \frac{e^{-i(p/\rho + \pi/4)}}{\sqrt{2 \pi^3}}   \left[ e^{i2p/\rho} -8i \left(\frac{p}{\rho} \right)^2  \right] \log\left( \frac{p}{k} \right)  \left( \frac{\rho}{p} \right)^{3/2} \,, \quad \Imaginary \left( (p/\rho)^2 \right) < 0 \,.  \label{eq:Phi_large_p}
\end{equation}
Then, the zeros of $\Phi(p)$ correspond to the solutions of the equation
\begin{equation}
 e^{i2p/\rho} = 8i \left(\frac{p}{\rho} \right)^2  \,,
\end{equation}
which turn out to be
\begin{equation}
\frac{s}{\rho^2} = -\mathcal W_n\left[ \frac{\varepsilon}{4}(1+i) \right]^2  \,, \qquad \varepsilon = \pm 1, \qquad n = -1,-2,-3, \cdots \,. \label{eq:p2Lambert}
\end{equation}
Here $\mathcal W_n$ is the $n$-th branch of  the Lambert function~\footnote{The Lambert function is the solution of the equation $z = \mathcal W(z) \, e^{\mathcal W(z)}$. In addition to the principal branch $n=0$, there are other infinite branches denoted by $\mathcal W_n(z)$ for integer $n$.}. We display as red dots in Fig.~\ref{fig:Phi_2D} (left) the results of Eq.~(\ref{eq:p2Lambert}). Moving from lighter to heavier resonances corresponds to taking $(n,\varepsilon)= (-1,+1), (-1,-1), (-2,+1), (-2,-1), \cdots$, in this order. Note the close agreement of the analytical results with the true zeros of $\Phi(p)$, even for the lightest resonances: the relative error for the values of $M/\rho$ and $\Gamma/M$ decreases with $M/\rho$, and it is $\lesssim 2\%$ except for the lightest resonance which is $\sim 15\%$. 
\begin{figure}[htb]
\centering
\includegraphics[width=4.8cm]{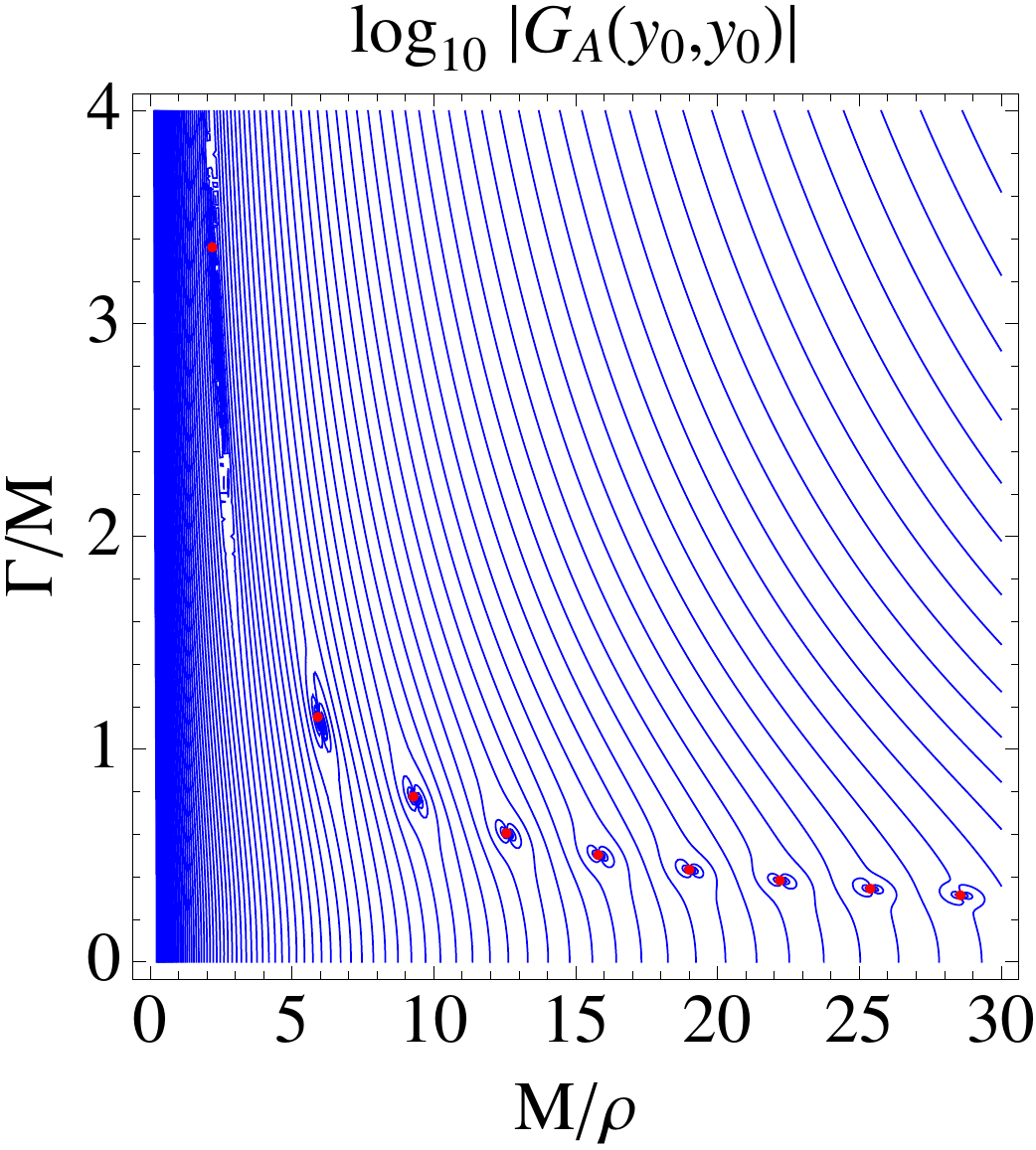} \hspace{0.10cm}
\includegraphics[width=4.8cm]{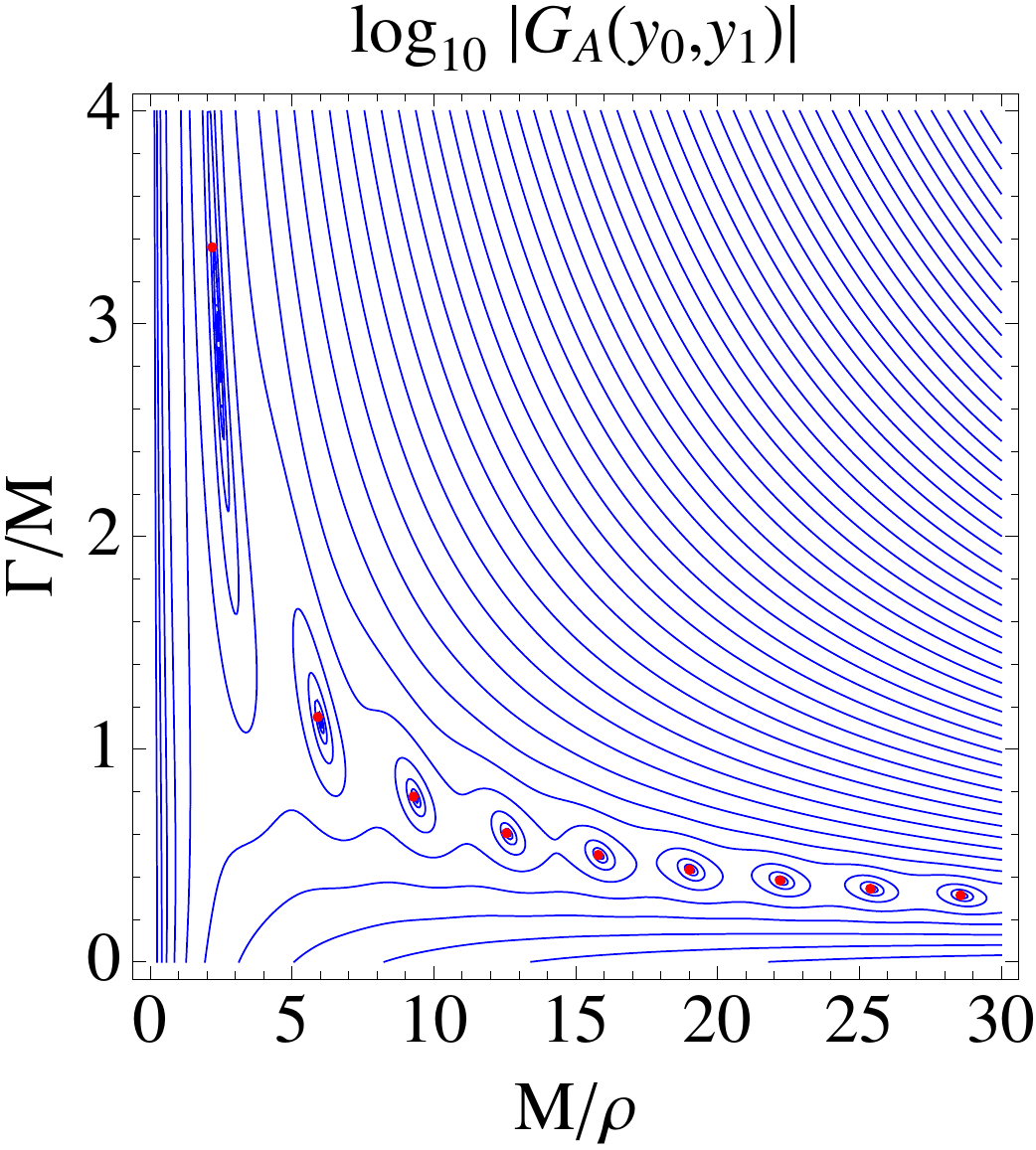} \hspace{0.10cm}
\includegraphics[width=4.8cm]{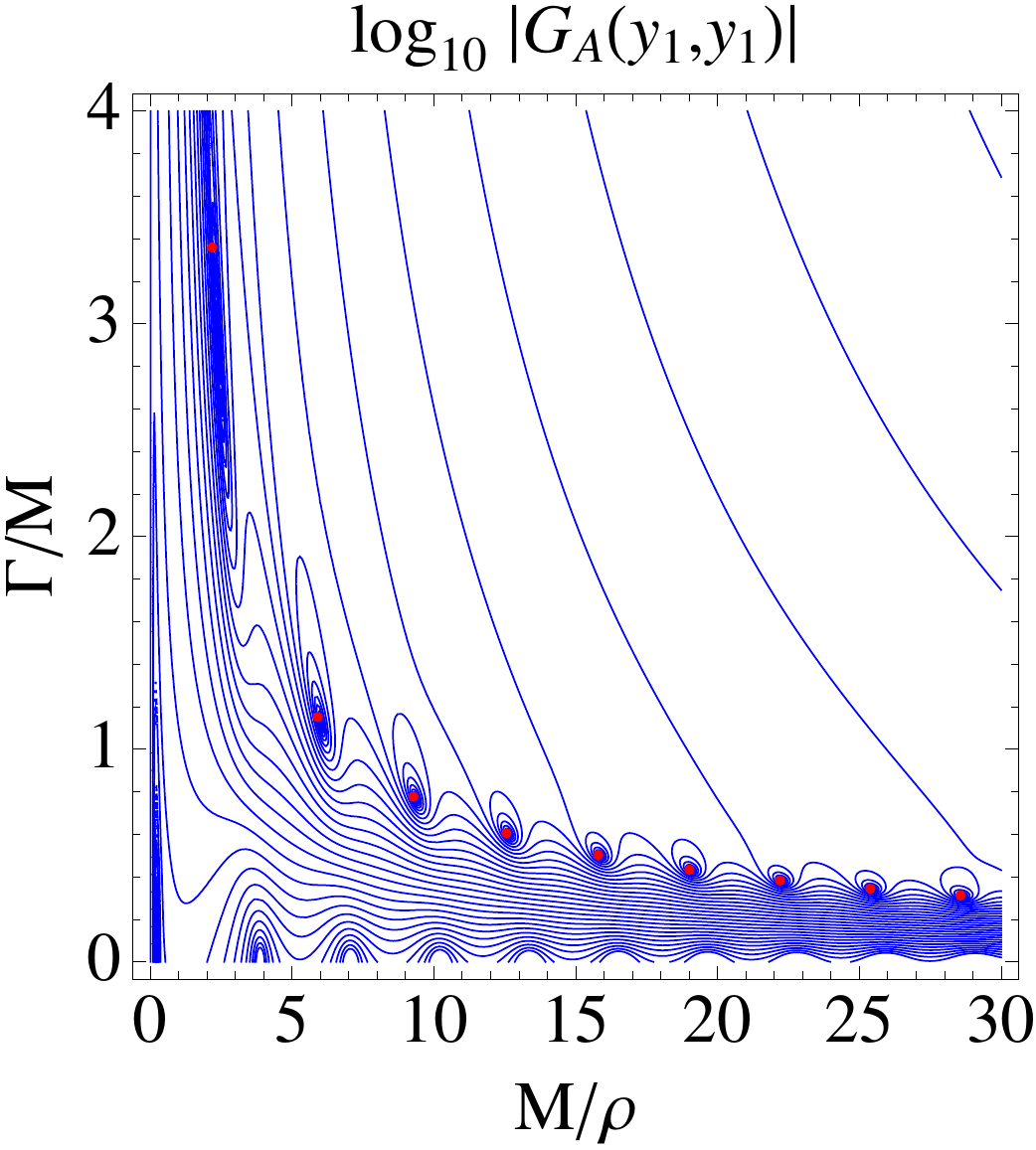}
\caption{\it Contour plot in the plane $(M/\rho,\Gamma/M)$ of the common logarithm of the absolute value of the Green's functions $\log_{10} |G_A(y_0,y_0)|$ (left panel), $\log_{10} |G_A(y_0,y_1)|$ (middle panel) and $\log_{10} |G_A(y_1,y_1)|$ (right panel). The (red) dots stand for the positions of the poles of the Green's functions as predicted by the analytical formula of Eq.~(\ref{eq:p2Lambert}). The contour lines form small circles around the poles of $G_A(y_\alpha,y_\beta)$ (those circles with red dots), or circles around the zeros of $G_A(y_\alpha,y_\beta)$ (those circles with no red dots). We have considered $A_1 = 35$.}
\label{fig:AbsGA_resonances}
\end{figure} 
\begin{figure}[h]
\centering
\includegraphics[width=7.5cm]{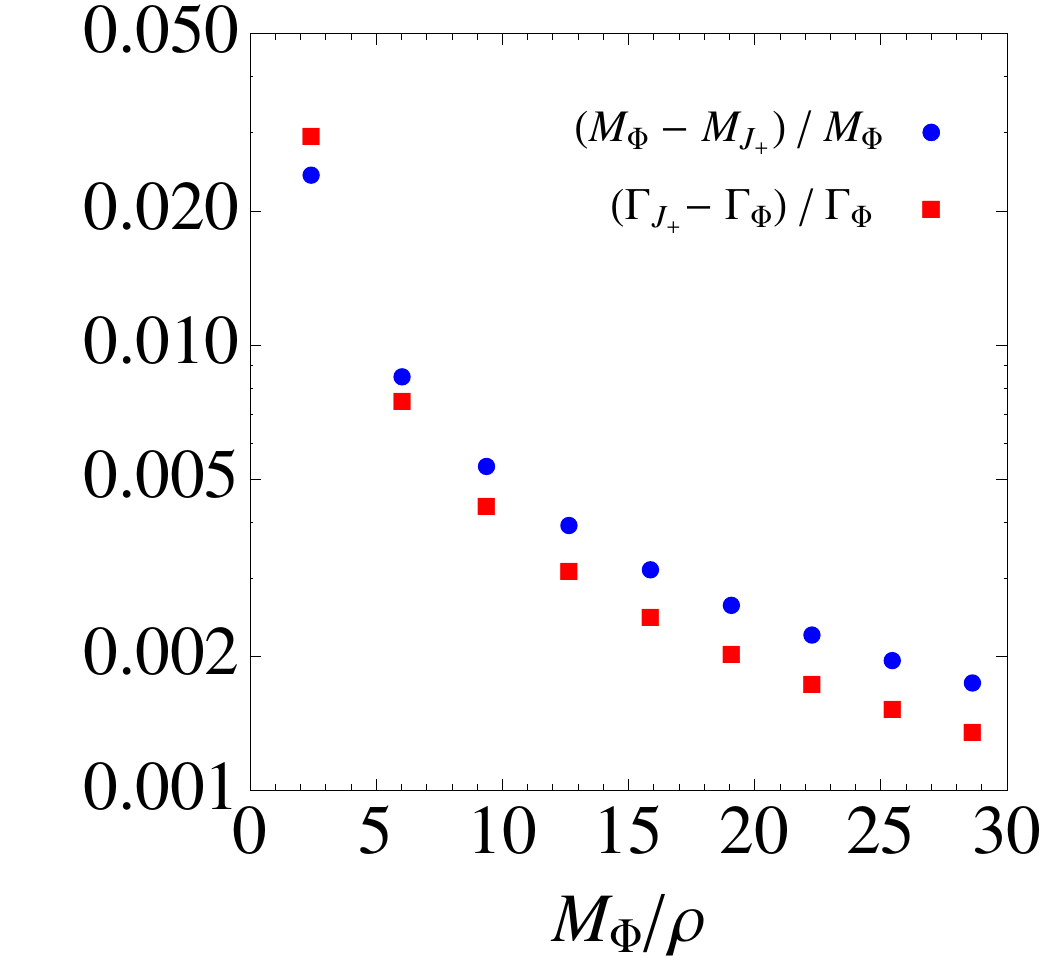}
\caption{\it Relative difference, in the plane $(M,\Gamma)$, between the positions of the zeros of $\Phi(p)$ and of $J_+(p)$ computed in the second Riemann sheet, corresponding to the poles and zeros of the Green's function $G_A(y_0,y_0)$, respectively. These zeros are denoted by $(M_\Phi,\Gamma_\Phi)$ and $(M_{J_+},\Gamma_{J_+})$, respectively.}
\label{fig:MGamma}
\end{figure} 

We display in Fig.~\ref{fig:AbsGA_resonances} contour plots in the
plane $(M/\rho,\Gamma/M)$ of the (common logarithm of the absolute
value of the) brane-to-brane Green's functions computed in the second
Riemann sheet. All Green's functions $G_A(y_\alpha,y_\beta;s)$ exhibit
poles in the complex $s$-plane located at the corresponding zeros of
the function $\Phi(p)$. In the case of $G_A(y_0,y_0)$, for each pole
of the Green's function there appears a zero close to it. The function
$J_+(p)$ in the numerator of $G_A(y_0,y_0)$, in
Eq.~(\ref{eq:GA_z0z0_asymp}), behaves, for large momentum, as
\be
J_+(p)\stackrel[\rho \ll |p|]{\simeq}{} \frac{ e^{-i(p/\rho+\pi/4)} }{ \sqrt{8\pi} } \left[ e^{i2p/\rho}-8i \left( \frac{p}{\rho} \right)^2 \right] \left(\frac{\rho}{p} \right)^{3/2}\,, \quad \Imaginary \left( (p/\rho)^2 \right) < 0 \,.
\ee
From a comparison with Eq.~(\ref{eq:Phi_large_p}), we conclude that the zeros of $G_A(y_0,y_0)$ turn out to be very close to its poles, and their residues approximately cancel. 
In order to quantitatively characterize the difference between the positions of these zeros and poles, we have displayed in Fig.~\ref{fig:MGamma} the relative difference in the plane $(M,\Gamma)$ of the location of the zeros of $\Phi(p)$ and of $J_+(p)$ in the second Riemann sheet, with $s_\Phi \equiv M_\Phi^2 - i M_\Phi \Gamma_\Phi$ and $s_{J_+} \equiv M_{J_+}^2 - i M_{J_+} \Gamma_{J_+}$, respectively. This difference is $\lesssim 3\%$, and rapidly decreases for heavier resonances.  Finally, let us notice that the Green's function $G_A(y_0,y_1)$ does not have any zero, while $G_A(y_1,y_1)$ has also zeros located in the real axis, corresponding to zeros in the denominator of Eq.~(\ref{eq:GA_z1z1_asymp}). Subsequently there is no suppression of the pole residues in these cases.

It is interesting to realize that the poles of the Green's function in the real axis of the complex $s$ plane correspond to eigenvalues of the EoM of the fluctuations~(\ref{eq:fAy}), a property that can be checked as follows.  The wave function $f_A(y)$ is subject to the following boundary condition in the UV brane and jumping conditions in the IR brane
\begin{equation}
C_{\rm UV}(p) \equiv \frac{\partial_y f_A(y)}{f_A(y)} \Bigg|_{y=0} = 0 \,, \qquad \Delta f_A(y_1) = 0  \,, \qquad \Delta  (\partial_y f_A)(y_1) = 0 \,.  \label{eq:fA_bc}
\end{equation}
In addition, for states with mass below the mass gap (the zero mode) regularity should be imposed at the singularity $y = y_s$ which implies $f_A(y) \stackrel[y \to y_s]{\simeq}{} (y_s - y)^{\frac{1}{2} \Delta_A^+}$, an IR behavior that can be written more explicitly as
\begin{equation}
f_A(z)  \stackrel[z_1 \ll z]{\propto}{} e^{-\sqrt{m_g^2 - p^2} z} \Theta(m_g^2 - p^2) + e^{i \sqrt{-m_g^2 + p^2} z} \Theta(p^2 - m_g^2) \,.  \label{eq:fA_largez}
\end{equation}
However, regularity in the IR singularity should not be imposed for states with mass above the mass gap. The general solution of the EoM of the fluctuations~(\ref{eq:fAy}) contains four integration constants $C_i^{I,II} \; (i = 1,2)$, i.e. two constants per region: i)~Region I: $0 < y < y_1$, and ii)~Region II: $y_1 < y < y_s$. The jumping conditions in the IR brane fix two of the constants. 

Let us first discuss the eigenvalue problem below the mass gap $(p < m_g)$. In this case, the regularity condition at $y = y_s$ fixes one of the integration constants, as the solution with $+$ and $-$ in the first and second exponents of Eq.~(\ref{eq:fA_largez}), respectively, is absent.  The remaining integration constant can only be fixed by normalization of the wave function. Finally, the UV boundary condition is fulfilled only for certain values of the momentum. From an explicit computation of $C_{\rm UV}(p)$, it turns out that
\begin{equation}
C_{\rm UV}(p) = G_A^{-1}(y_0,y_0;p) \,, \label{eq:CUV_Gy0y0}
\end{equation}
where the explicit expression of the UV-to-UV Green's function is given by Eq.~(\ref{eq:GA_z0z0_asymp}). Then, we conclude that the values of the momenta $p$ fulfilling the UV boundary condition $(C_{\rm UV}(p) = 0)$ correspond exactly to the poles of the Green's function, in our case to the zero mode $(p^2 = 0)$. Let us point out that only the zero mode (and not the resonances discussed in this section) corresponds to a genuine single bound state, and then to the solution of an eigenvalue problem for a Hermitian Hamiltonian $(p^2 \in \mathbb R)$~\footnote{An analysis similar to the one presented above was also performed for the radion field within the LDM in Ref.~\cite{Megias:2021mgj}. In this case, the condition~(\ref{eq:CUV_Gy0y0}) (with $G_A$ replaced by the Green's function of the radion) was also obtained, and it correctly predicted the mass of the radion corresponding to a single bound state below the mass gap.}.

Regarding the states with mass above the mass gap $(p > m_g)$, the three conditions in Eq.~(\ref{eq:fA_bc})  fix three of the integration constants.  As in the case of the bound state, the remaining integration constant can only be fixed by normalization of the wave function in the continuum, something that can be done, for instance, as $\langle f_{p} | f_{p^\prime} \rangle = \delta(p^2 - p^{\prime \, 2})$. It is precisely the absence of the regularity condition at $y = y_s$ which gives rise to a continuum spectrum, analogous to scattering states in quantum mechanics.

\subsection{Spectral functions}
\label{subsec:spectral_function}

\begin{figure}[t]
\centering
\includegraphics[width=4.5cm]{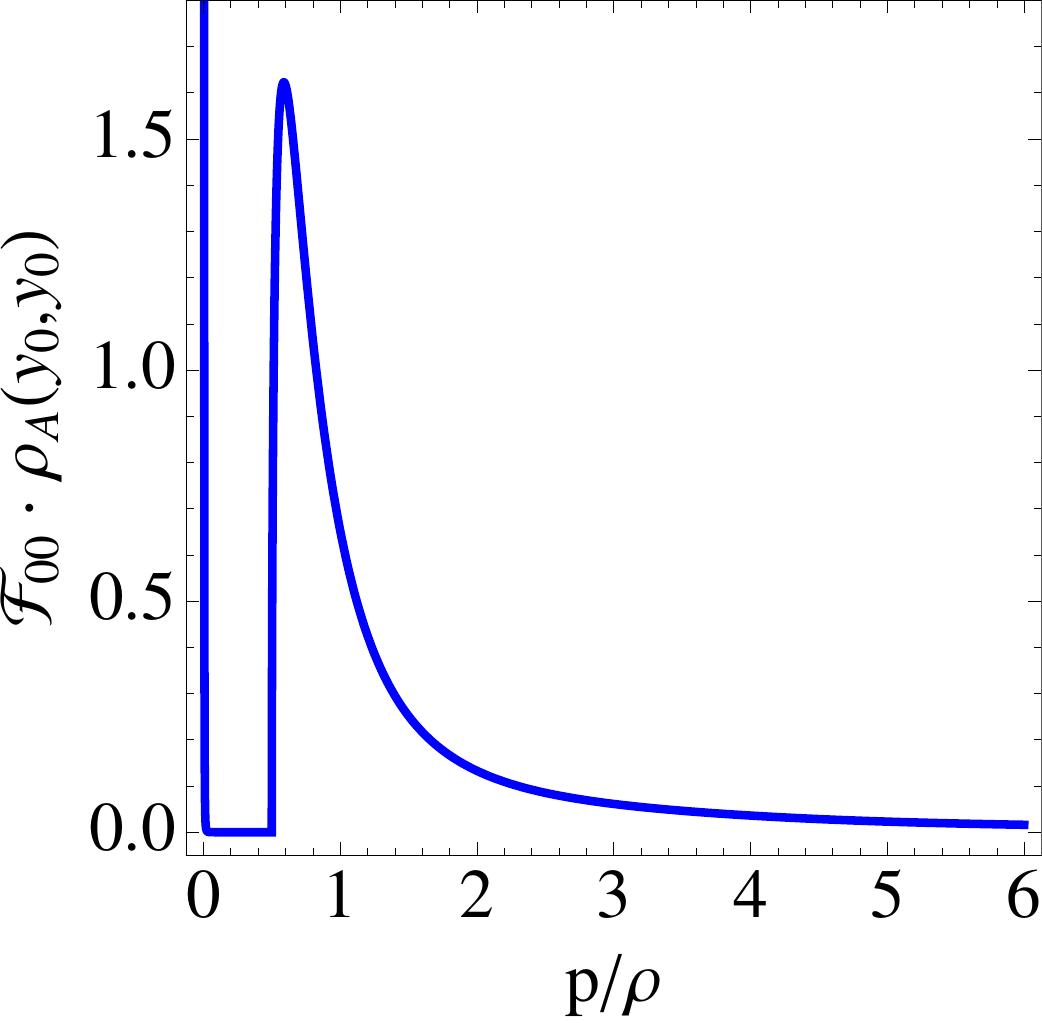} \hspace{0.10cm}
\includegraphics[width=4.7cm]{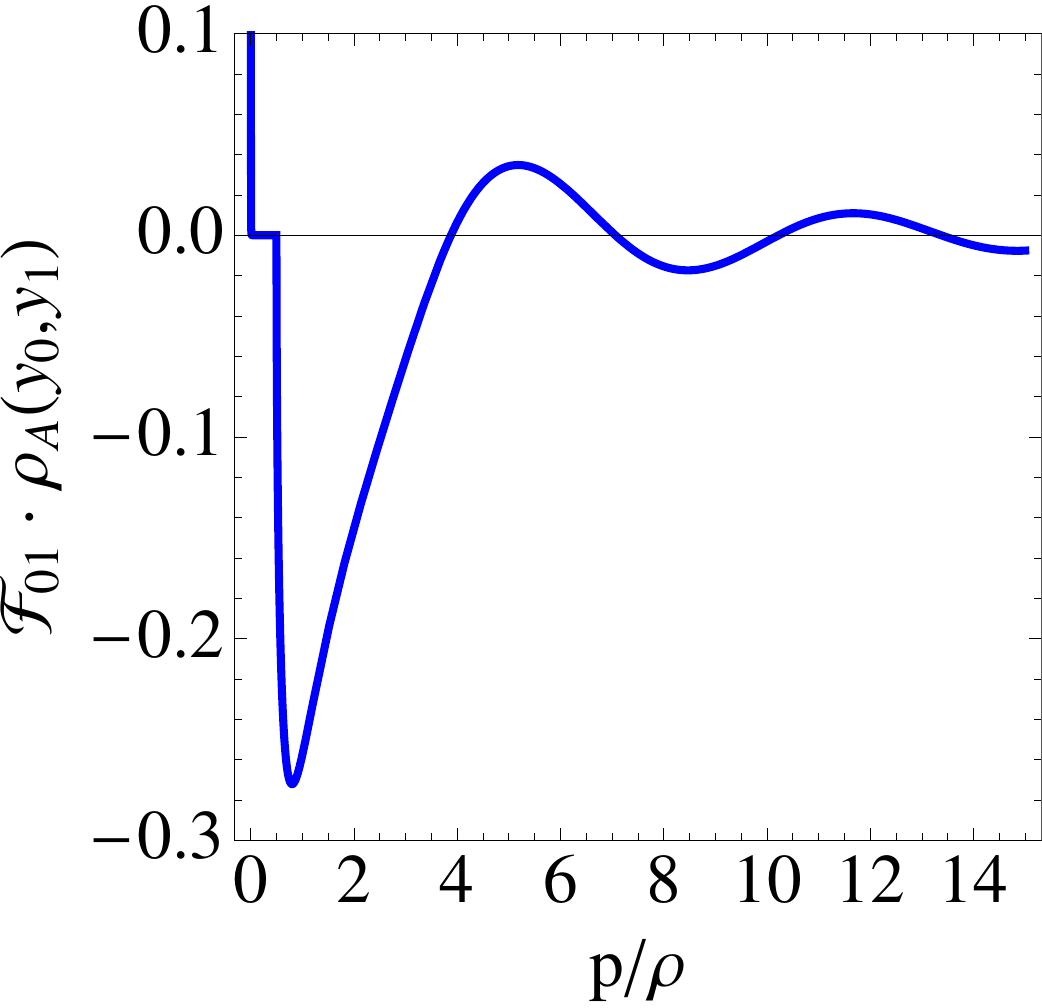} \hspace{0.10cm}
\includegraphics[width=4.7cm]{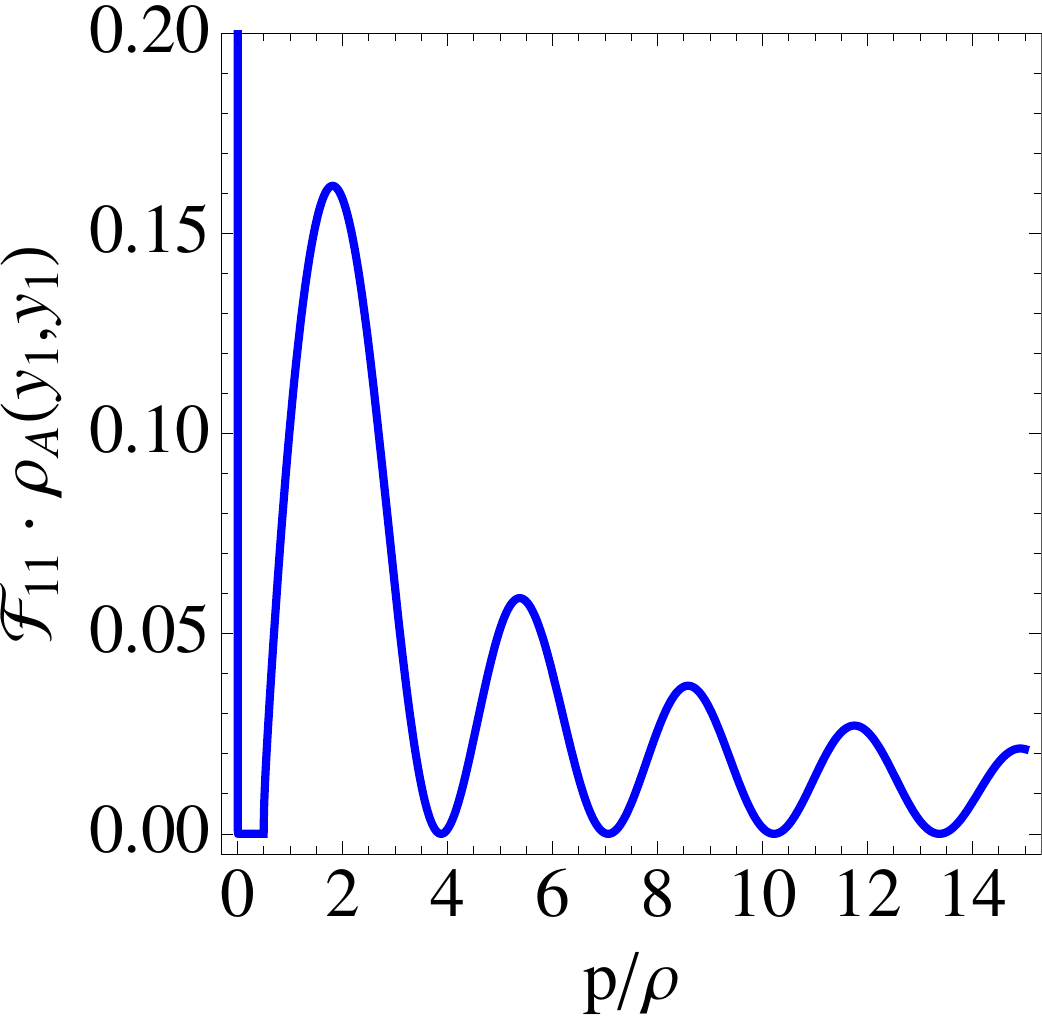} 
\caption{\it Rescaled spectral functions $\mathcal F_{00} \cdot \rho_A(y_0,y_0;p)$ (left panel), $\mathcal F_{01} \cdot \rho_A(y_0,y_1;p)$ (middle panel) and $\mathcal F_{11} \cdot \rho_A(y_1,y_1;p)$ (right panel) as a function of $p/\rho$, for a continuum gauge boson. We have used $A_1 = 35$ in all panels and assume time-like momenta $p^2>0$.
}
\label{fig:spectral}
\end{figure}

For time-like momenta, $p^2>0$, all Green's functions have imaginary contributions for values of $p> m_g=\rho/2$, which is not associated to a particle threshold decay, an intrinsic property of e.g.~unparticle theories. In this way we can define the corresponding spectral functions as
\be
\rho_A(y,y^\prime;s) \equiv - \frac{1}{\pi} \textrm{Im }G_A(y,y^\prime;s + i\epsilon) \,, \qquad s \equiv p^2 \,.
\label{eq:spectral}
\ee 
In Fig.~\ref{fig:spectral} we show the spectral functions $\rho_A(y_0,y_0;p)$, $\rho_A(y_0,y_1;p)$ and $\rho_A(y_1,y_1;p)$ as functions of $p/\rho$  where the prefactors, defined by Eq.~(\ref{eq:Fab}), make them approximately invariant for $p>0$ under a rescaling of the form of Eq.~(\ref{eq:prhok_rescaling}), i.e. under shifts of the value of $ky_s$~\cite{Megias:2019vdb}. By using the identity
\begin{equation}
\lim_{\epsilon \to 0^+} \frac{1}{x + i\epsilon} = \PP \frac{1}{x} - i\pi \delta(x) \,, \label{eq:PP}
\end{equation}
one can see that the small $p$ behavior of the Green's functions provided in Sec.~\ref{subsec:brane_to_brane} implies the existence of a Dirac delta behavior in the spectral functions at $p=0$,
\begin{equation}
\rho_A(y,y^\prime;s) = \frac{1}{y_s} \delta(s) + \cdots \,.
\end{equation}
This delta function appears in all the spectral functions of Fig.~\ref{fig:spectral}. 

Notice that while the spectral functions $\rho_A(y_0,y_0)$ and $\rho_A(y_1,y_1)$ are positive definite (as they are 4D spectral functions in the corresponding 4D branes), the UV-to-IR brane spectral function $\rho_A(y_0,y_1)$ is not, a fact that challenges the physical interpretation of the spectral function in 4D Quantum Field Theory, as it is positive definite by its probabilistic interpretation. This apparent contradiction was already noticed and addressed for the graviton field in the context of LDM, cf. Ref.~\cite{Megias:2021mgj}. Following similar ideas, we will briefly explain below how the positivity of the spectral function in our theory is understood.

From the 4D point of view, the spectral function $\rho_A(y,y^\prime;s)$ can be considered as the matrix element $(y,y^\prime)$ of an operator $\hat \rho_A$, i.e.~\footnote{We could also use a Dirac notation for the matrix element, $\rho_A(y,y^\prime;s) = \langle y | \hat \rho_A | y^\prime \rangle$. We thank Prof. L.L.~Salcedo for a private communication on the meaning of the spectral operator and its matrix elements.}
\begin{equation}
(\hat \rho_A)_y^{y^\prime} \equiv \rho_A(y,y^\prime;s)  \,.
\end{equation}
This operator acts on the infinite dimensional space parametrized by the coordinate~$y$.  We can similarly define for the Green's functions $G_A(y,y^\prime;s)$ the operator $\hat G_A$ such that
\begin{equation}
\hat \rho_A = - \frac{1}{\pi} \Imaginary \; \hat G_A \,, \qquad \textrm{where} \qquad \Imaginary \; \hat G_A = \frac{1}{2i} \left( \hat G_A - \hat G_A^\dagger \right) \,.
\end{equation}
Let us clarify at this point that it is expected that the operator $\hat\rho_A$ is positive semidefinite, but this does not imply that every matrix element is positive semidefinite. The elements of $\hat \rho_A$ form an infinite (continuous) dimensional matrix. Using the explicit expressions for the Green's function given by Eq.~(\ref{eq:GAypy1}), and taking into account the properties of Eqs.~(\ref{eq:ImAB}) and (\ref{eq:prop_conj}), it is possible to check that the determinant of any $2\times 2$ submatrix is vanishing, i.e.
\begin{equation}
(\hat\rho_A)_{2\times 2} = \left( \begin{array}{cc}
(\hat\rho_A)^y_{y}       &   (\hat\rho_A)^y_{y^\prime} \\
(\hat\rho_A)^{y^\prime}_{y} &   (\hat\rho_A)^{y^\prime}_{y^\prime}
\end{array}\right)  \ \Longrightarrow \ \det (\hat\rho_A)_{2\times 2} = (\hat\rho_A)^y_{y} (\hat\rho_A)^{y^\prime}_{y^\prime} - (\hat\rho_A)^y_{y^\prime} (\hat\rho_A)^{y^\prime}_y = 0 \,.
\end{equation}
This property, together with $(\hat\rho_A)^{y^\prime}_y = (\hat\rho_A)^{y}_{y^\prime}$, cf. Eq.~(\ref{eq:GA_sim}), implies that the matrix $\hat\rho_A$ turns out to have a factorizable form, i.e. any matrix element can be written in the form
\begin{equation}
\mathcal (\hat\rho_A)^y_{y^\prime} = \rho_y \rho_{y^\prime} \qquad \textrm{where} \qquad \rho_y = \sqrt{(\hat\rho_A)_y^y} \,. \label{eq:rho_ab}
\end{equation}
Given this factorization property, it turns out that the operator $\hat\rho_A$ is positive semidefinite, and all its eigenvalues are zero except one $\lambda(p)$, which is given by the trace of the matrix, i.e.
\begin{equation}
\lambda(p) = \tr \hat \rho_A =  \int_0^{y_s} dy \, \rho_A(y,y;p) \,.
\end{equation}
In particular, note that $(\hat\rho_A)^y_y = \rho_y^2 \ge 0$ implies that $\lambda(p) \ge 0$~\footnote{Notice that a symmetric matrix is positive semidefinite if and only if all its eigenvalues are non-negative.}.

\begin{figure}[htb]
\centering
\includegraphics[width=7cm]{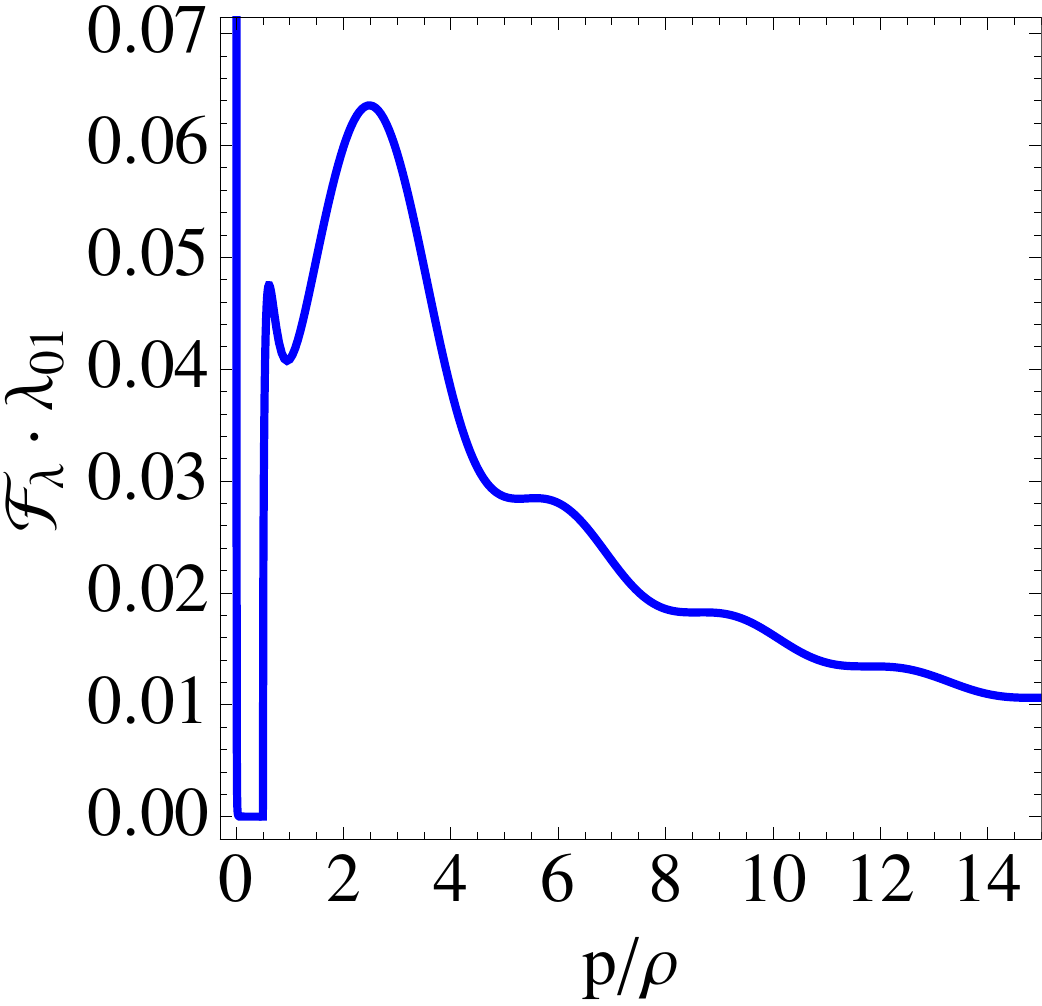}
\caption{\it The UV-to-IR brane contribution to the rescaled 4D spectral function:  $\mathcal F_\lambda \cdot \lambda_{01}(p)$. We have used $A_1 = 35$ and assume time-like momenta $p^2>0$.}
\label{fig:spectralPhi}
\end{figure}

In order to perform this integral, let us split it into two domains,
\begin{equation}
\lambda(p) = \lambda_{01}(p) + \lambda_{1s}(p)       \,,
\end{equation}
where
\begin{equation}
\lambda_{01}(p) \equiv  \int_0^{y_1} dy \, \rho_A(y,y;p) \,, \qquad  \lambda_{1s}(p) \equiv  \int_{y_1}^{y_s} dy \, \rho_A(y,y;p)  \,.
\end{equation}
The integral of $\lambda_{01}$ can easily be performed, and the result
is plotted in Fig.~\ref{fig:spectralPhi}.
The prefactor, defined as
\begin{equation}
\mathcal F_\lambda = \rho^2 \,, \label{eq:F_lambda}
\end{equation}
makes it almost invariant under shifts of $ky_s$. Notice the appearance of softened peaks, at values $p/\rho \simeq n \pi$, i.e. at the positions of the resonances obtained in Sec.~\ref{sec:Resonances}. There is also a Dirac delta behavior at $p=0$. Focusing now on $\lambda_{1s}$, it turns out to be divergent due to the term $\propto 1/(y_s-y)$ in Eq.~(\ref{eq:Q}), so it needs to be regularized. We will do it by introducing the cutoff $\bar\epsilon$ in the integral, so integrating up to $ky_s-\bar\epsilon$. The integral will then be dominated by its value at $ky_s-\bar\epsilon$ giving a term proportional to $-\log\bar\epsilon$. The final result is then
\be
\lambda(s)=\delta(s)+\left[ -\frac{\log\bar\epsilon}{2\pi\rho}\lambda_{\rm cont}(s)+\mathcal O (\bar\epsilon^0) \right] \Theta(s-m_g^2) \,, \qquad 
\lambda_{\rm cont}(s) = (s-m_g^2)^{-1/2} \,,
\label{eq:desclambda}
\ee
where $\delta(s)$ is the contribution of the zero mode, $\lambda_{\rm cont}(s)$ the contribution from the continuum, and the $\mathcal O(\bar\epsilon^0)$ term denotes the contribution from resonances. As we can see the contribution from the continuum is the dominant one and comes entirely from the singularity at $y_s$. 

Finally, let us point out that the Green's function and spectral function can be written also in the form
\begin{eqnarray}
G_A(y,y^\prime;s) &=& \frac{f_0(y) f_0(y^\prime)}{|| f_0 ||^2}\frac{1}{s + i \epsilon}  +  \int_{m_g^2}^\infty dm^2 \sigma(m^2) \frac{f_{m^2}(y) f_{m^2}(y^\prime)}{s - m^2 + i \epsilon} \,, \label{eq:GA_mass} \\
\rho_A(y,y^\prime;s) &=& \frac{f_0(y) f_0(y^\prime)}{ || f_0||^2} \delta(s)  +  \sigma(s) f_{s}(y) f_{s}(y^\prime) \Theta(s - m_g^2) \,,  \label{eq:rhoA_mass}
\end{eqnarray}
respectively, where $f_0(y)$ is the zero mode eigenfunction, $f_{m^2}(y)$ are continuum eigenfunctions, $|| f_0 ||^2 \equiv \int_0^{y_s} dy \, f_0(y)^2$ is the squared norm, and $\sigma(s)$ is a spectral density in the continuum which is finite and whose relation with $\lambda(s)$ can be obtained by integrating Eq.~(\ref{eq:rhoA_mass}). These formulas generalize well-known expressions for the discrete case. A more detailed analysis of $\lambda(s)$ as well as the decomposition of Eqs.~(\ref{eq:GA_mass})-(\ref{eq:rhoA_mass}) will be studied in a forthcoming publication~\cite{Megias:prep}.

\subsection{Unparticle vs.~resonant contribution to Green's functions}
\label{sec:unpVscont}

\begin{figure}[t]
\centering
\includegraphics[width=4.7cm]{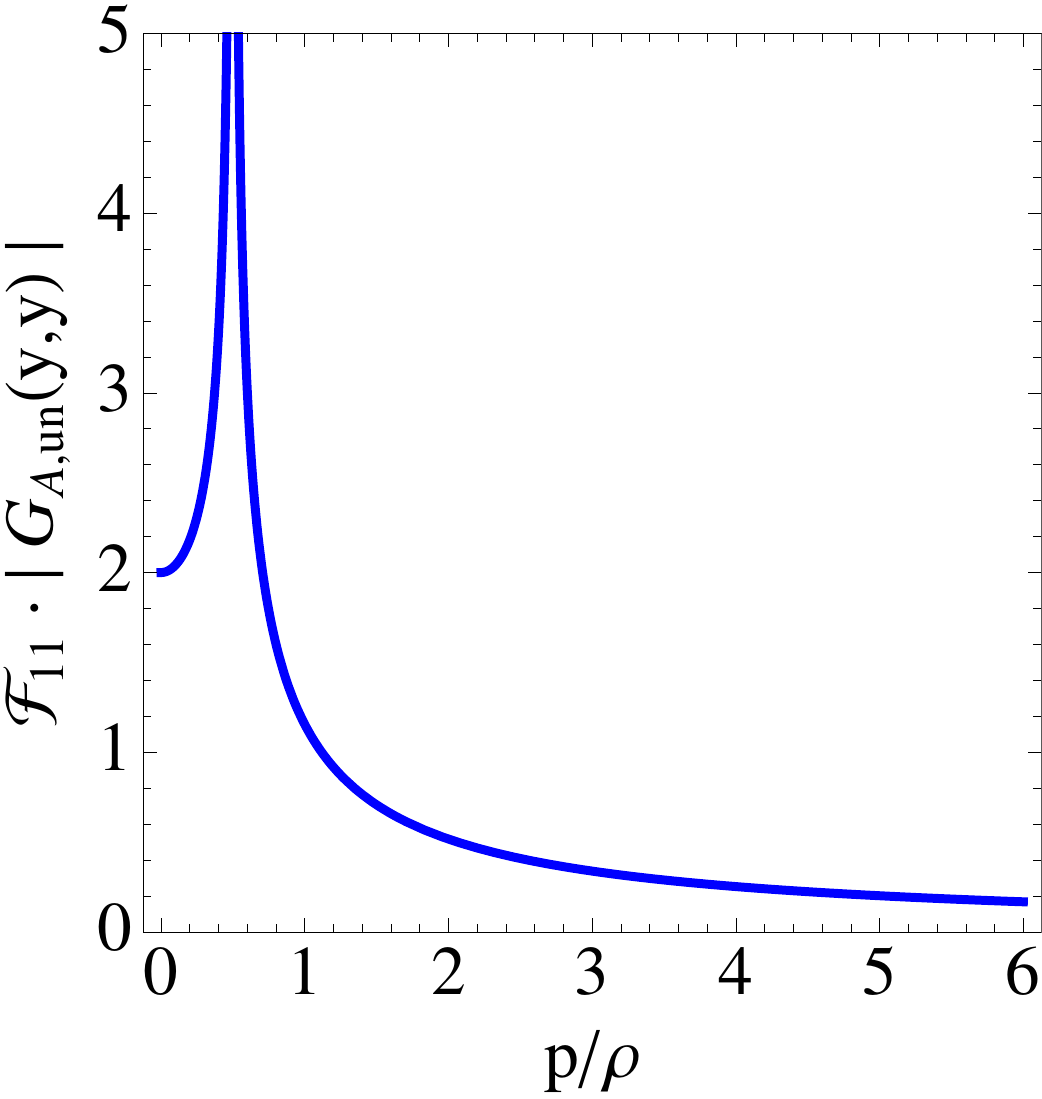}  \hspace{0.10cm}
\includegraphics[width=4.9cm]{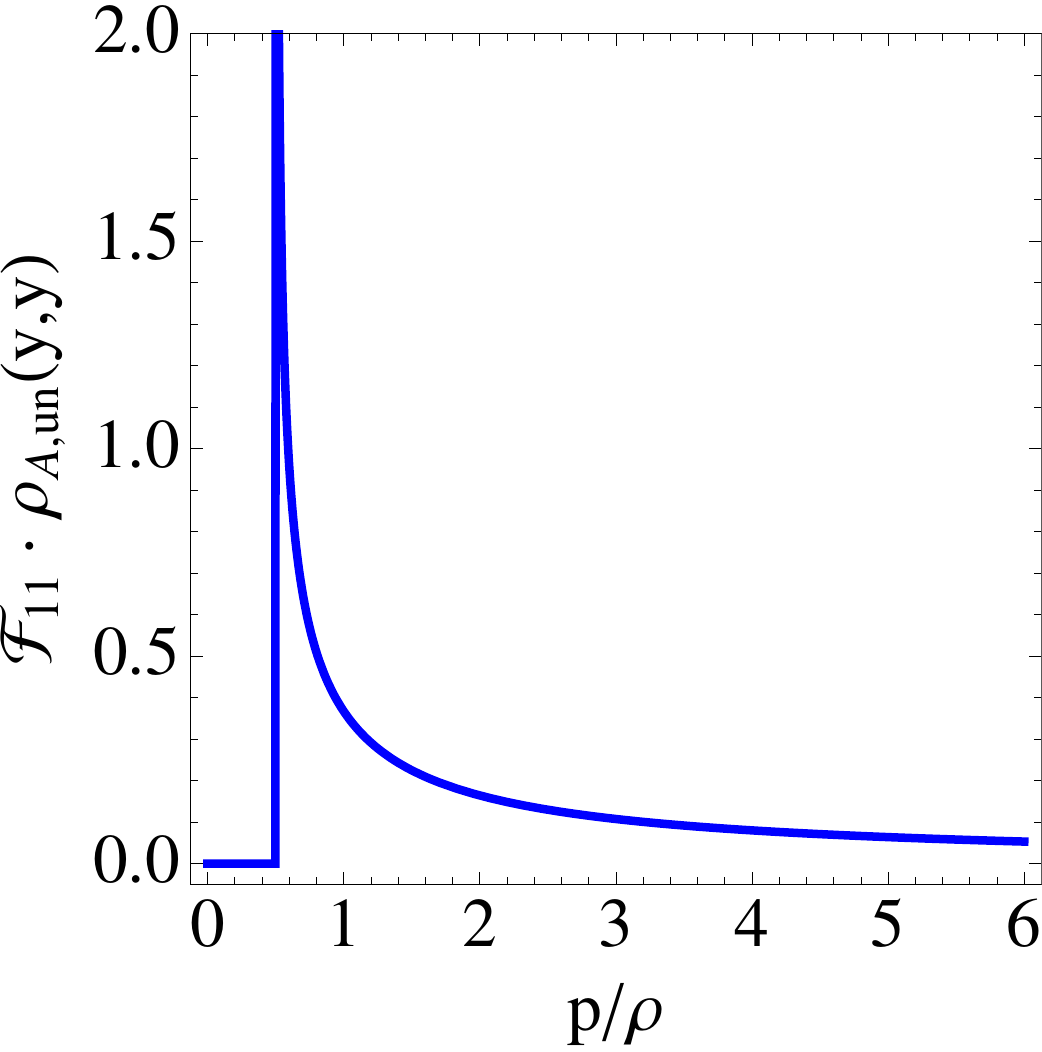}  \hspace{0.10cm}
\includegraphics[width=5.1cm]{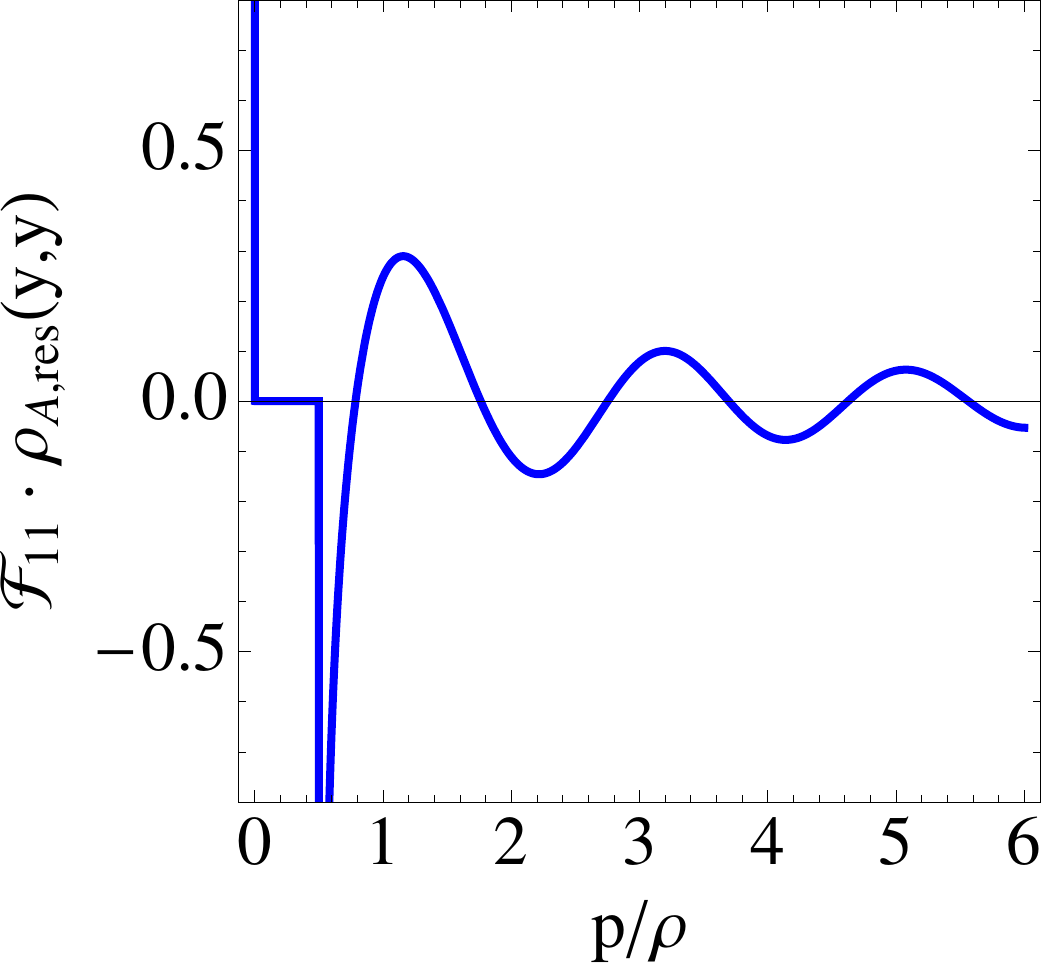}
\caption{\it Plots of the unparticle and resonant contributions to the Green's function. We display the result for $\mathcal F_{11} \cdot |G_{A,\un}(y,y^\prime;p)|$ (left panel), $\mathcal F_{11} \cdot \rho_{A,\un}(y,y^\prime;p)$ (middle panel) and $\mathcal F_{11} \cdot \rho_{A,\res}(y,y^\prime;p)$ (right panel), as functions of $p/\rho$. We have used $y = y^\prime = (y_1 + y_s)/2$ and $A_1 = 35$ in all panels, and assume time-like momenta $p^2>0$.
}
\label{fig:G_un}
\end{figure}

Once we have studied the resonances appearing in the Green's functions, related to the zeros of $\Phi(p)$, one question arises: can the Green's functions be considered just as summations of resonant contributions or, is there any genuine continuous contribution on top of the resonances? To answer this question, let us have a look at the explicit expression of the Green's function given by Eq.~(\ref{eq:GAypy1}). Notice that in the three regions ($y_\downarrow , y_\uparrow \le y_1$, $y_\downarrow \le y_1 < y_\uparrow$ and $y_1 < y_\downarrow , y_\uparrow$), the function $1/\Phi(p)$ multiplies the full expression. Taking into account the structure of the function $\Q(y)$ given by the last line of Eq.~(\ref{eq:Q}), we can see that only in the last region one can split the Green's function into two terms
\begin{equation}
G_A(y,y^\prime;p) = G_{A,\un}(y,y^\prime;p) + G_{A,\res}(y,y^\prime;p) \,, \qquad y_1 < y_\downarrow , y_\uparrow \,, \label{eq:GA_unres}
\end{equation}
where
\begin{align}
G_{A,\un}(y,y^\prime;p) &= -\frac{k}{\rho^2} \frac{1}{k(y_s - y_\downarrow)} \left( \frac{y_s - y_\uparrow}{ y_s - y_\downarrow} \right)^{\Delta_A^+/2} \frac{1}{\delta_A(p)} \,, \label{eq:G_un} \\
G_{A,\res}(y,y^\prime;p) &= \frac{k}{\rho^2} \left[ k(y_s - y_\downarrow) \cdot k(y_s - y_\uparrow) \right]^{\Delta_A^+/2} \frac{1}{\delta_A(p)}  \cdot \frac{\Psi(p)}{\Phi(p)} \,, \label{eq:G_res}
\end{align}
corresponding to a continuous and a resonant contribution, respectively. The first contribution $G_{A,\un} \propto \delta_A^{-1}$ is free of resonances, as $\Phi(p)$ does not appear in its expression, and it has a structure similar to the propagator of gapped unparticles~\cite{Delgado:2008gj}. We display in Fig.~\ref{fig:G_un} 
the results for $|G_{A,\un}(y,y;p)|$, for the corresponding spectral function $\rho_{A,\un}(y,y;p)$, and for the resonant contribution to the spectral function $\rho_{A,\res}(y,y;p)$, with $y = (y_1 + y_s)/2$. Note that the unparticle contribution does not have any zero mode, and both the Green's function and the spectral function have divergent behaviors when the momentum approaches the mass gap from above, i.e. $\rho_{A,\un}(y,y^\prime,p) \stackrel[p^2 \to m_g^{2\, +}]{\longrightarrow}{} +\infty$. 

An interesting property of the functions $\Phi(p)$ and $\Psi(p)$ is that~\footnote{The property of Eq.~(\ref{eq:PhiPsiI_II}) follows from $(\Delta_A^{\pm})^{\rm II} = (\Delta_A^{\mp})^{\rm I}$.}
\begin{equation}
\Phi^{\rm II}(p) = \Psi^{\rm I}(p) \,, \label{eq:PhiPsiI_II}
\end{equation}
where the superindexes ${\rm I}$ and ${\rm II}$ stand for the first and second Riemann sheet, respectively. This property implies that while $G_{A,\res}(y,y^\prime;p)$ has poles in the second Riemann sheet, it has zeros in the first Riemann sheet that are located in the complex $s$ plane at the same positions as the poles of the second Riemann sheet. Note, however, that these are not zeros of the full Green's function $G_{A}(y,y^\prime;p)$, but only of the resonant part.

Finally notice that the (divergent) continuous eigenvalue of the 4D spectral function $\lambda(p)$, which was computed in Sec.~\ref{subsec:spectral_function}, $\lambda_{\rm cont}(s)$, corresponds to a contribution from unparticles with a dimension $d_U=3/2$ and mass gap $m_g$~\cite{Delgado:2008gj,Landshoff:1963}.

\section{Gauge bosons with Dirichlet boundary condition}
\label{sec:Dirichlet}

In the considered extension of the SM we will use Dirichlet boundary condition for the extra gauge bosons ($W_R,Z_R$) on the UV brane, see Sec.~\ref{sec:custodial_model}. To study this, one should start from the general solution of Eq.~(\ref{eq:GAy}), which is given by Eq.~(\ref{eq:solGA_homogeneous}). While the Neumann boundary condition in the UV brane, used in Eq.~(\ref{eq:GA_bc}), leads to $(\partial_y G_A)(y_0) = 0$, the Dirichlet boundary condition is given by
\begin{equation}
G_A^{(-+)}(y_0) = 0 \,. \label{eq:Dirichlet_bc}
\end{equation}
This condition is supplemented by the other conditions in Eq.~(\ref{eq:GA_bc}). As an example, the integration constants in Region I turn out to fulfill the relation
\begin{equation}
C_1^I = - \frac{Y_\alpha\left( \frac{p}{k} \right)}{J_\alpha\left( \frac{p}{k} \right)}  C_2^I  \qquad \textrm{with} \qquad \alpha = \left\{ 
\begin{array}{cl}
0  & \quad \textrm{(Neumann)}  \\
1  & \quad \textrm{(Dirichlet)}  
\end{array} \,. \right. \label{eq:CI1}
\end{equation}
As a consequence, the difference between the Green's functions with Neumann and Dirichlet boundary conditions will be in the indexes of some of the Bessel functions. Following a procedure similar to the one explained in Sec.~\ref{sec:gauge_bosons_z0z1}, the Green's functions with Dirichlet boundary condition turn out to be
\begin{equation}
G_A^{(-+)}(y,y^\prime;p) =  \left\{ 
\begin{array}{cc}
\frac{\pi}{2k} e^{k (y + y^\prime)}  \frac{\bar\PP(y_\downarrow) \Z(y_\uparrow)}{\Omega(p)}  & \quad  y_\downarrow, y_\uparrow \le y_1   \\
-\frac{2}{\rho} e^{k y_\downarrow} \left( k(y_s-y_\uparrow) \right)^{\Delta_A^+/2} \frac{\bar\PP(y_\downarrow)}{\Omega(p)} & \quad  y_ \downarrow \le y_1 < y_\uparrow  \\
 \left( \frac{y_s - y_\uparrow}{y_s - y_\downarrow} \right)^{\Delta_A^+/2} \delta_A^{-1}  \cdot \frac{\bar\Q(y_\downarrow) }{\Omega(p)}   &  \quad y_1 < y_\downarrow , y_\uparrow 
\end{array} \,. \right. \label{eq:GAmp}
\end{equation}
We have used the notation
\begin{align}
\Omega(p) &= \Z(0) \,, \qquad \bar \Psi(p) =  Y_1( p/k) J_-(p/\rho)  -  J_1(p/k) Y_-(p/\rho) \,,  \nonumber \\ 
\bar \PP(y) &=  Y_1 \left( p/k \right) J_1\left( e^{ky} p/k  \right)   - J_1\left( p/k \right)  Y_1 \left( e^{ky} p/k \right)  \,,  \nonumber \\
\bar{\Q}(y) &= - \frac{k}{\rho^2}\frac{1}{k(y_s-y)} \left[\Omega(p) -   (k(y_s-y))^{\delta_A} \bar\Psi(p)\right]   \,, \label{eq:Qbar} 
\end{align}
where the function $\Z(y)$ is defined in Eq.~(\ref{eq:Q}). Note that in the limit $y \to y_0$ one has $\bar\PP(y_0) = 0$, so that
\begin{eqnarray}
G_A^{(-+)}(y_0,y^\prime;p) &=& 0\,, \label{eq:GAz0zpmp}
\end{eqnarray}
a property that is consequence of the boundary condition of Eq.~(\ref{eq:Dirichlet_bc}). As in the case of gauge bosons with Neumann boundary condition in the UV brane, this Green's function also fulfills the property $G_A^{(-+)}(y,y^\prime;p) = G_A^{(-+)}(y^\prime,y;p)$.

The expression for the IR-to-IR Green's function $G_A^{(-+)}(y_1,y_1;p)$ reduces to
\begin{equation}
G_A^{(-+)}(y_1,y_1;p)^{-1} = -\frac{\rho^2}{2k}  \cdot \frac{\Omega(p)}{\bar \PP(y_1)} \simeq -\frac{\rho^2}{2k} \left[ \Delta_A^+  + 2 \frac{p}{\rho} \frac{J_0\left( \frac{p}{\rho} \right)}{ J_1\left( \frac{p}{\rho} \right) }  \right] \,, \label{eq:GA_z1z1_asymp_Dirichlet} 
\end{equation}
where in the second equality we have assumed $p \ll k$, while the other brane-to-brane Green's functions, $G_A^{(-+)}(y_0,y_0;p)$ and  $G_A^{(-+)}(y_0,y_1;p)$, are vanishing as a consequence of Eq.~(\ref{eq:Dirichlet_bc}), cf. Eq.~(\ref{eq:GAz0zpmp}). The limit $p \ll \rho$ of the IR-to-IR Green's function is
\begin{equation}
G_A^{(-+)}(y_1,y_1;p)^{-1} \stackrel[p \ll \rho]{\simeq}{} -2 \frac{\rho^2}{k} + \frac{5}{4} \frac{p^2}{k} + \mathcal{O}(p^4)   \,,  \label{eq:GAmp_z1z1_p0} 
\end{equation}
while its behaviors in the regime $\rho \ll p$  for time-like momenta, $p^2>0$, and space-like momenta, $p^2 < 0$, (and $p \ll k$) are the same as for the Green's function $G_A(y_1,y_1;p)$, cf. Eqs.~(\ref{eq:GA_z1z1_largep}) and (\ref{eq:GA_z1z1sl_largep}). 

\begin{figure}[t]
\centering
\includegraphics[width=5.1cm]{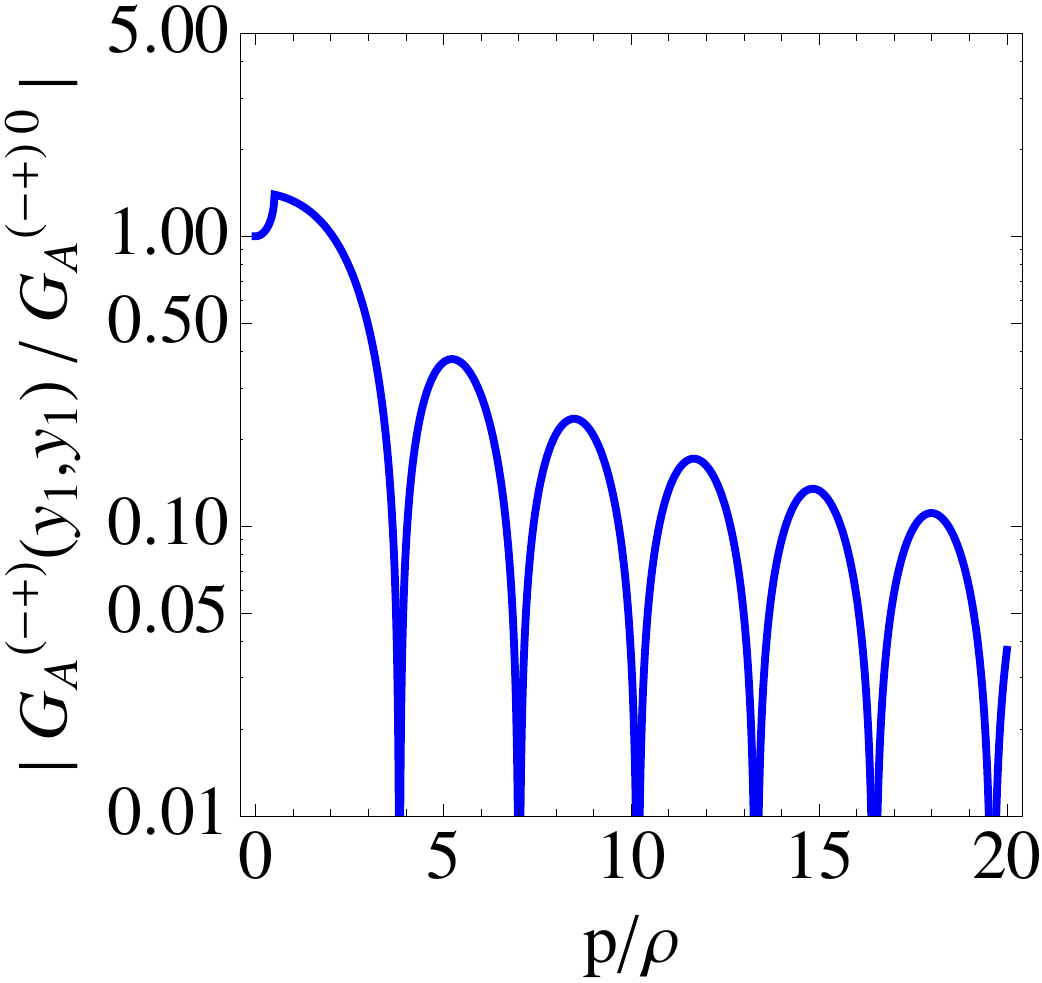}
\includegraphics[width=4.7cm]{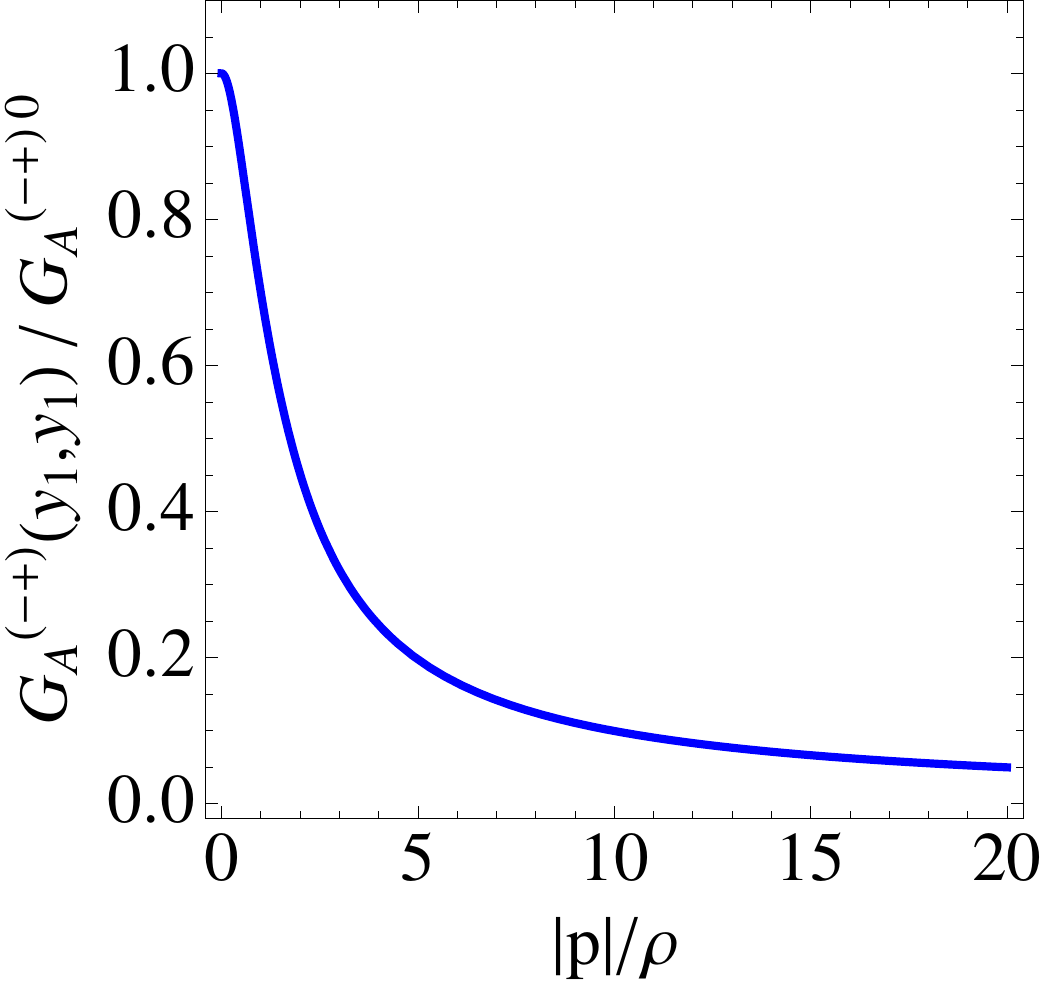}
\includegraphics[width=4.9cm]{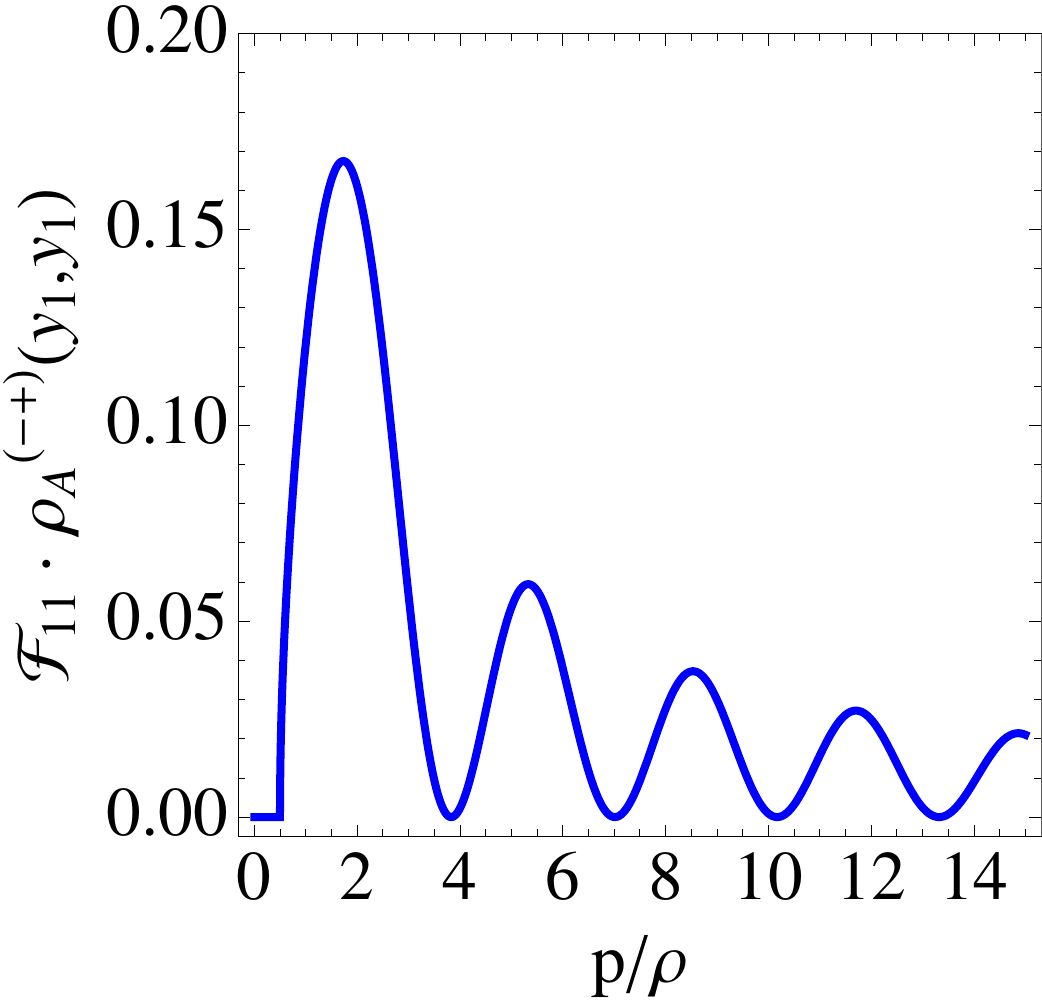}
\caption{\it Plots of the IR-to-IR Green's function $|G_A^{(-+)}(y_1,y_1;p)/G_A^{(-+) \, 0}|$ for $p^2 > 0$ (left panel), $G_A^{(-+)}(y_1,y_1;|p|)/G_A^{(-+) \, 0}$ for $p^2 < 0$ (middle panel), and the rescaled spectral function $\mathcal F_{11} \cdot \rho_A^{(-+)}(y_1,y_1;p)$ (right panel) as functions of $p/\rho$. We have used $A_1 = 35$ in all panels.
}
\label{fig:Green_gauge_Dirichlet}
\end{figure}

In the following we denote the zero momentum limit of the IR-to-IR Green's function as $G_A^{(-+)\, 0} = -k/(2\rho^2)$. We display in Fig.~\ref{fig:Green_gauge_Dirichlet} 
the results for the normalized IR-to-IR Green's function $|G_A^{(-+)}(y_1,y_1;p)/G_A^{(-+)\,0}|$ for time-like momenta $p^2 > 0$ (left panel), and $G_A^{(-+)}(y_1,y_1;|p|)/G_A^{(-+)\, 0}$ for space-like momenta $p^2 < 0$ (middle panel). In the latter case, the Green's function is purely real, and $G_A^{(-+)}(y_1,y_1,|p|)$ decreases like the inverse power of $|p|$ for momenta $|p| \gg \rho$, i.e. $\sim \rho/|p|$. 

Finally, it is displayed in the right panel of Fig.~\ref{fig:Green_gauge_Dirichlet} the result for the rescaled spectral function $\mathcal F_{11} \cdot \rho_A^{(-+)}(y_1,y_1;p)$, where the prefactor $\mathcal F_{11}$ is defined in Eq.~(\ref{eq:Fab}). Notice that gauge bosons with Dirichlet boundary conditions do not have zero modes, so that no Dirac delta behavior in the spectral function at $p=0$ is present in this case.

\begin{figure}[htp]
\centering
\includegraphics[width=5.8cm]{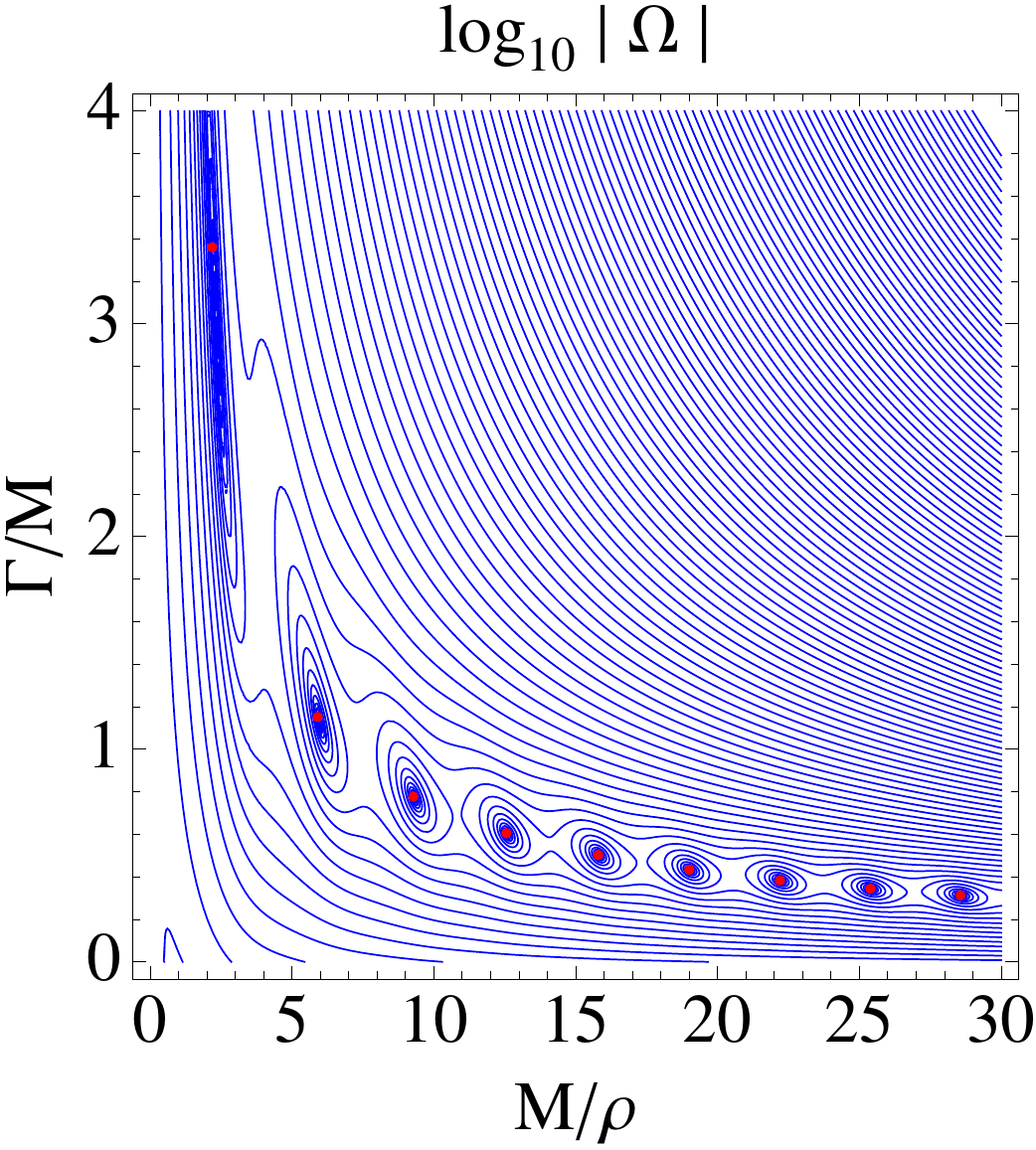} \hspace{0.50cm}
\includegraphics[width=5.8cm]{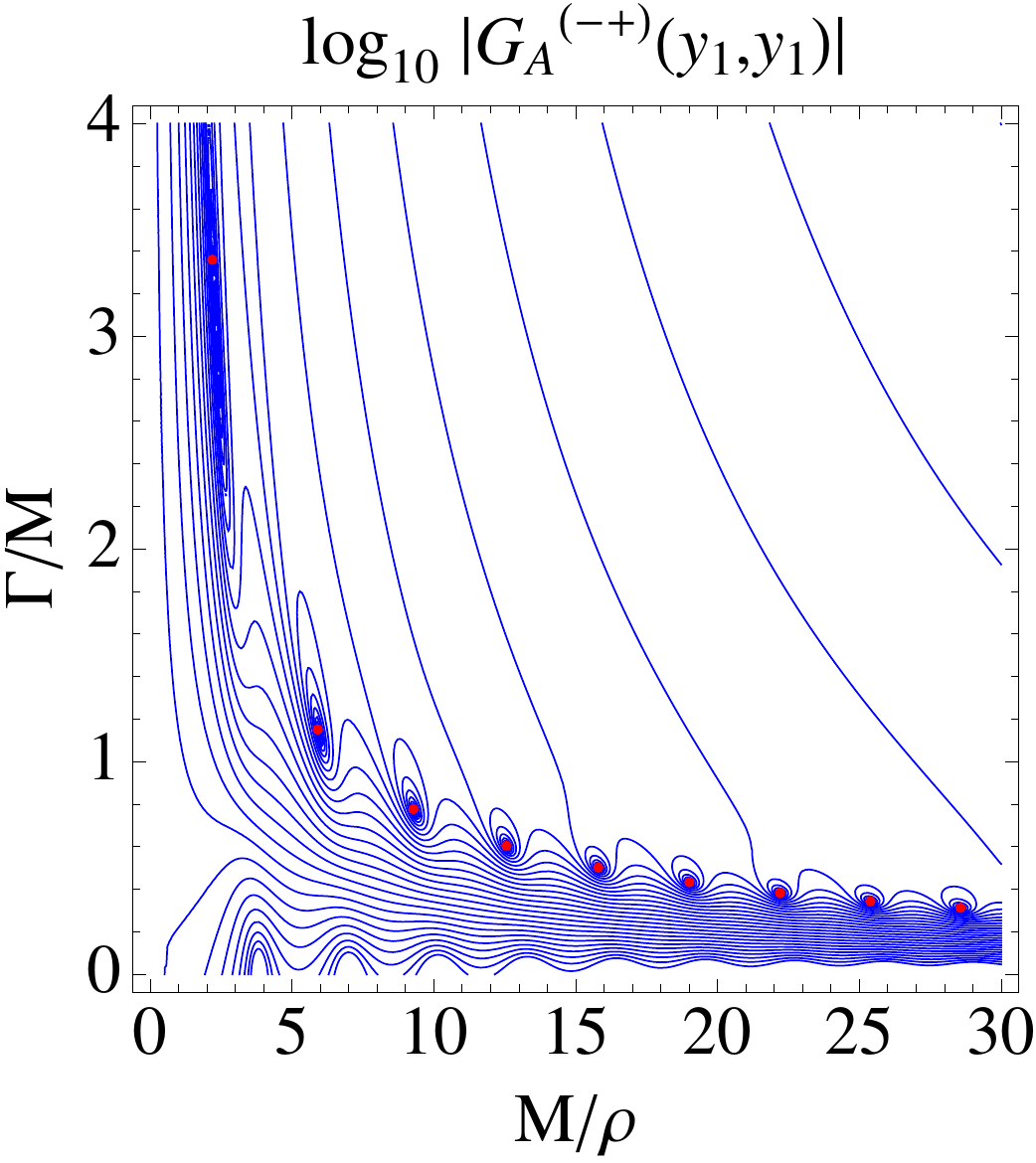}
\caption{\it Contour plot in the plane $(M/\rho,\Gamma/M)$ of $\log_{10}|\Omega(p)|$ cf. Eq.~(\ref{eq:Q}) (left panel), and the absolute value of the IR-to-IR Green's function $\log_{10} |G_A^{(-+)}(y_1,y_1)|$ (right panel). The (red) dots stand for the positions of the poles of the Green's function with Dirichlet boundary condition as predicted by Eq.~(\ref{eq:Z0_large_p}), and given by the analytical formula of Eq.~(\ref{eq:p2Lambert}). We have considered $A_1 = 35$.}
\label{fig:AbsGAmp_resonances}
\end{figure}

We now study the Green's function in the complex plane. One can see from Eq.~(\ref{eq:GAmp}) that the pole structure of the Green's function $G_A^{(-+)}(y,y^\prime;p)$ corresponds to the zeros of $\Omega(p)$. We display in the left panel of Fig.~\ref{fig:AbsGAmp_resonances} a contour plot of $\log_{10}|\Omega(p)|$ in the second Riemann sheet, the lightest resonances appearing in the complex plane at
\begin{equation}
(M/\rho,\Gamma/M) = (2.36,2.95), (5.98,1.13), (9.32,0.771), (12.59,0.603), (15.82,0.501) , \cdots  \,.
\end{equation}

We can analytically study these zeros in a way similar to the procedure explained in Sec.~\ref{sec:Resonances}. The expansion of $\Omega(p)$ at large momentum $\rho \ll |p|$ $(|p| \ll k)$ leads to
\begin{equation}
\Omega(p) \stackrel[\rho \ll |p|]{\simeq}{}  \frac{e^{-i(p/\rho + \pi/4)}}{\sqrt{2 \pi^3}}   \left[ e^{i2p/\rho} -8i \left(\frac{p}{\rho} \right)^2  \right]  \frac{k}{\rho} \left( \frac{\rho}{p} \right)^{5/2} \,, \quad \Imaginary \left( (p/\rho)^2 \right) < 0 \,. \label{eq:Z0_large_p}
\end{equation}
This asymptotic behavior is similar to the one of $\Phi(p)$, cf. Eq.~(\ref{eq:Phi_large_p}), hence we find that at this order of the computation the zeros of $\Omega(p)$ are located at the same positions as the zeros of~$\Phi(p)$, and they are given by Eq.~(\ref{eq:p2Lambert}).  We conclude that the poles of the Green's function for gauge bosons  $G_A^{(-+)}(y,y^\prime;p)$ are located very close to the poles of $G_A(y,y^\prime;p)$. The relative difference between the location of the poles in both cases is $\lesssim 2\%$ for the lightest resonances, and $\lesssim 0.5\%$ for resonances $M/\rho \gtrsim 10$.  Finally, let us mention that the relative error of the approximate formula of Eq.~(\ref{eq:p2Lambert}) with respect to the true zeros of $\Omega(p)$  is a factor $2$--$3$ better than for the case of $\Phi(p)$, except for the lightest resonance, for which is similar, cf. Sec.~\ref{sec:Resonances}. 

The Green's function for gauge bosons with Dirichlet boundary condition in the UV brane also has unparticle and resonant contributions, as for the massless case in Eq.~(\ref{eq:GA_unres}). The unparticle contribution turns out to be identical as for massless gauge bosons, and it is given by Eq.~(\ref{eq:G_un}).

Finally, we can study as well the positivity of the spectral operator $\hat \rho_A^{(-+)}$. The procedure is similar to the one presented in Sec.~\ref{subsec:spectral_function}, leading to a single non-vanishing eigenvalue $\lambda^{(-+)}(p)$ given by the trace of the matrix
$(\hat\rho_A^{(-+)})^{y'}_y$. As for the case of gauge bosons with Neumann boundary conditions, the integral over the $y$ coordinate has to be regularized with a cutoff $\bar\epsilon$, leading to a divergent unparticle contribution from the region $y_1<y<y_s$, and other finite contribution from the resonances. The result is given by Eq.~(\ref{eq:desclambda}) but without the pole at the origin, i.e.
\be
\lambda^{(-+)}(s)=\left[ -\frac{\log\bar\epsilon}{2\pi\rho}\lambda_{\rm un}(s)+\mathcal O (\bar\epsilon^0) \right] \Theta(s-m_g^2) \,, \qquad 
\lambda_{\rm un}(s) = (s-m_g^2)^{-1/2} \,,
\label{eq:desclambdaD}
\ee
where $\lambda_{\rm un}(s)$ is the spectral function of an unparticle with a mass gap $m_g$ and dimension $d_{U}=3/2$.

\section{Electroweak precision observables}
\label{sec:EWPO}

Even if the observable $T$ is protected by the custodial symmetry, as it is well known the rest of observables are unprotected and, as our model departs from the usual RS models since resonances have a (broad) width, it is worth doing a detailed analysis of (oblique) electroweak observables.

When the electroweak symmetry is broken there is a mixing between the SM fields $W_L$ and $Z_L$ and the heavy modes of $W_{L,R}$ and $Z_{L,R}$ induced by the Lagrangian
\be
\mathcal L=\tr|g_L^5 W_L^aT_L^a \mathcal H-g_R^5 \mathcal H W_R^a T_R^a |^2 \,,
\label{eq:mixing}
\ee
where we are indicating with the script $g_5$ the 5D gauge couplings, related to the 4D couplings $g_4$ by $g_5 = g_4 \sqrt{y_s}$.

After putting the Higgs bi-doublet $\mathcal H$, which we assume to be localized on the IR brane,
\be
\mathcal H=\begin{pmatrix} H_2^0 & H_1^-\\
H_2^- & H_1^0
\end{pmatrix}
\ee
at its minimum, $\langle H_{1,2}^0\rangle=v_{1,2}$, with $v_1^2+v_2^2=v^2$ and $v=246.22$ GeV, the Lagrangian (\ref{eq:mixing}) gives rise to the quadratic terms
\begin{align}
\mathcal L =& \frac{v^2}{4} y_s \bigg[ g_L^2 W_L(y_1,x) W_L(y_1,x) + g_R^2 W_R(y_1,x) W_R(y_1,x) - \frac{2 v_1 v_2}{v^2} g_L g_R W_L(y_1,x) W_R(y_1,x)  \nonumber\\
+&\frac{1}{2}\frac{g_L^2}{c_L^2}Z_L(y_1,x)Z_L(y_1,x) +  \frac{1}{2}g_R^2c_R^2Z_R(y_1,x)Z_R(y_1,x)  - g_L g_R \frac{c_R}{c_L}Z_L(y_1,x)Z_R(y_1,x) \bigg] \,,
\end{align}
where $W_X W_X \equiv W_X^- W_X^+$ for $X = L, R$, and $W_L W_R \equiv W_L^- W_R^+ + W_L^+ W_R^-$. One has $v_1 = v \cdot \cos\beta$ and $v_2 = v \cdot \sin\beta$, and then $2 v_1 v_2 / v^2 = 2 t_\beta / (1 + t_\beta^2)$ where we have defined $t_\beta \equiv \tan\beta$. In the custodial limit $t_\beta = 1$ and $v_1 = v_2 = v/\sqrt{2}$.

As we have seen in the previous sections and in the Appendix~\ref{sec:massive_gauge_bosons}, the fields $W_L$ and $Z_L$ have a zero mode, which is the corresponding SM field, and a gapped continuum of states, while the fields $W_R$ and $Z_R$ do not possess zero mode, but only the continuum of states above the mass gap. For the electroweak observables contributing to the new physics, the oblique $T$, $S$ and $U$ parameters are defined as~\cite{Peskin:1991sw}
\begin{align}
\alpha T&=\frac{\Pi_{WW}(0)}{m_W^2}-\frac{\Pi_{ZZ}(0)}{m_Z^2} \,, \nonumber\\
\alpha S&= 4 s_L^2 c_L^2 \Pi^\prime_{ZZ}(0) \,, \nonumber \\
\alpha (S + U) &= 4 s_L^2 \Pi^\prime_{WW}(0) \,.  \label{eq:TSU}
\end{align}

For the computation of these parameters, we need to select as external fields the zero modes of either $W_L$ and $Z_L$ and only propagate the continuum of states. We will then define the Green's functions propagating only the continuum of states as
\begin{align}
\mathcal G_{W_L,Z_L}(y_1,y_1;p) &= G_{W_L,Z_L}(y_1,y_1;p)-G^0_{W_L,Z_L}(p) \,, \nonumber\\
\mathcal G_{W_R,Z_R}(y_1,y_1;p) &= G_{W_R,Z_R}(y_1,y_1;p) \,,
\end{align}
where $G_{W_L,Z_L}$ and $G_{W_R,Z_R}$ are, respectively, the Green's functions $G_A$ and $G_A^{(-+)}$ computed in Secs.~\ref{sec:gauge_bosons} and \ref{sec:Dirichlet}~\footnote{In this section we are neglecting the finite mass effects of $W$ and $Z$ bosons, an approximation which is valid as long as $m_{W,Z} \ll \rho$. These effects could have been easily considered by using the propagators of Appendix~\ref{sec:massive_gauge_bosons} and evaluating the expressions of Eq.~(\ref{eq:TSU}) at the pole of the Green's function $G_{A,M}(y_1,y_1;p^2)$, i.e. at $p^2 \simeq m_{W,Z}^2$. However these effects should be negligible in view of the mass hierarchy $m_A\ll\rho$.}. By using the notation $\mathcal G(0)\equiv \lim_{p \to 0} \mathcal G(y_1,y_1;p)$, a straightforward calculation yields~\cite{Carena:2018cow}
\be
\alpha T=m_W^2 y_s\left[\mathcal G_{W_L}(0)+ \frac{4 t_\beta^2}{(1+t_\beta^2)^2} \frac{g_R^2}{g_L^2}\mathcal G_{W_R} (0) \right]-
m_Z^2 y_s\left[\mathcal G_{Z_L}(0)+\frac{g_R^2}{g_L^2}c_L^2 c_R^2\mathcal G_{Z_R} (0) \right] \,,
\ee
where $t_\beta=v_2/v_1$. Using the results of previous sections we find
\begin{align}
\mathcal G_{W_L} (0)&=\mathcal G_{Z_L}(0)=\frac{-2 (k y_s)^2+ 6(k y_s)-9}{4(k y_s) y_s\rho^2}\equiv \mathcal G_L(0) \,, \nonumber\\
\mathcal G_{W_R}(0) &=\mathcal G_{Z_R}(0)=-\frac{(k y_s)}{2 y_s \rho^2}\equiv \mathcal G_R(0) \,,
\end{align}
so that
\be
\alpha T=m_W^2 y_s\frac{s_L^2}{c_L^2}\left[ \left( 1-\frac{1}{s_R^2}\frac{(1-t_\beta^2)^2}{(1+t_\beta^2)^2} \right) \mathcal G_R(0)-\mathcal G_L(0)\right] \,.
\ee

A similar calculation yields
\begin{align}
\alpha S &= 4 m_Z^4 y_s s_L^2 c_L^2  \left[  \mathcal G^\prime_L(0)+\frac{s_L^2 c_R^2}{s_R^2} \mathcal G^\prime_R(0) \right] \,, \\
\alpha U &= 4 m_Z^4 y_s  s_L^4 c_L^2 \left[ \left( 1-\frac{1}{s_R^2}\frac{(1-t_\beta^2)^2}{(1+t_\beta^2)^2} \right) \mathcal G^\prime_R(0) - \mathcal G^\prime_L(0) \right] \,,
\end{align}
where the prime stands for $\frac{d}{dp^2}$, and $\mathcal G'_{W_L}(0)=\mathcal G^\prime_{Z_L}(0)\equiv \mathcal G^\prime_L(0)$, $\mathcal G'_{W_R}(0)=\mathcal G^\prime_{Z_R}(0)\equiv \mathcal G^\prime_R(0)$. Then we find
\begin{align}
\mathcal G^\prime_L(0)&= \frac{-40 (k y_s)^3+324 (k y_s)^2-977 (k y_s)+648}{128 (k y_s)^2 y_s \rho^4} \,, \nonumber\\
\mathcal G^\prime_R(0) &= -\frac{5 (k y_s)}{16 y_s \rho^4} \,.
\end{align}
We can see that, in the limit of large value of $k y_s$, $\mathcal
G_{L,R}(0), \mathcal G_{L,R}^\prime(0)=\mathcal O(k y_s)$, while
$\mathcal G_R(0)-\mathcal G_L(0), \mathcal G_R^\prime(0)-\mathcal
G_L^\prime(0)=\mathcal O(k y_s)^0$, which is the cancellation which
appears on the observables $T$ and $U$ in the custodial limit
$t_\beta=1$. However, still the observable $S$ gives a sizable
contribution, which is partly cancelled if we introduce a small
breaking of the custodial symmetry, i.e.~when we introduce a small
value of $t_\beta-1$.

\begin{figure}[t]
\centering
\includegraphics[width=7cm]{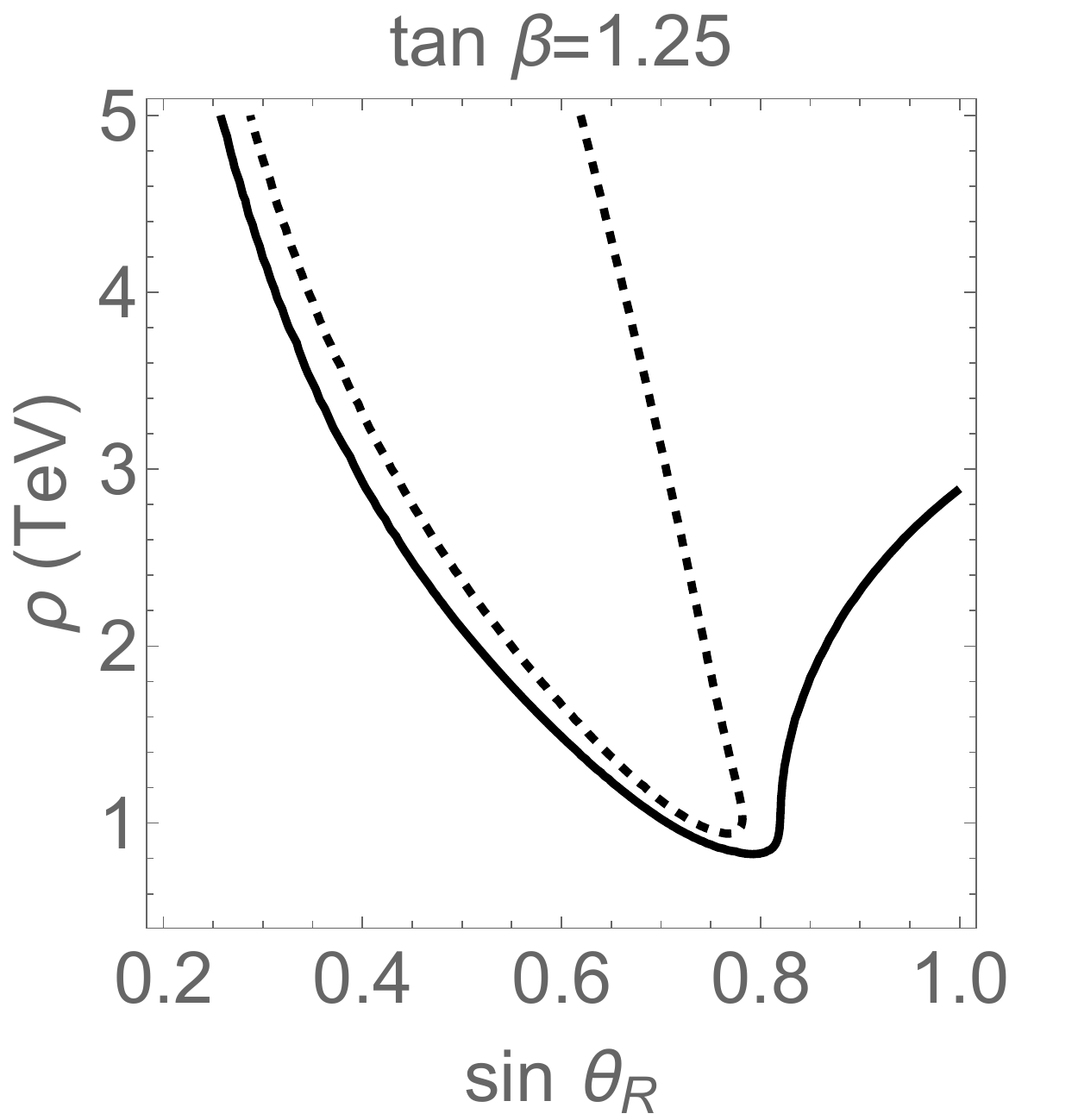}
\includegraphics[width=7cm]{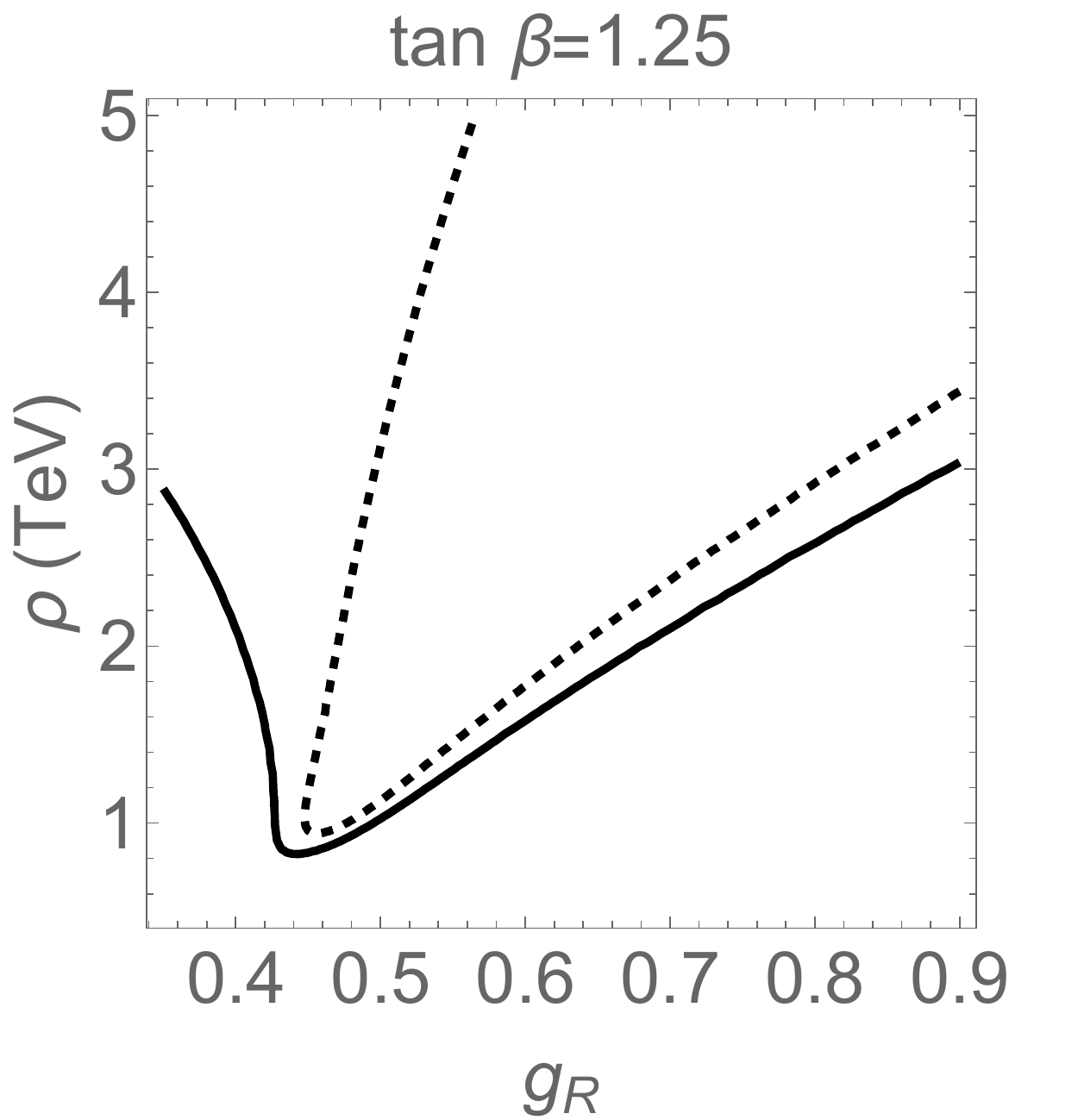}
\caption{\it $\chi^2$ distribution in the plane $(s_R,\rho)$ (left panel) and $(g_R,\rho)$ (right panel) for the value of $t_\beta=1.25$. The solid (dashed) lines are the corresponding 95\% (67\%) C.L. limits. }
\label{fig:chi2}
\end{figure}

The most recent experimental constraints for the oblique $S$, $T$ and $U$ parameters~\cite{Zyla:2020zbs} gives
\begin{equation} 
 S = -0.01 \pm 0.10  \,, \qquad  T =  0.03 \pm 0.12 \,, \qquad U = 0.02 \pm 0.11 \,, \label{eq:STU}
\end{equation}
with correlations
\begin{equation}
\textrm{corr}(S,T) = 92\% \,, \qquad \textrm{corr}(S,U) = -80\% \,, \qquad \textrm{corr}(T,U) = -93\% \,.
\end{equation}
We display in Fig.~\ref{fig:chi2} the $\chi^2$ distribution in the plane $(s_R,\rho)$ (left panel) and $(g_R,\rho)$ (right panel) for the value of $t_\beta=1.25$. The solid (dashed) lines are the corresponding 95\% (67\%) C.L. limits. 
As we can see the lowest value of the parameter $\rho$ is $\rho_{\rm min}\sim 1$ TeV, which corresponds to a value of the coupling $g_R\sim 0.45$, in turn corresponding to $s_R\sim 0.8$, well in the perturbative region. For other values of $t_\beta$ the result smoothly changes. For instance in the custodial limit, i.e.~for $t_\beta=1$ we find that $\rho_{\rm min}\sim 2.5$ TeV, corresponding to a value of $g_R\sim 5.6$, well inside the non-perturbative region.

\section{Conclusions and outlook}
\label{sec:conclusions}

In this paper we have studied a 5D model which naturally leads to gapped continuum spectra. The model is defined in terms of a metric which has an AdS$_5$ behavior in the UV, and a linear behavior in the IR for the scalar field in conformal coordinates, and constitutes a faithful enough approximation of the model presented in Ref.~\cite{Megias:2019vdb}, having the advantage of allowing for analytical expressions for the Green's functions. In this paper we have concentrated on the case of bulk propagating gauge bosons $A_M$, with Green's functions $G_A(y,y^\prime)$. The spectrum of SM massless gauge bosons (the photon and gluon) is a continuum of KK modes with a mass gap equal to $m_g = \rho/2$ where $\rho \sim \TeV$, and an isolated massless pole which corresponds to the corresponding 4D gauge boson. In the case of SM massive gauge bosons (the $W$ and $Z$ bosons) the isolated pole becomes massive. For the case of gauge bosons with Dirichlet boundary conditions on the UV brane, the spectrum is a continuum with a mass gap equal to $m_g$, but without any isolated pole, which has been projected out of the spectrum by the boundary conditions.

We have considered the Green's functions in the complex $s$ plane, and found the existence of poles in the second Riemann sheet. We have computed the masses and widths of the associated resonances, and found that, while the former are close to the masses of the KK modes in the RS model, the latter are quite large $(\Gamma/\rho \gtrsim  7)$ indicating the presence of broad resonances. Their relative widths $\Gamma/M$, however, decrease with increasing energy, so that they tend to a distribution closer to Dirac delta functions in this regime. The behavior of the Green's functions $G_A(y,y^\prime)$ with $y_1 < y,y^\prime$, i.e.~between the IR brane and the singularity, is then explained as a summation of two contributions: a contribution which is purely continuous, and turns out to be related to gapped unparticle propagators, and other contribution that contains the resonances. We have extended theses analyses to gauge bosons with Dirichlet boundary condition in the UV brane, as well as to massive gauge bosons, leading to similar conclusions. In all the cases the resonances are present: in the Dirichlet case the positions of the poles in the complex plane are very close to the poles for massless gauge bosons, while in the massive case the values of the widths turn out to decrease with increasing values of $m_A/\rho$. 

Notice that while the RS model leads to a discrete KK spectrum with zero widths~\cite{Randall:1999ee}, the linear dilaton model has a purely continuous spectrum above the mass gap, apart from the possible existence of isolated zero modes~\cite{Megias:2021mgj} (see also~\cite{Megias:2020cpw}). As our model shares both features, a RS metric between the UV and IR branes, and a linear dilaton model metric between the IR brane and the singularity, the result is that the RS resonances are endowed with a width, as a result of the effect of the linear dilaton metric while there still remains a pure unparticle contribution to the Green's functions. In fact our explicit Green's functions exhibit a pole structure in the second Riemann of the complex $s$ plane, with a broad width, which widely depart from an infinite series of Breit-Wigner resonances. Notice that in spite of being a 4D Green's function the particle width is incorporated \textit{ab initio} and is not associated to particle production, a characteristic feature of unparticles. We plan to go deeper into this issue in the future. 

This present study can be extended to the computation of Green's functions of other fields, i.e.~fermions, Higgs bosons, the graviton and the radion. Regarding phenomenological applications, the brane-to-brane Green's functions can be used to study the excess with respect to the SM prediction of some processes at the LHC, in particular the cross-section of $pp$ collisions where a continuum KK gluon is produced by Drell-Yan processes and decays into a pair of light/heavy fermions localized in the UV/IR brane. For other phenomenological applications in particle physics, it would be interesting to study the couplings of the continuum KK modes with the SM fields, and provide values for the Wilson coefficients of the corresponding effective field theory. Other possible applications include the study of dark matter as a weakly interacting continuum~\cite{Chaffey:2021tmj,Csaki:2021gfm}. Some of these issues will be addressed in a forthcoming publication~\cite{Megias:prep}.

\vspace{0.5cm}
\section*{Acknowledgments}
We would like to thank A.~Carmona, M.~P\'erez-Victoria and
L.L.~Salcedo for fruitful discussions. The authors thank the ICTP
South American Institute for Fundamental Research (SAIFR), Sao Paulo,
Brazil, and its Program on Particle Physics, September 30-November 30,
2019, where part of this work was done, for hospitality. The work of
EM is supported by the Spanish MINEICO under Grants
FIS2017-85053-C2-1-P and PID2020-114767GB-I00, by the FEDER/Junta de
Andaluc\'{\i}a-Consejer\'{\i}a de Econom\'{\i}a y Conocimiento
2014-2020 Operational Programme under Grant A-FQM-178-UGR18, by Junta
de Andaluc\'{\i}a under Grant FQM-225, and by the Consejer\'{\i}a de
Conocimiento, Investigaci\'on y Universidad of the Junta de
Andaluc\'{\i}a and European Regional Development Fund (ERDF) under
Grant SOMM17/6105/UGR. The research of EM is also supported by the
Ram\'on y Cajal Program of the Spanish MINEICO under Grant
RYC-2016-20678. The work of MQ is partly supported by Spanish MINEICO
under Grant FPA2017-88915-P, by the Catalan Government under Grant
2017SGR1069, and by Severo Ochoa Excellence Program of MINEICO under
Grant SEV-2016-0588. IFAE is partially funded by the CERCA program of
the Generalitat de Catalunya.

\appendix

\section{Standard model massive gauge bosons}
\label{sec:massive_gauge_bosons}

In this Appendix we will study the Green's functions $G_{A,M}$ for massive SM gauge bosons. As we are considering the Higgs sector localized on the IR brane, in the case of the SM massive gauge bosons $A_\mu$ there are extra terms in the 5D Lagrangian, Eq.~(\ref{eq:Lagrangian_GB}), as
\be
\Delta\mathcal L_5=\left(-\frac{1}{2}  M_Z^2 Z_\mu^2-  M_W^2 |W_\mu|^2 \right) \delta(y-y_1),\quad M_A^2=y_s m_A^2 \,,  \label{eq:L5_mass}
\ee
(for $A=W,Z$), which leads to a modification of the EoM for the Green's function, as
\be
\left[ p^2- y_s m_A^2\,\delta(y-y_1) \right] G_{A,M}(y,y^\prime;p)  + \partial_y \left( e^{-2A} \partial_y G_{A,M}(y,y^\prime;p) \right) = \delta(y-y^\prime) \,.  \label{eq:GAymassive}
\ee
Using Eq.~(\ref{eq:GAymassive}), the derivative of the Green's functions turns out to be discontinuous at $y = y_{1}$, with a jump given by~\footnote{In the following we will assume that $y^\prime \ne y_{1}$, hence $\int_{y_{1} - \epsilon}^{y_{1} + \epsilon} dy \, \delta(y-y^\prime) = 0$ and the term in the right-hand side of Eq.~(\ref{eq:GAymassive}) does not contribute to Eq.~(\ref{eq:GAmassivejump}). Then, the Green's functions involving the IR brane are computed as $G_{A,M}(y,y_1)=\lim_{y^\prime\to y_1}G_{A,M}(y,y^\prime)$. Alternatively we could directly compute the Green's function $G_{A,M}(y,y_1)$ by considering a jump at $y=y_1$ given by 
$$\left.\Delta (\partial_y G_{A,M})(y,y_1)\right|_{y=y_1} = m_A^2 y_s e^{2A(y_{1})} G_{A,M}(y_{1},y_1)+e^{2A(y_{1})} \,,$$ 
where the first and second term in the right-hand side correspond to the contribution of the terms $\propto \delta(y-y_1)$ and $\propto \delta(y-y^\prime)$ in Eq.~(\ref{eq:GAymassive}), respectively. We have checked that both procedures lead to the same result.}
\begin{equation}
\left.\Delta (\partial_y G_{A,M})(y,y^\prime)\right|_{y=y_1} = m_A^2 y_s e^{2A(y_{1})} G_{A,M}(y_{1},y^\prime) \,.  \label{eq:GAmassivejump}
\end{equation}

One can now solve the EoM by dividing the $y$ space into three regions, as explained in Sec.~\ref{sec:gauge_bosons_z0z1}. When doing that, we find a general solution identical to Eq.~(\ref{eq:solGA_homogeneous}) which is subject to the same boundary and matching conditions as in Eq.~(\ref{eq:GA_bc}), except for $\Delta \left( \partial_y G_{A,M} \right)(y_{1})$ which is given by Eq.~(\ref{eq:GAmassivejump}). Finally, one finds that the Green's function for massive gauge bosons is given by Eq.~(\ref{eq:GAypy1}) with the replacements $\Phi(p)\to \Phi_M(p)$, $\Z(y)\to \Z_M(y)$ and $\Q(y)\to \Q_M(y)$, where the functions $\Phi_M(p)$, $\Z_M(y)$ and $\Q_M(y)$ are given by
\begin{align}
\Phi_M(p) &= Y_0(p/k) \cdot J_{M+}(p/\rho) - J_0(p/k) \cdot Y_{M+}(p/\rho) \,, \nonumber \\
\Psi_M(p) &= Y_0(p/k)  \cdot J_{M-}(p/\rho) - J_0(p/k) \cdot Y_{M-}(p/\rho) \,,  \nonumber\\
\mathcal Z_M(y) &= J_{M+}(p/\rho) \cdot Y_1\left( e^{k y}p/k \right) - Y_{M+}(p/\rho) \cdot J_1\left( e^{k y}p/k \right) \,, \nonumber\\
\Q_M(y) &= - \frac{k}{\rho^2}\frac{1}{k(y_s-y)} \left[\Phi_M(p) -   (k(y_s-y))^{\delta_A} \Psi_M(p)\right] \,,  \label{eq:QM}
\end{align}
with
\begin{equation}
J_{M\pm}(p/\rho) = 2\frac{p}{\rho}J_0(p/\rho)+\Xi_A^\pm J_1(p/\rho),\quad Y_{M\pm}(p/\rho) =  2\frac{p}{\rho}Y_0(p/\rho) + \Xi_A^\pm Y_1(p/\rho) \,,
\end{equation}
and we have used the notation
\begin{equation}
\Xi_A^{\pm} = \Delta_A^{\pm} + 2k y_s \cdot (m_A/\rho)^2 \,. \label{eq:Xipm}
\end{equation}
The approximate expressions of $\Phi_M(p)$ and $\Psi_M(p)$ for $p \ll k$ turn out to be
\begin{equation}
\Phi_M(p) =\mathcal K J_{M+}(p/\rho)-Y_{M+}(p/\rho) \,, \qquad \Psi_M(p)=\mathcal K J_{M-}(p/\rho)-Y_{M-}(p/\rho) \,.
\end{equation}
In the limit $y \to y_0$ one finds the same expression as Eq.~(\ref{eq:GAz0zp}) with the replacements for $\Phi$ and $\Z$ mentioned above. As it is obvious from Eq.~(\ref{eq:Xipm}), the present Green's function tends to the result for massless gauge bosons given by Eq.~(\ref{eq:GAypy1}), when considering the limit $m_A \to 0$ ($\Xi_A^{\pm} \to \Delta_A^{\pm}$). As in the massless case, the Green's function for massive gauge bosons fulfills the property $G_{A,M}(y,y^\prime;p) = G_{A,M}(y^\prime,y;p)$.

\begin{figure}[t]
\centering
\includegraphics[width=7cm]{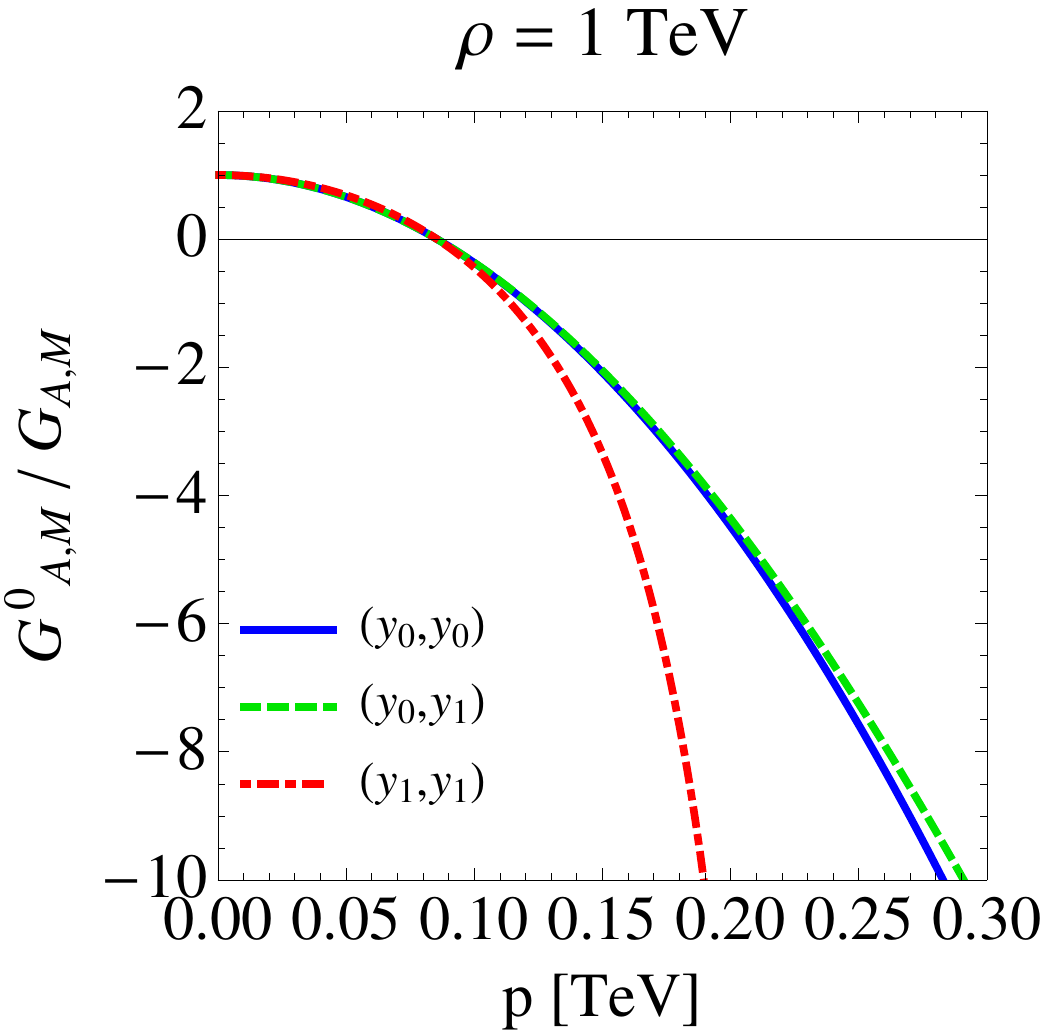} \hspace{0.5cm} 
\includegraphics[width=7cm]{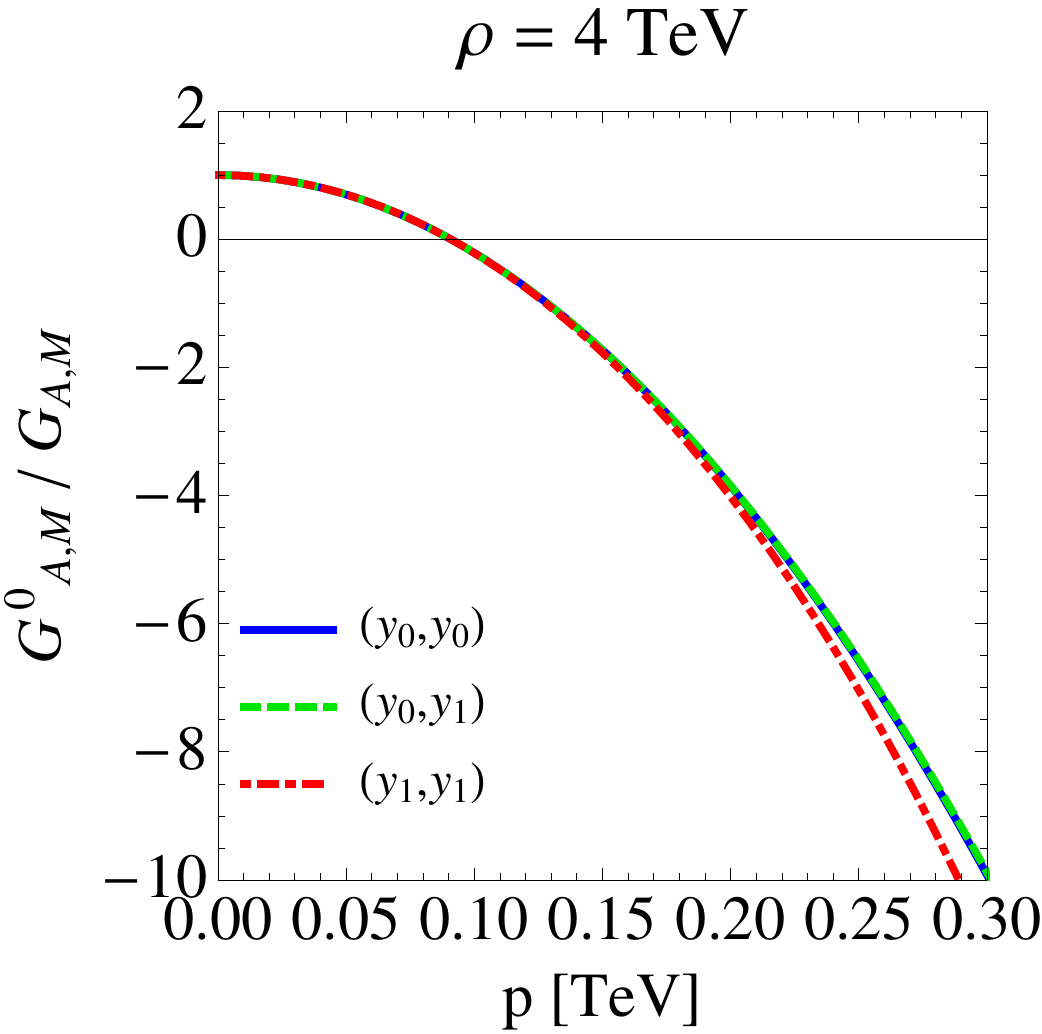} 
\caption{\it Inverse $(y_0,y_0)$, $(y_0,y_1)$ and $(y_1,y_1)$ Green's functions for $A=Z$, $m_Z= 0.091$ TeV. They are normalized by their values in $p=0$. We display the results for $\rho = 1$ TeV (left panel) and $\rho = 4$ TeV (right panel). We have used $A_1 = 35$.}
\label{fig:invGAmassive}
\end{figure}

The analytical expressions for the brane-to-brane Green's functions are, respectively,
\begin{eqnarray}
G_{A,M}^{-1}(y_0,y_0;p) &=& \frac{p \Phi_M(p)}{ \mathcal Z_M(y_0)}  \simeq  -\frac{\pi p^2}{2k} \cdot  \frac{  \Phi_M(p)  }{J_{M+}(p) }    \label{eq:GA_z0z0_asympM}  \,, \\
G_{A,M}^{-1}(y_0,y_1;p) &=& -\frac{\pi}{4} \frac{\rho}{k} p \Phi_M(p)  \,, \label{eq:GA_z0z1_asympM}  \\
G_{A,M}^{-1}(y_1,y_1;p) &=& -\frac{\rho^2}{2k} \cdot \frac{\Phi_M(p)}{\PP(y_1)}  \simeq -\frac{\rho^2}{2k}\cdot \frac{\Phi_M(p)}{ \K \cdot J_1\left(\frac{p}{\rho} \right) - Y_1\left( \frac{p}{\rho} \right) }   \,, \label{eq:GA_z1z1_asympM} 
\end{eqnarray}
where in the second equality of Eqs.~(\ref{eq:GA_z0z0_asympM}) and (\ref{eq:GA_z1z1_asympM}) we have assumed $p \ll k$, and in the limit $p , m_A \ll \rho$ they are
\begin{eqnarray}
G_{A,M}^{-1}(y_0,y_0;p) &\simeq& -\left[ \frac{1}{k y_s \left( m_A/\rho \right)^2} + \frac{1}{2} \right]^{-1} \frac{\rho^2}{k}  \label{eq:GAz0z0massiveana} \\
&&+ \left[ 1 - \frac{3}{2} \left( m_A/\rho \right)^2 \right]  y_s p^2 + {\cal O}\left( (m_A/\rho)^4, (p/\rho)^4 \right) \,, \nonumber \\
G_{A,M}^{-1}(y_0,y_1;p) &\simeq& - y_s m_A^2  \label{eq:GAz0z1massiveana} \\
&&+ \left[ 1 + \frac{1}{2} \left( - \frac{3}{2} + k y_s \right) \left( m_A/\rho \right)^2 \right]  y_s p^2 + {\cal O}\left( (m_A/\rho)^4, (p/\rho)^4 \right) \,, \nonumber   \\
G_{A,M}^{-1}(y_1,y_1;p)&\simeq& - y_s m_A^2 +  y_s p^2 + {\cal O}\left( (m_A/\rho)^4, (p/\rho)^4 \right) \,.  \label{eq:GAz1z1massiveana}
\end{eqnarray}
These functions have poles at
\begin{eqnarray}
p^2 \Big|_{(y_0,y_0)} &\simeq& m_A^2 \left[ 1 + \left( 3 - k y_s \right) \frac{m_A^2}{2\rho^2} + \mathcal O\left( (m_A/\rho)^{4}\right) \right] \,, \\
p^2 \Big|_{(y_0,y_1)} &\simeq& m_A^2 \left[ 1 + \left( \frac{3}{2}  - k y_s \right) \frac{m_A^2}{2\rho^2} + \mathcal O\left( (m_A/\rho)^{4}\right) \right] \,, \\
p^2 \Big|_{(y_1,y_1)} &\simeq&  m_A^2 \left[ 1 + \mathcal O\left( (m_A/\rho)^{4}\right) \right] \,.
\end{eqnarray}
We show in Fig.~\ref{fig:invGAmassive} 
the inverse Green's functions $G_{A,M}^{-1}(y_0,y_0;p)$, $G_{A,M}^{-1}(y_0,y_1;p)$ and $G_{A,M}^{-1}(y_1,y_1;p)$ as functions of $p$, for $\rho = 1\, \textrm{GeV}$ (left panel), $\rho = 4\, \textrm{GeV}$ (right panel) and $m_A = m_Z$. The position of the zero in these panels corresponds to the pole in $G_A(y_\alpha,y_\alpha;p)$. Notice that the Green's functions are real in the range of momenta $0 \le p < m_g$ (except for a Dirac delta behavior at $p \simeq m_A$ as we will see below). We have normalized the plots by the values of the Green's functions at $p=0$, i.e.
\begin{equation}
G_{A,M}^0= \lim_{p\to 0}G_{A,M}(y,y^\prime;p) \simeq - \frac{1}{y_s m_A^2} \,. \label{eq:GAM0}
\end{equation}
The asymptotic behaviors for the inverse Green's functions with time-like momenta $p^2>0$, $p\gg \rho$, and with space-like momenta $p^2<0$, $|p|\gg \rho$, are given, respectively, by Eqs.~(\ref{eq:GA_z0z0_largep})-(\ref{eq:GA_z1z1_largep}) and (\ref{eq:GA_z0z0sl_largep})-(\ref{eq:GA_z1z1sl_largep}).

\begin{figure}[htb]
\centering
\includegraphics[width=5cm]{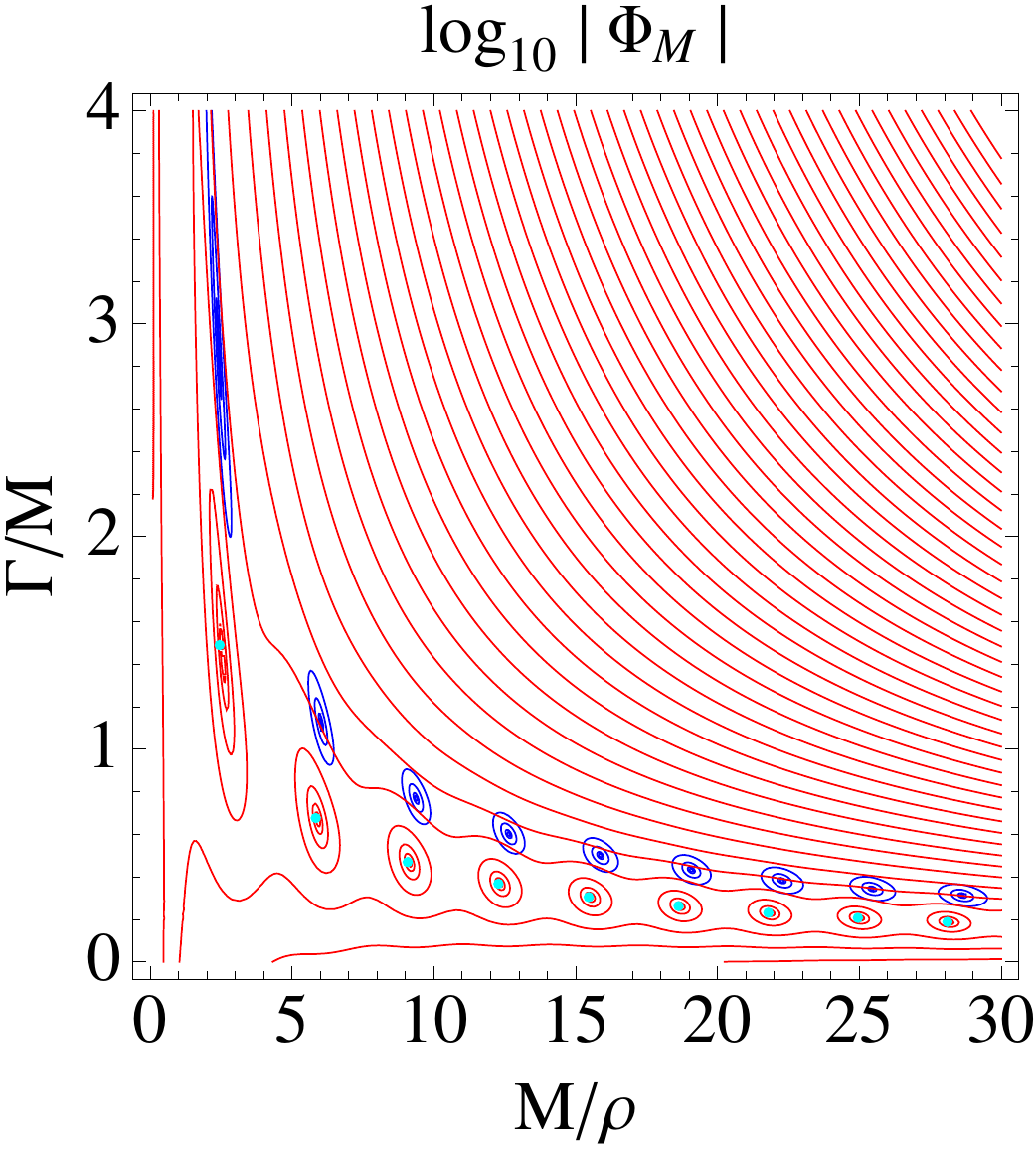} \hspace{1.00cm} \includegraphics[width=5cm]{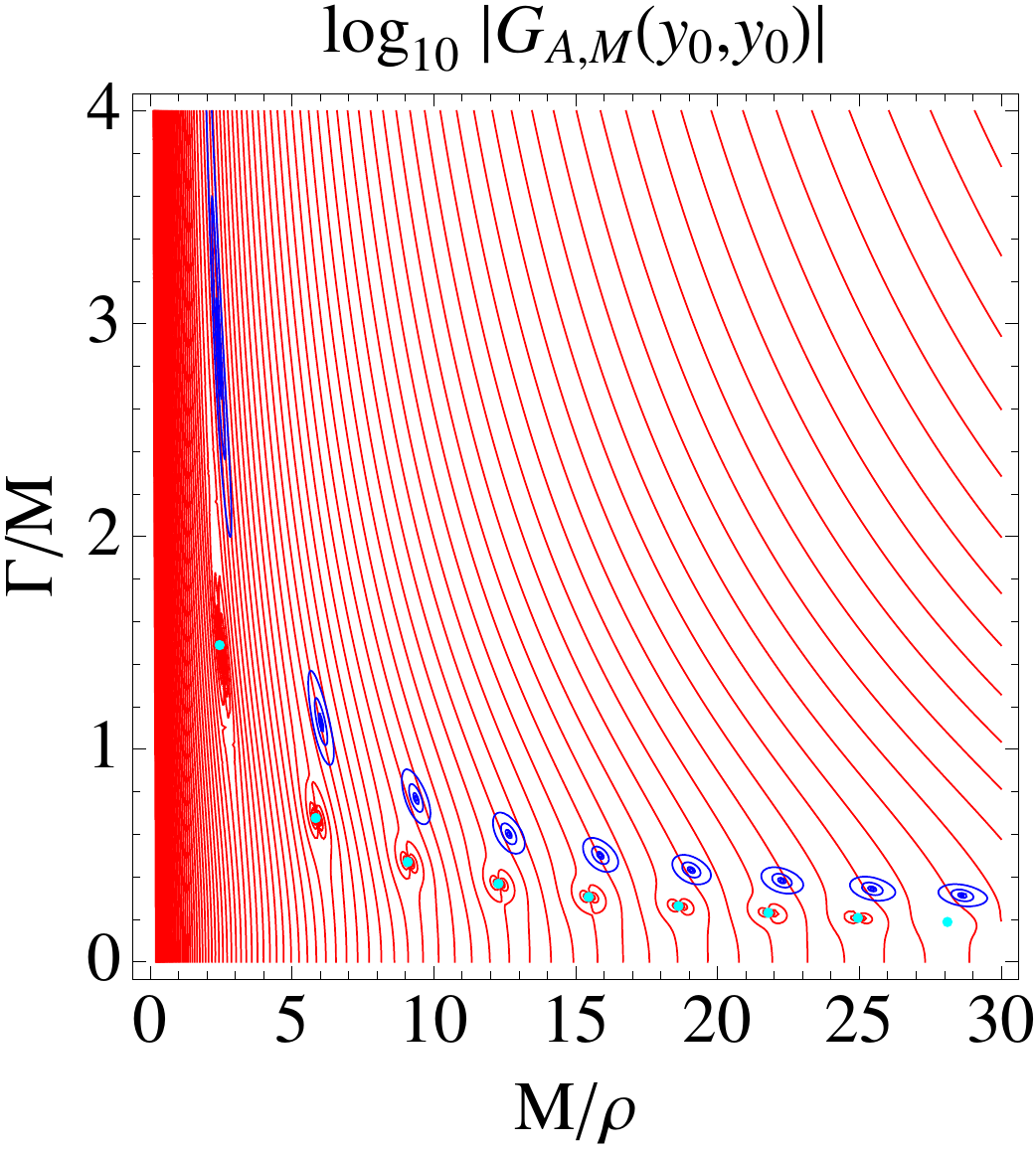} \\
\includegraphics[width=5cm]{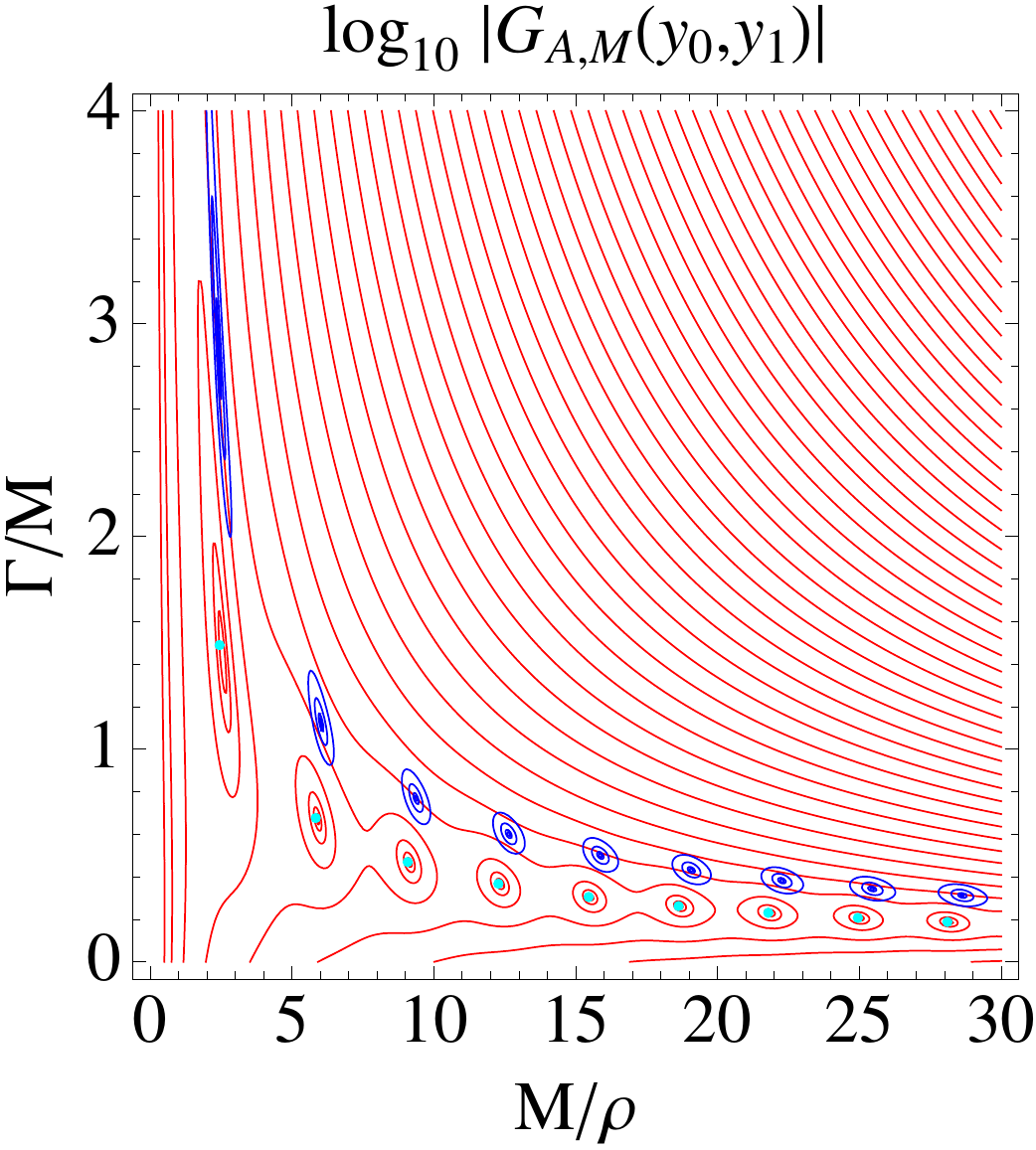} \hspace{1.00cm}
\includegraphics[width=5.15cm]{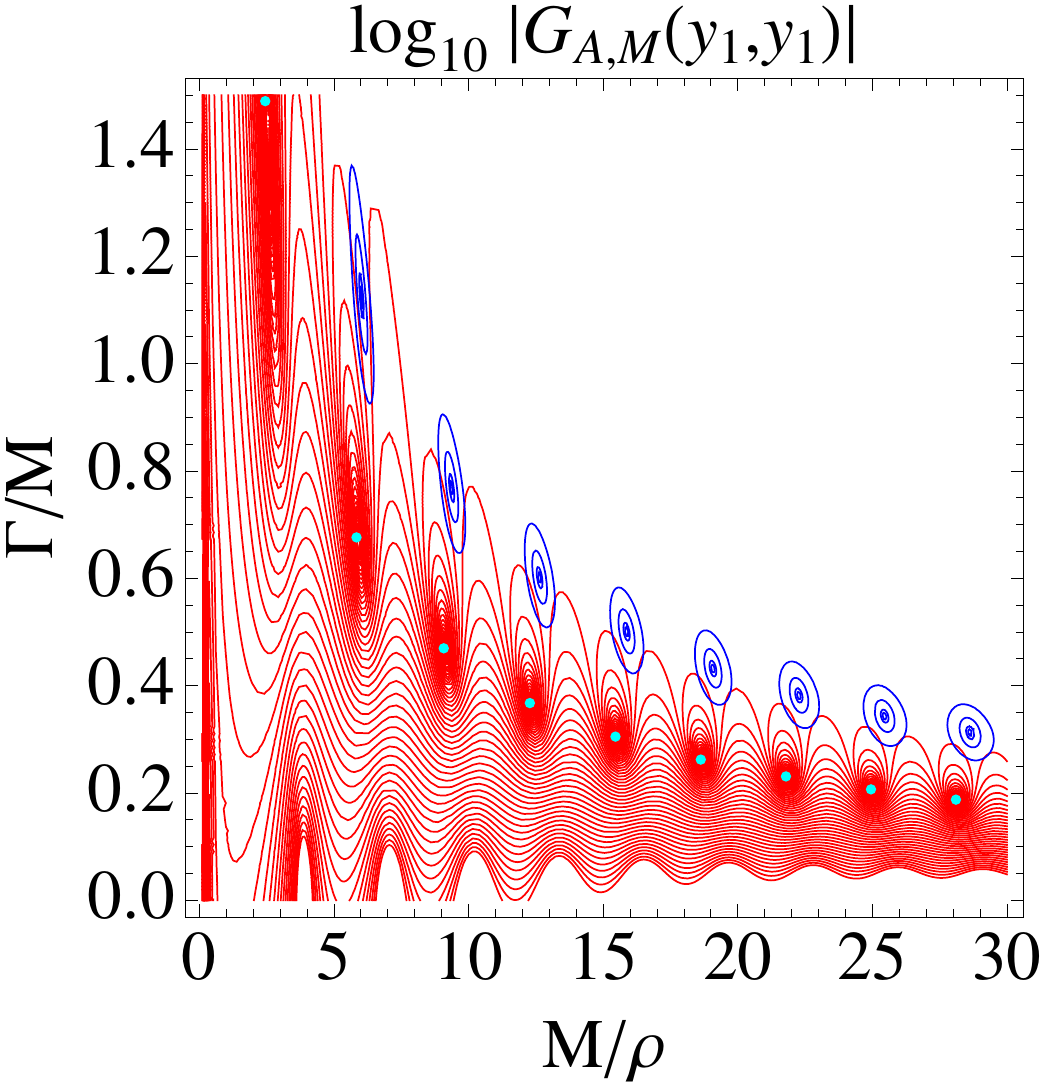}
\caption{\it Upper left panel: Contour plot in the plane $(M/\rho,\Gamma/M)$ of $\log_{10}|\Phi_M(p)|$ cf. Eq.~(\ref{eq:QM}) (red lines). Upper right and lower panels: Common logarithm of the absolute value of the Green's function $\log_{10} |G_{A,M}(y_0,y_0)|$ (upper right), $\log_{10} |G_{A,M}(y_0,y_1)|$ (lower left) and $\log_{10} |G_{A,M}(y_1,y_1)|$ (lower right) . The (cyan) dots stand for the positions of the poles of the Green's function as predicted by the analytical formula of Eq.~(\ref{eq:p2Lambert_M}).  For comparison, the positions of the zeros of $\Phi(p)$ in the massless case, cf. Fig.~\ref{fig:Phi_2D}, are displayed in blue in each panel. We have considered $\rho = 1 \, \textrm{TeV}$, $m_Z = 0.091\,\TeV$  and $A_1 = 35$.}
\label{fig:AbsGAM_resonances}
\end{figure}

In order to study the poles in the (second Riemann sheet of the) complex $s$ plane of the Green's functions, we display in Fig.~\ref{fig:AbsGAM_resonances} 
\begin{figure}[htb]
\centering
\includegraphics[width=7cm]{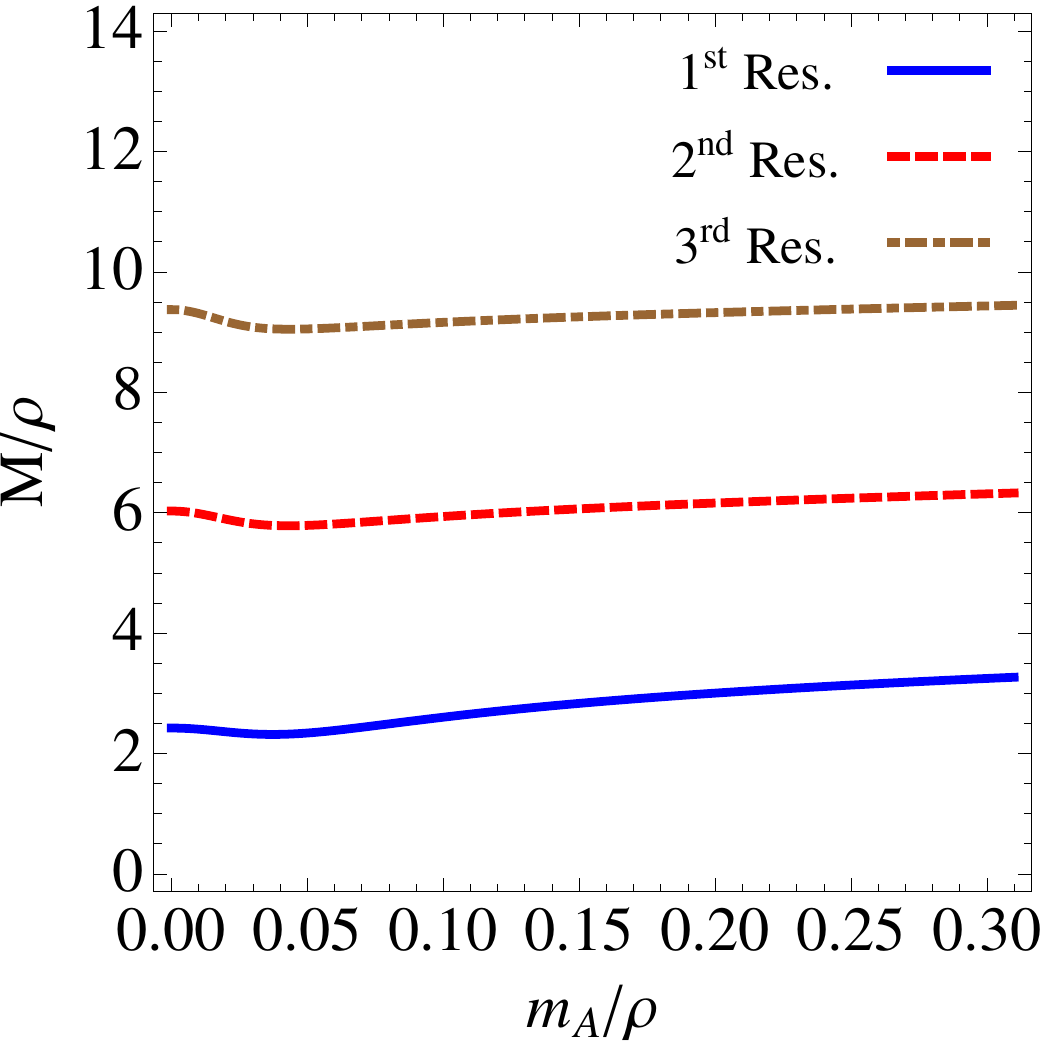} \hspace{0.5cm} \includegraphics[width=7cm]{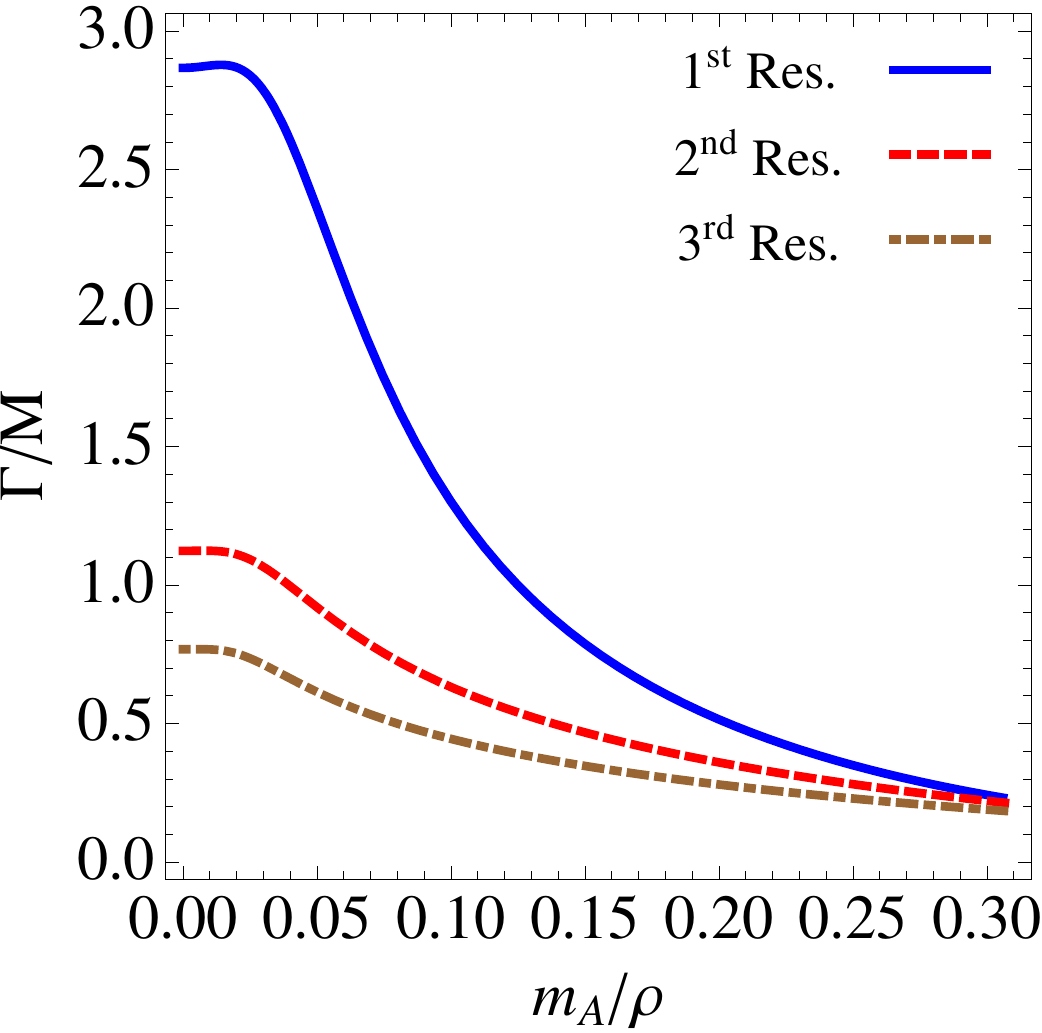} 
\caption{\it $M/\rho$ (left panel) and $\Gamma/M$ (right panel) as a function of the gauge boson mass $m_A/\rho$. It is displayed the results for the first (solid blue), second (dashed red) and third (dotdashed brown) resonances. We have considered $A_1 = 35$.}
\label{fig:MassWidth}
\end{figure} 
a contour plot of $\log_{10} |\Phi_M(p)|$, where the structure of zeros of this function, corresponding to resonances, can be seen. The lightest resonances appear at
\begin{eqnarray}
(M/\rho,\Gamma/M)_{\rho = 1 \,\TeV} &=& (2.55,1.44), (5.91,0.670), (9.14,0.467), (12.34,0.366), \nonumber \\
&& (15.52,0.304) , \cdots  \,, \\
(M/\rho,\Gamma/M)_{\rho = 4 \,\TeV} &=& (2.35,2.86), (5.87,1.10), (9.14,0.744), (12.35,0.574), \nonumber \\
&& (15.54,0.472) , \cdots  \,.
\end{eqnarray}
As it can be seen in Fig.~\ref{fig:AbsGAM_resonances}, and from a comparison with Eq.~(\ref{eq:MG}), the effect of the mass for the zero modes of the gauge bosons is to reduce the width of the resonances, but the masses of the resonances are not much affected. We display in Fig.~\ref{fig:MassWidth} the dependence of the ratios $M/\rho$ (left panel) and $\Gamma/M$ (right panel) with the gauge boson mass for the three lightest resonances. Notice that for  $\rho = 1 \, \TeV$ the physical values for the $Z$ and $W$ masses correspond to $m_A/\rho \simeq 0.1$, and this leads to a reduction of the widths of the resonances by a factor $\sim 0.6$. A physical interpretation consistent with this property is that the KK modes are quasi bound states with a finite probability to tunnel to the continuum region $z > z_1$.  As the brane mass of Eq.~(\ref{eq:L5_mass}) is taken larger, it tends to produce almost Dirichlet boundary conditions at the IR brane and the KK resonances become more stable, i.e. their widths tend to zero~\footnote{We thank the (anonymous) referee for a useful comment on this issue.}.

We can study analytically the zeros of $\Phi_M(p)$ following the procedure of Sec.~\ref{sec:Resonances}. The expansion of the function $\Phi_M(p)$ at large momentum $\rho \ll |p|$ $(|p| \ll k)$  and $1 \ll k y_s (m_A/\rho)^2 \cdot |p|/\rho \ll  |p|^2/\rho^2 $ leads to~\footnote{The large momentum expansion of the function $\Phi_M(p)$ is
\begin{equation}
\Phi_M(p) \stackrel[\rho \ll |p|]{\propto}{}  e^{i2p/\rho} \left(1 + 4i k y_s (m_A/\rho)^2 \cdot p/\rho \right) + 4 p/\rho \left( -2i p/\rho + k y_s (m_A/\rho)^2 \right) \,, \label{eq:PhiM_propto}
\end{equation}
for $\Imaginary \left( (p/\rho)^2 \right) < 0$, so that one can see that in the limit $m_A/\rho \to 0$ the massless case formula of Eq.~(\ref{eq:Phi_large_p}) is recovered. However, the zeros of this formula do not admit a direct analytical expression unless some of the terms are neglected. This is why in getting Eq.~(\ref{eq:PhiM_large_p}) we have assumed $k y_s (m_A/\rho)^2  \ll \frac{|p|}{\rho}$ and $1 \ll k y_s (m_A/\rho)^2  \cdot \frac{|p|}{\rho}$, so that two of the terms in Eq.~(\ref{eq:PhiM_propto}) have been neglected. The disadvantage of this approximation is that the massless limit can no longer be recovered.}
\begin{eqnarray}
\Phi_M(p) &\stackrel[\rho \ll |p|]{\simeq}{}&   2 \sqrt{\frac{2}{\pi^3}}  e^{-i(p/\rho - \pi/4)}   \left[ k y_s (m_A/\rho)^2 e^{i2p/\rho} - 2 \frac{p}{\rho}  \right] \times  \nonumber \\
&& \qquad \times \log\left( \frac{p}{k} \right)\left( \frac{\rho}{p} \right)^{1/2} \,, \qquad\qquad \Imaginary \left( (p/\rho)^2 \right) < 0 \,.  \label{eq:PhiM_large_p}
\end{eqnarray}
Then, the zeros of $\Phi_M(p)$ correspond to the solutions of the equation
\begin{equation}
k y_s (m_A/\rho)^2 e^{i2p/\rho} =  2 \frac{p}{\rho}   \,,
\end{equation}
which turn out to be
\begin{equation}
\left( \frac{p}{\rho} \right)^2 = - \frac{1}{4}  \mathcal W_n\left[ - i k y_s (m_A/\rho)^2 \right]^2  \,, \quad n = -1,-2, \cdots \,. \label{eq:p2Lambert_M}
\end{equation}

We display in Fig.~\ref{fig:AbsGAM_resonances} as cyan dots the results of Eq.~(\ref{eq:p2Lambert_M}) with $m_A = m_Z$ and $\rho = 1 \, \textrm{TeV}$. The relative error of Eq.~(\ref{eq:p2Lambert_M}) with respect to the true zeros of $\Phi_M(p)$ decreases with $M/\rho$, and it is $\lesssim 0.5\%$ except for the lightest resonance which is $\sim 4\%$. The error of the approximate formula~(\ref{eq:p2Lambert_M}) increases when $m_A/\rho$ decreases as in this case the approximation $1 \ll k y_s (m_A/\rho)^2  \cdot \frac{|p|}{\rho}$ is no longer valid. In this regime of very small gauge boson masses the formula of Eq.~(\ref{eq:p2Lambert}) is a much better approximation. 

Let us notice that the Green's function for massive gauge bosons can be split also into unparticle and resonant contributions, as for the massless case in Eq.~(\ref{eq:GA_unres}). The unparticle contribution for massive gauge bosons turns out to be identical as for massless gauge bosons, and it is given by Eq.~(\ref{eq:G_un}). 

\begin{figure}[htb]
\centering
\includegraphics[width=4.5cm]{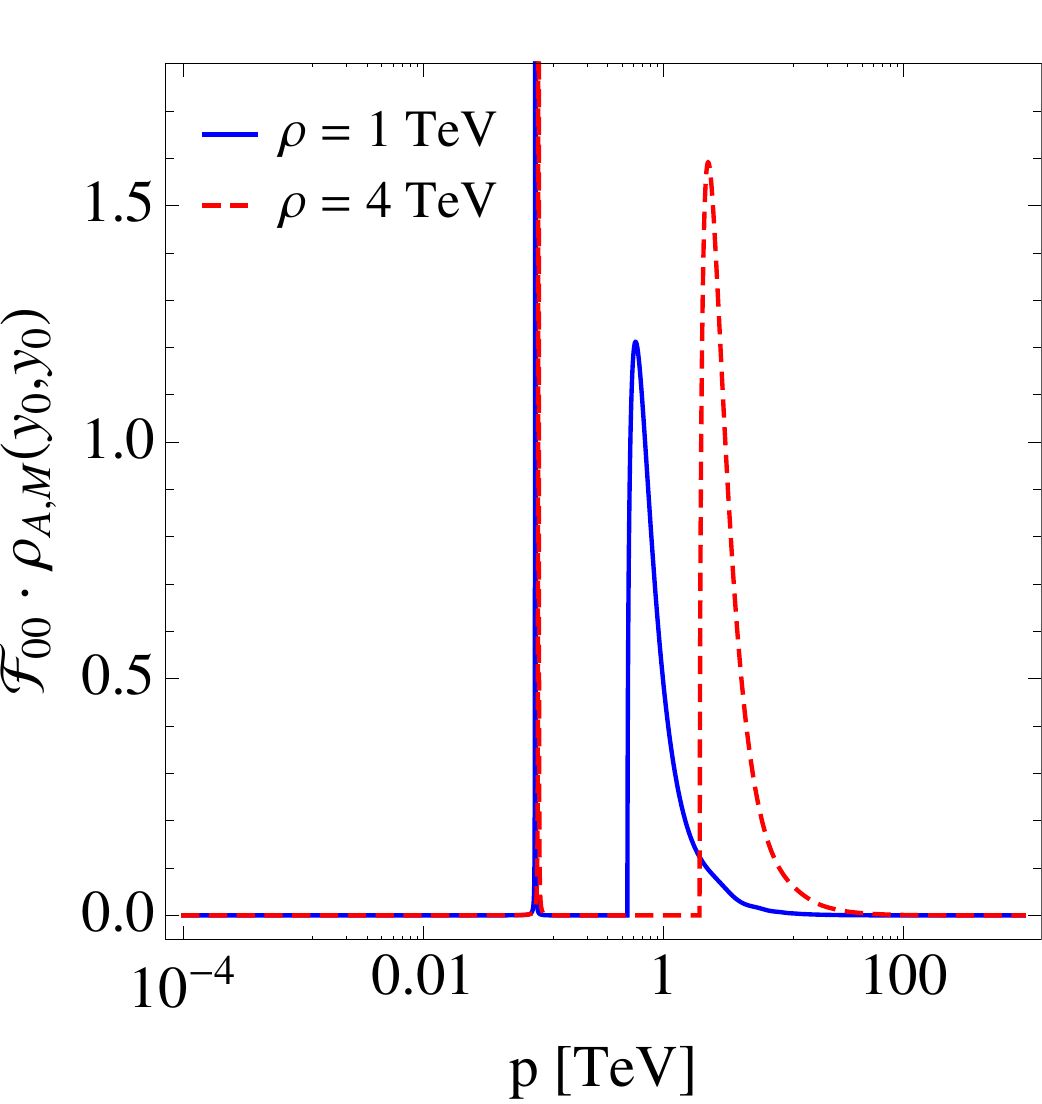} \hspace{0.10cm}
\includegraphics[width=4.7cm]{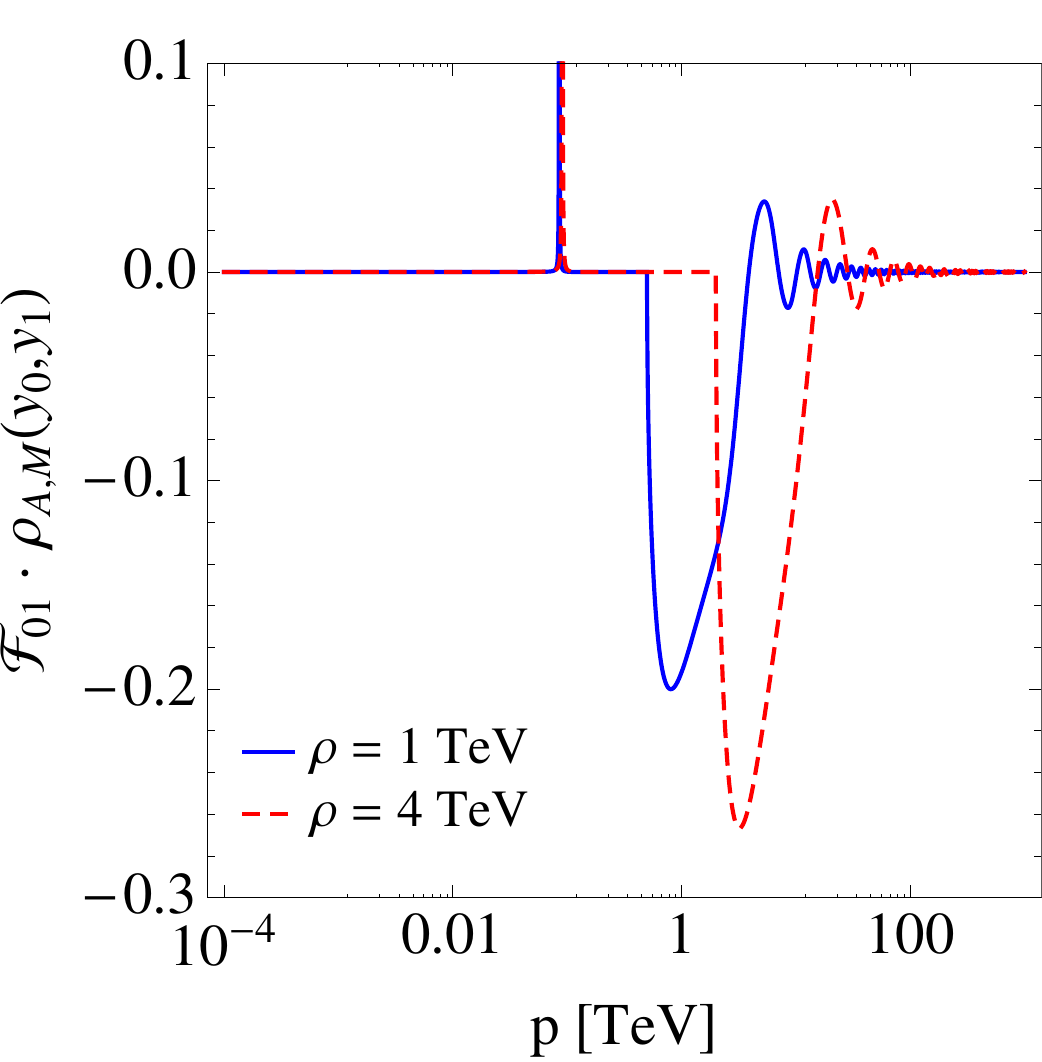} \hspace{0.10cm}
\includegraphics[width=4.7cm]{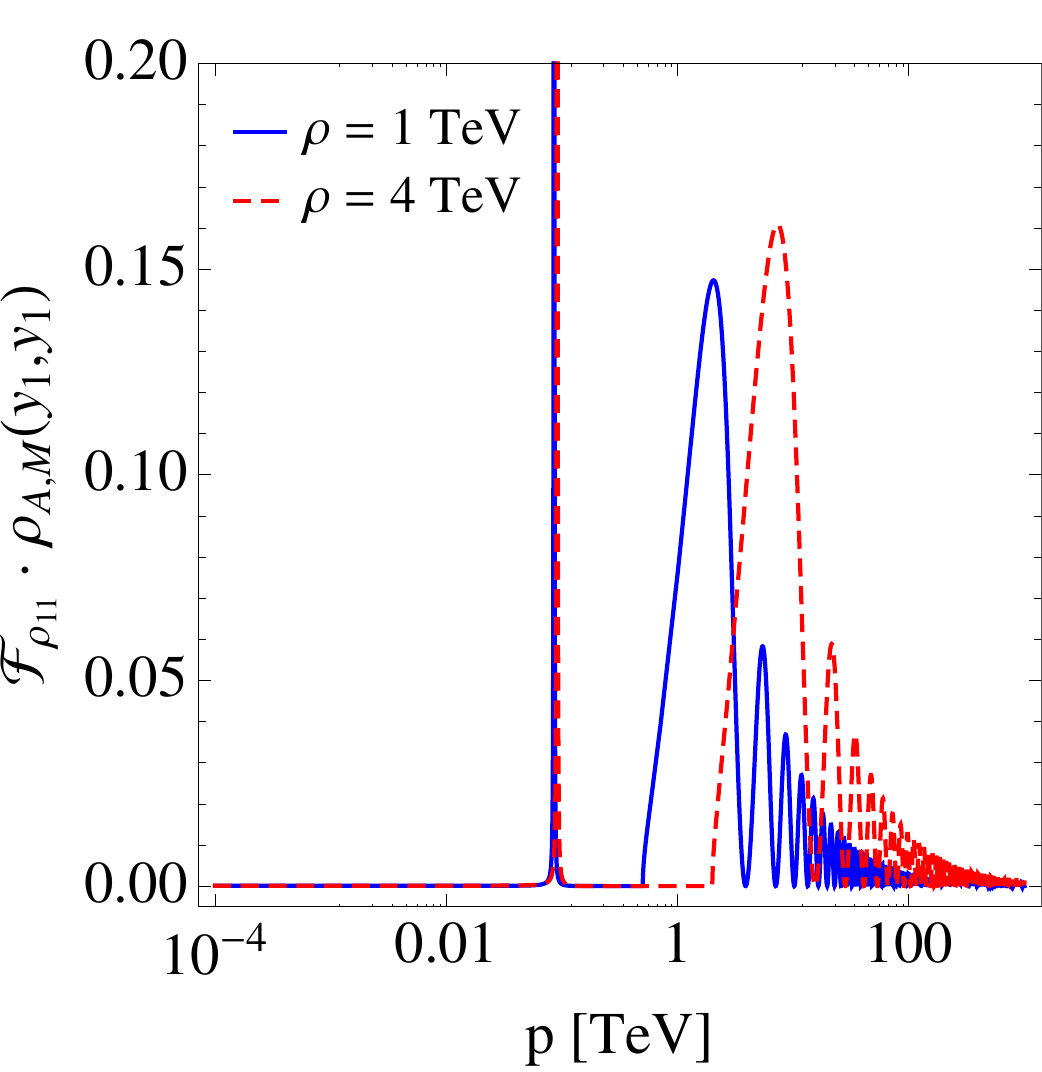} 
\caption{\it Rescaled spectral functions $\mathcal F_{00} \cdot \rho_{A,M}(y_0,y_0;p)$ (left panel), $\mathcal F_{01} \cdot \rho_{A,M}(y_0,y_1;p)$ (middle panel) and $\mathcal F_{11} \cdot \rho_{A,M}(y_1,y_1;p)$ (right panel) for $A = Z$, $m_Z= 0.091$ TeV. We display the results for $\rho = 1 $ TeV (solid blue lines) and $\rho = 4$ TeV (dashed red lines). We have used $A_1 = 35$ in all panels and assume time-like momenta $p^2>0$.
}
\label{fig:spectralM}
\end{figure}

Finally, we show in Fig.~\ref{fig:spectralM} the brane-to-brane spectral functions $\rho_{A,M}(y_\alpha,y_\beta;p)$ as functions of $p$, for $\rho = 1$ TeV and $\rho = 4$ TeV. The prefactors, defined in Eq.~(\ref{eq:Fab}), make them approximately invariant under shifts of $ky_s$. In contrast to the results obtained in Sec.~\ref{subsec:spectral_function} for massless gauge bosons, where it appears a Dirac delta behavior at $p=0$, in the massive case this behavior is found at $p \simeq m_A$. In all the cases the continuum spectrum starts at $p = 0.5\,\textrm{TeV} \; (2 \, \textrm{TeV})$ for $\rho = 1 \,\textrm{TeV} \; (4 \, \textrm{TeV})$. 

We can study as well the positivity of the spectral operator $\hat \rho_{A,M}$. The procedure is  similar to the one presented in Sec.~\ref{subsec:spectral_function} and the eigenvalue $\lambda_M(p)$ is given by
\be
\lambda_M(s)=\delta(s-m_A^2)+\left[ -\frac{\log\bar\epsilon}{2\pi\rho}\lambda_{\rm un}(s)+\mathcal O (\bar\epsilon^0) \right] \Theta(s-m_g^2) \,, \qquad 
\lambda_{\rm un}(s) = (s-m_g^2)^{-1/2} \,,
\ee
reflecting the existence of the isolated zero mode with squared mass $m_A^2\ll m_g^2$. 

\bibliographystyle{JHEP}
\bibliography{refs}

\providecommand{\href}[2]{#2}\begingroup\raggedright\begin{thebibliography}{10}

\bibitem{ALEPH:2005ab}
{\scshape SLD Electroweak Group, DELPHI, ALEPH, SLD, SLD Heavy Flavour Group,
  OPAL, LEP Electroweak Working Group, L3} collaboration, S.~Schael et~al.,
  \emph{{Precision electroweak measurements on the $Z$ resonance}},
  \href{http://dx.doi.org/10.1016/j.physrep.2005.12.006}{\emph{Phys. Rept.}
  {\bfseries 427} (2006) 257--454},
  [\href{https://arxiv.org/abs/hep-ex/0509008}{{\ttfamily hep-ex/0509008}}].

\bibitem{Olive:2016xmw}
{\scshape Particle Data Group} collaboration, C.~Patrignani et~al.,
  \emph{{Review of Particle Physics}},
  \href{http://dx.doi.org/10.1088/1674-1137/40/10/100001}{\emph{Chin. Phys.}
  {\bfseries C40} (2016) 100001}.

\bibitem{Randall:1999ee}
L.~Randall and R.~Sundrum, \emph{{A Large mass hierarchy from a small extra
  dimension}}, \href{http://dx.doi.org/10.1103/PhysRevLett.83.3370}{\emph{Phys.
  Rev. Lett.} {\bfseries 83} (1999) 3370--3373},
  [\href{https://arxiv.org/abs/hep-ph/9905221}{{\ttfamily hep-ph/9905221}}].

\bibitem{Sirunyan:2018ryr}
{\scshape CMS} collaboration, A.~M. Sirunyan et~al., \emph{{Search for resonant
  $ \mathrm{t}\overline{\mathrm{t}} $ production in proton-proton collisions at
  $ \sqrt{s}=13 $ TeV}},
  \href{http://dx.doi.org/10.1007/JHEP04(2019)031}{\emph{JHEP} {\bfseries 04}
  (2019) 031}, [\href{https://arxiv.org/abs/1810.05905}{{\ttfamily
  1810.05905}}].

\bibitem{Aaboud:2019roo}
{\scshape ATLAS} collaboration, M.~Aaboud et~al., \emph{{Search for heavy
  particles decaying into a top-quark pair in the fully hadronic final state in
  $pp$ collisions at $\sqrt{s} =$ 13 TeV with the ATLAS detector}},
  \href{http://dx.doi.org/10.1103/PhysRevD.99.092004}{\emph{Phys. Rev.}
  {\bfseries D99} (2019) 092004},
  [\href{https://arxiv.org/abs/1902.10077}{{\ttfamily 1902.10077}}].

\bibitem{Escribano:2021jne}
R.~Escribano, M.~Mendizabal, M.~Quir\'os and E.~Royo, \emph{{On Broad
  Kaluza-Klein Gluons}},
  \href{http://dx.doi.org/10.1007/JHEP05(2021)121}{\emph{JHEP} {\bfseries 05}
  (2021) 121}, [\href{https://arxiv.org/abs/2102.11241}{{\ttfamily
  2102.11241}}].

\bibitem{Giudice:2017suc}
G.~F. Giudice and M.~McCullough, \emph{{Comment on "Disassembling the Clockwork
  Mechanism"}},  \href{https://arxiv.org/abs/1705.10162}{{\ttfamily
  1705.10162}}.

\bibitem{Giudice:2017fmj}
G.~F. Giudice, Y.~Kats, M.~McCullough, R.~Torre and A.~Urbano,
  \emph{{Clockwork/linear dilaton: structure and phenomenology}},
  \href{http://dx.doi.org/10.1007/JHEP06(2018)009}{\emph{JHEP} {\bfseries 06}
  (2018) 009}, [\href{https://arxiv.org/abs/1711.08437}{{\ttfamily
  1711.08437}}].

\bibitem{Antoniadis:2011qw}
I.~Antoniadis, A.~Arvanitaki, S.~Dimopoulos and A.~Giveon, \emph{{Phenomenology
  of TeV Little String Theory from Holography}},
  \href{http://dx.doi.org/10.1103/PhysRevLett.108.081602}{\emph{Phys. Rev.
  Lett.} {\bfseries 108} (2012) 081602},
  [\href{https://arxiv.org/abs/1102.4043}{{\ttfamily 1102.4043}}].

\bibitem{Cox:2012ee}
P.~Cox and T.~Gherghetta, \emph{{Radion Dynamics and Phenomenology in the
  Linear Dilaton Model}},
  \href{http://dx.doi.org/10.1007/JHEP05(2012)149}{\emph{JHEP} {\bfseries 05}
  (2012) 149}, [\href{https://arxiv.org/abs/1203.5870}{{\ttfamily 1203.5870}}].

\bibitem{Antoniadis:2001sw}
I.~Antoniadis, S.~Dimopoulos and A.~Giveon, \emph{{Little string theory at a
  TeV}}, \href{http://dx.doi.org/10.1088/1126-6708/2001/05/055}{\emph{JHEP}
  {\bfseries 05} (2001) 055},
  [\href{https://arxiv.org/abs/hep-th/0103033}{{\ttfamily hep-th/0103033}}].

\bibitem{Csaki:2018kxb}
C.~Csaki, G.~Lee, S.~J. Lee, S.~Lombardo and O.~Telem, \emph{{Continuum
  Naturalness}}, \href{http://dx.doi.org/10.1007/JHEP03(2019)142}{\emph{JHEP}
  {\bfseries 03} (2019) 142},
  [\href{https://arxiv.org/abs/1811.06019}{{\ttfamily 1811.06019}}].

\bibitem{Megias:2019vdb}
E.~Meg\'{\i}as and M.~Quir\'os, \emph{{Gapped Continuum Kaluza-Klein
  spectrum}}, \href{http://dx.doi.org/10.1007/JHEP08(2019)166}{\emph{JHEP}
  {\bfseries 08} (2019) 166},
  [\href{https://arxiv.org/abs/1905.07364}{{\ttfamily 1905.07364}}].

\bibitem{Megias:2020cpw}
E.~Meg\'{\i}as and M.~Quir\'os, \emph{{On gapped continuum resonance spectra}},
   in \emph{{8th International Conference on New Frontiers in Physics (ICNFP
  2019) Kolymbari, Crete, Greece, August 21-29, 2019}}, 2020.
\newblock \href{https://arxiv.org/abs/2002.11756}{{\ttfamily 2002.11756}}.

\bibitem{Megias:2021mgj}
E.~Meg\'{\i}as and M.~Quir\'os, \emph{{The Continuum Linear Dilaton}},
  \href{http://dx.doi.org/10.5506/APhysPolB.52.711}{\emph{Acta Phys. Polon. B}
  {\bfseries 52} (2021) 711},
  [\href{https://arxiv.org/abs/2104.10260}{{\ttfamily 2104.10260}}].

\bibitem{Csaki:2021gfm}
C.~Csaki, S.~Hong, G.~Kurup, S.~J. Lee, M.~Perelstein and W.~Xue,
  \emph{{Continuum Dark Matter}},
  \href{https://arxiv.org/abs/2105.07035}{{\ttfamily 2105.07035}}.

\bibitem{Cabrer:2010si}
J.~A. Cabrer, G.~von Gersdorff and M.~Quir\'os, \emph{{Warped Electroweak
  Breaking Without Custodial Symmetry}},
  \href{http://dx.doi.org/10.1016/j.physletb.2011.01.058}{\emph{Phys. Lett.}
  {\bfseries B697} (2011) 208--214},
  [\href{https://arxiv.org/abs/1011.2205}{{\ttfamily 1011.2205}}].

\bibitem{Cabrer:2011fb}
J.~A. Cabrer, G.~von Gersdorff and M.~Quir\'os, \emph{{Suppressing Electroweak
  Precision Observables in 5D Warped Models}},
  \href{http://dx.doi.org/10.1007/JHEP05(2011)083}{\emph{JHEP} {\bfseries 05}
  (2011) 083}, [\href{https://arxiv.org/abs/1103.1388}{{\ttfamily 1103.1388}}].

\bibitem{Cabrer:2011vu}
J.~A. Cabrer, G.~von Gersdorff and M.~Quir\'os, \emph{{Improving Naturalness in
  Warped Models with a Heavy Bulk Higgs Boson}},
  \href{http://dx.doi.org/10.1103/PhysRevD.84.035024}{\emph{Phys. Rev.}
  {\bfseries D84} (2011) 035024},
  [\href{https://arxiv.org/abs/1104.3149}{{\ttfamily 1104.3149}}].

\bibitem{Cabrer:2011mw}
J.~A. Cabrer, G.~von Gersdorff and M.~Quir\'os, \emph{{Warped 5D Standard Model
  Consistent with EWPT}},
  \href{http://dx.doi.org/10.1002/prop.201100054}{\emph{Fortsch. Phys.}
  {\bfseries 59} (2011) 1135--1138},
  [\href{https://arxiv.org/abs/1104.5253}{{\ttfamily 1104.5253}}].

\bibitem{Cabrer:2011qb}
J.~A. Cabrer, G.~von Gersdorff and M.~Quir\'os, \emph{{Flavor Phenomenology in
  General 5D Warped Spaces}},
  \href{http://dx.doi.org/10.1007/JHEP01(2012)033}{\emph{JHEP} {\bfseries 01}
  (2012) 033}, [\href{https://arxiv.org/abs/1110.3324}{{\ttfamily 1110.3324}}].

\bibitem{Cabrer:2009we}
J.~A. Cabrer, G.~von Gersdorff and M.~Quir\'os, \emph{{Soft-Wall
  Stabilization}},
  \href{http://dx.doi.org/10.1088/1367-2630/12/7/075012}{\emph{New J. Phys.}
  {\bfseries 12} (2010) 075012},
  [\href{https://arxiv.org/abs/0907.5361}{{\ttfamily 0907.5361}}].

\bibitem{Georgi:2007ek}
H.~Georgi, \emph{{Unparticle physics}},
  \href{http://dx.doi.org/10.1103/PhysRevLett.98.221601}{\emph{Phys. Rev.
  Lett.} {\bfseries 98} (2007) 221601},
  [\href{https://arxiv.org/abs/hep-ph/0703260}{{\ttfamily hep-ph/0703260}}].

\bibitem{Georgi:2007si}
H.~Georgi, \emph{{Another odd thing about unparticle physics}},
  \href{http://dx.doi.org/10.1016/j.physletb.2007.05.037}{\emph{Phys. Lett.}
  {\bfseries B650} (2007) 275--278},
  [\href{https://arxiv.org/abs/0704.2457}{{\ttfamily 0704.2457}}].

\bibitem{Delgado:2007dx}
A.~Delgado, J.~R. Espinosa and M.~Quir\'os, \emph{{Unparticles Higgs
  Interplay}},
  \href{http://dx.doi.org/10.1088/1126-6708/2007/10/094}{\emph{JHEP} {\bfseries
  10} (2007) 094}, [\href{https://arxiv.org/abs/0707.4309}{{\ttfamily
  0707.4309}}].

\bibitem{Delgado:2008rq}
A.~Delgado, J.~R. Espinosa, J.~M. No and M.~Quir\'os, \emph{{The Higgs as a
  Portal to Plasmon-like Unparticle Excitations}},
  \href{http://dx.doi.org/10.1088/1126-6708/2008/04/028}{\emph{JHEP} {\bfseries
  04} (2008) 028}, [\href{https://arxiv.org/abs/0802.2680}{{\ttfamily
  0802.2680}}].

\bibitem{Delgado:2008gj}
A.~Delgado, J.~R. Espinosa, J.~M. No and M.~Quir\'os, \emph{{A Note on
  Unparticle Decays}},
  \href{http://dx.doi.org/10.1103/PhysRevD.79.055011}{\emph{Phys. Rev.}
  {\bfseries D79} (2009) 055011},
  [\href{https://arxiv.org/abs/0812.1170}{{\ttfamily 0812.1170}}].

\bibitem{Delgado:2008px}
A.~Delgado, J.~R. Espinosa, J.~M. No and M.~Quir\'os, \emph{{Phantom Higgs from
  Unparticles}},
  \href{http://dx.doi.org/10.1088/1126-6708/2008/11/071}{\emph{JHEP} {\bfseries
  11} (2008) 071}, [\href{https://arxiv.org/abs/0804.4574}{{\ttfamily
  0804.4574}}].

\bibitem{Stancato:2008mp}
D.~Stancato and J.~Terning, \emph{{The Unhiggs}},
  \href{http://dx.doi.org/10.1088/1126-6708/2009/11/101}{\emph{JHEP} {\bfseries
  11} (2009) 101}, [\href{https://arxiv.org/abs/0807.3961}{{\ttfamily
  0807.3961}}].

\bibitem{Falkowski:2008yr}
A.~Falkowski and M.~P\'erez-Victoria, \emph{{Holographic Unhiggs}},
  \href{http://dx.doi.org/10.1103/PhysRevD.79.035005}{\emph{Phys. Rev.}
  {\bfseries D79} (2009) 035005},
  [\href{https://arxiv.org/abs/0810.4940}{{\ttfamily 0810.4940}}].

\bibitem{Falkowski:2009uy}
A.~Falkowski and M.~P\'erez-Victoria, \emph{{Electroweak Precision Observables
  and the Unhiggs}},
  \href{http://dx.doi.org/10.1088/1126-6708/2009/12/061}{\emph{JHEP} {\bfseries
  12} (2009) 061}, [\href{https://arxiv.org/abs/0901.3777}{{\ttfamily
  0901.3777}}].

\bibitem{Bellazzini:2015cgj}
B.~Bellazzini, C.~Csaki, J.~Hubisz, S.~J. Lee, J.~Serra and J.~Terning,
  \emph{{Quantum Critical Higgs}},
  \href{http://dx.doi.org/10.1103/PhysRevX.6.041050}{\emph{Phys. Rev.}
  {\bfseries X6} (2016) 041050},
  [\href{https://arxiv.org/abs/1511.08218}{{\ttfamily 1511.08218}}].

\bibitem{Agashe:2003zs}
K.~Agashe, A.~Delgado, M.~J. May and R.~Sundrum, \emph{{RS1, custodial isospin
  and precision tests}},
  \href{http://dx.doi.org/10.1088/1126-6708/2003/08/050}{\emph{JHEP} {\bfseries
  08} (2003) 050}, [\href{https://arxiv.org/abs/hep-ph/0308036}{{\ttfamily
  hep-ph/0308036}}].

\bibitem{Gubser:1999vj}
S.~S. Gubser, \emph{{AdS / CFT and gravity}},
  \href{http://dx.doi.org/10.1103/PhysRevD.63.084017}{\emph{Phys. Rev.}
  {\bfseries D63} (2001) 084017},
  [\href{https://arxiv.org/abs/hep-th/9912001}{{\ttfamily hep-th/9912001}}].

\bibitem{York:1972sj}
J.~W. York, Jr., \emph{{Role of conformal three geometry in the dynamics of
  gravitation}},
  \href{http://dx.doi.org/10.1103/PhysRevLett.28.1082}{\emph{Phys. Rev. Lett.}
  {\bfseries 28} (1972) 1082--1085}.

\bibitem{Gibbons:1976ue}
G.~W. Gibbons and S.~W. Hawking, \emph{{Action Integrals and Partition
  Functions in Quantum Gravity}},
  \href{http://dx.doi.org/10.1103/PhysRevD.15.2752}{\emph{Phys. Rev.}
  {\bfseries D15} (1977) 2752--2756}.

\bibitem{Megias:2018sxv}
E.~Meg\'{\i}as, G.~Nardini and M.~Quir\'os, \emph{{Cosmological Phase
  Transitions in Warped Space: Gravitational Waves and Collider Signatures}},
  \href{http://dx.doi.org/10.1007/JHEP09(2018)095}{\emph{JHEP} {\bfseries 09}
  (2018) 095}, [\href{https://arxiv.org/abs/1806.04877}{{\ttfamily
  1806.04877}}].

\bibitem{DeWolfe:1999cp}
O.~DeWolfe, D.~Z. Freedman, S.~S. Gubser and A.~Karch, \emph{{Modeling the
  fifth-dimension with scalars and gravity}},
  \href{http://dx.doi.org/10.1103/PhysRevD.62.046008}{\emph{Phys. Rev.}
  {\bfseries D62} (2000) 046008},
  [\href{https://arxiv.org/abs/hep-th/9909134}{{\ttfamily hep-th/9909134}}].

\bibitem{Csaki:2000zn}
Csaki, M.~L. Graesser and G.~D. Kribs, \emph{{Radion dynamics and electroweak
  physics}}, \href{http://dx.doi.org/10.1103/PhysRevD.63.065002}{\emph{Phys.
  Rev.} {\bfseries D63} (2001) 065002},
  [\href{https://arxiv.org/abs/hep-th/0008151}{{\ttfamily hep-th/0008151}}].

\bibitem{Csaki:2004ay}
C.~Csaki, \emph{{TASI lectures on extra dimensions and branes}},  in
  \emph{{From fields to strings: Circumnavigating theoretical physics. Ian
  Kogan memorial collection (3 volume set)}}, pp.~605--698, 2004.
\newblock \href{https://arxiv.org/abs/hep-ph/0404096}{{\ttfamily
  hep-ph/0404096}}.

\bibitem{Mohapatra:1974hk}
R.~N. Mohapatra and J.~C. Pati, \emph{{Left-Right Gauge Symmetry and an
  Isoconjugate Model of CP Violation}},
  \href{http://dx.doi.org/10.1103/PhysRevD.11.566}{\emph{Phys. Rev.} {\bfseries
  D11} (1975) 566--571}.

\bibitem{Mohapatra:1974gc}
R.~N. Mohapatra and J.~C. Pati, \emph{{A Natural Left-Right Symmetry}},
  \href{http://dx.doi.org/10.1103/PhysRevD.11.2558}{\emph{Phys. Rev.}
  {\bfseries D11} (1975) 2558}.

\bibitem{Senjanovic:1975rk}
G.~Senjanovic and R.~N. Mohapatra, \emph{{Exact Left-Right Symmetry and
  Spontaneous Violation of Parity}},
  \href{http://dx.doi.org/10.1103/PhysRevD.12.1502}{\emph{Phys. Rev.}
  {\bfseries D12} (1975) 1502}.

\bibitem{Carena:2018cow}
M.~Carena, E.~Meg\'{\i}as, M.~Quir\'os and C.~Wagner, \emph{{$
  {R}_{D^{\left(*\right)}} $ in custodial warped space}},
  \href{http://dx.doi.org/10.1007/JHEP12(2018)043}{\emph{JHEP} {\bfseries 12}
  (2018) 043}, [\href{https://arxiv.org/abs/1809.01107}{{\ttfamily
  1809.01107}}].

\bibitem{Wolkanowski:2013qca}
T.~Wolkanowski, \emph{{Resonances and poles in the second Riemann sheet}},
  \href{https://arxiv.org/abs/1303.4657}{{\ttfamily 1303.4657}}.

\bibitem{Son:2002sd}
D.~T. Son and A.~O. Starinets, \emph{{Minkowski space correlators in AdS/CFT
  correspondence: Recipe and applications}},
  \href{http://dx.doi.org/10.1088/1126-6708/2002/09/042}{\emph{JHEP} {\bfseries
  09} (2002) 042}, [\href{https://arxiv.org/abs/hep-th/0205051}{{\ttfamily
  hep-th/0205051}}].

\bibitem{Costantino:2020vdu}
A.~Costantino and S.~Fichet, \emph{{Opacity from Loops in AdS}},
  \href{http://dx.doi.org/10.1007/JHEP02(2021)089}{\emph{JHEP} {\bfseries 02}
  (2021) 089}, [\href{https://arxiv.org/abs/2011.06603}{{\ttfamily
  2011.06603}}].

\bibitem{Landshoff:1963}
P.~V. Landshoff, \emph{{Poles and Thresholds and Unstable Particles}},
  \href{http://dx.doi.org/https://link.springer.com/article/10.1007/BF02806056}{\emph{Il
  Nuovo Cimento} {\bfseries 28} (1963) 123--131}.

\bibitem{Megias:prep}
E.~Meg\'{\i}as, M.~P\'erez-Victoria and M.~Quir\'os{\emph{, work in progress}
  (2021) }.

\bibitem{Peskin:1991sw}
M.~E. Peskin and T.~Takeuchi, \emph{{Estimation of oblique electroweak
  corrections}}, \href{http://dx.doi.org/10.1103/PhysRevD.46.381}{\emph{Phys.
  Rev.} {\bfseries D46} (1992) 381--409}.

\bibitem{Zyla:2020zbs}
{\scshape Particle Data Group} collaboration, P.~A. Zyla et~al., \emph{{Review
  of Particle Physics}},
  \href{http://dx.doi.org/10.1093/ptep/ptaa104}{\emph{PTEP} {\bfseries 2020}
  (2020) 083C01}.

\bibitem{Chaffey:2021tmj}
I.~Chaffey, S.~Fichet and P.~Tanedo, \emph{{Continuum-Mediated Self-Interacting
  Dark Matter}}, \href{http://dx.doi.org/10.1007/JHEP06(2021)008}{\emph{JHEP}
  {\bfseries 06} (2021) 008},
  [\href{https://arxiv.org/abs/2102.05674}{{\ttfamily 2102.05674}}].

\end{thebibliography}\endgroup

\end{document}